\documentstyle[epsfig,natbib2,natbibmnfix]{mn}

\newcommand{\be}{\begin{equation}}
\newcommand{\ee}{\end{equation}}
\newcommand{\bea}{\begin{eqnarray}}
\newcommand{\eea}{\end{eqnarray}}
\newcommand{\bc}{\begin{center}}
\newcommand{\ec}{\end{center}}

\renewcommand{\vec}[1]{ {\bmath #1} } 

\newcommand{\dd}{{\rm d}}

\title{Cosmological SPH simulations: A hybrid multi-phase model for
star formation}

\author[V.~Springel and L.~Hernquist]{\parbox{18cm}{Volker
Springel$^{1}$\footnotemark[1] and Lars
Hernquist$^2$\footnotemark[3]}\vspace{0.3cm}\\ $^1$Max-Planck-Institut
f\"{u}r Astrophysik, Karl-Schwarzschild-Stra\ss{}e 1, 85740 Garching
bei M\"{u}nchen, Germany\\ $^2$Harvard-Smithsonian Center for
Astrophysics, 60 Garden Street, Cambridge, MA 02138, USA}

\begin{document}

\maketitle
\begin{abstract}

We present a model for star formation and supernova feedback that
describes the multi-phase structure of star forming gas on scales that
are typically not resolved in cosmological simulations.  Our approach
includes radiative heating and cooling, the growth of cold clouds
embedded in an ambient hot medium, star formation in these clouds,
feedback from supernovae in the form of thermal heating and cloud
evaporation, galactic winds and outflows, and metal enrichment.
Implemented using smoothed particle hydrodynamics (SPH), our scheme is
a significantly modified and extended version of the grid-based method
of Yepes et al. (1997), and enables us to achieve high dynamic range
in simulations of structure formation.

We discuss properties of the feedback model in detail and show that it
predicts a self-regulated, quiescent mode of star formation, which, in
particular, stabilises the star forming gaseous layers of disk
galaxies.  The parameterisation of this mode can be reduced to a
single free quantity which determines the overall timescale for star
formation.  We fix this parameter numerically to match the observed
rates of star formation in local disk galaxies.  When normalised in
this manner, cosmological simulations employing our model nevertheless
overproduce the observed cosmic abundance of stellar material.  We are
thus motivated to extend our feedback model to include galactic winds
associated with star formation.

Using small-scale simulations of individual star-forming disk
galaxies, we show that these winds produce either galactic fountains
or outflows, depending on the depth of the gravitational potential.
In low-mass haloes, winds can greatly suppress the overall efficiency
of star formation.  When incorporated into cosmological simulations,
our combined model for star formation and winds predicts a cosmic star
formation density that is consistent with observations, provided that
the winds are sufficiently energetic.  Moreover, outflows from
galaxies in these simulations drive chemical enrichment of the
intergalactic medium, in principle accounting for the presence of
metals in the Lyman alpha forest.

\end{abstract}

\section{Introduction}

\renewcommand{\thefootnote}{\fnsymbol{footnote}}
\footnotetext[1]{E-mail: volker@mpa-garching.mpg.de}
\footnotetext[3]{\hspace{0.03cm}E-mail: lars@cfa.harvard.edu}%

The past two decades have witnessed the emergence of a successful
class of theoretical models to explain the formation of structure in
the Universe.  In these ``hierarchical'' scenarios, cold dark matter
(CDM) forms the dominant mass component, and structure grows via
gravitational instability from primordial perturbations seeded by
inflation.  Galaxy formation proceeds ``bottom-up'' in this picture,
with smaller galaxies forming first, and larger objects being
assembled through merging of less massive building blocks.

Numerical simulations have become an important tool for exploring the
detailed predictions of these models.  For example, the clustering
properties of matter on large scales, where gasdynamical forces are
negligible compared to gravity, can now be computed with high
precision.  However, extending these studies to understand the
formation of luminous components of galaxies is considerably more
difficult.  Cosmological simulations that attempt to include the
essential physical processes associated with galaxy formation, namely
radiative cooling and star formation, invariably encounter formidable
difficulties.  For this reason, the most popular approach to date for
describing hierarchical galaxy formation has been the so-called
``semi-analytic'' technique
\citep[e.g.][]{Wh91,Lac91,Kau93a,Kau94,Col94}, in which highly
simplified prescriptions for the baryonic physics are combined with
Monte-Carlo methods to construct merging history trees.  In recent
times, these Monte-Carlo trees have been replaced with merging
histories obtained directly from high-resolution N-body simulations
\citep{Kau99,Kau99b,Spr99b}.

Some impressive successes in direct numerical simulation of
cosmological hydrodynamics have been obtained since this method was
pioneered more than a decade ago
\citep{Ev88,Ev90,He89b,He89,Nav91,Ba91,Ka91,Hi91,Th92,KaHeWe92,Ce93}.
For example, these studies improved our understanding of quasar
absorption line systems \citep[e.g.][]{CMOR94,Zh95,He96}, the
intergalactic medium
\citep[e.g.][]{Ce93,Ce00,Ka96,Kat99,We97,Wein2000,St95,Bry94,Bry98,Bl99,
Pea99,Pea2000}, and galaxy formation
\citep[e.g.][]{Ka91,Nav94,St95,Mi96, Wa96,Na97,St99}.

However, the numerical modelling of key processes related to star
formation and associated feedback by supernovae and stellar winds has
remained problematic.  Because radiative cooling is particularly
efficient in dense, low-mass haloes at high redshift, simulations are
confronted with an overcooling problem, which is typically manifested
through an overproduction of stars.  Semi-analytic treatments indicate
that a physical handling of feedback processes is required to regulate
star formation and reduce the fraction of baryons that collapse in the
centres of haloes.  Unfortunately, it is not clear how best to
incorporate feedback processes into simulations of galaxy formation,
although they appear to be crucially important to a successful
description of hierarchical CDM universes.

Here, we present a model for star formation and feedback that goes
beyond a purely phenomenological description.  In most attempts to
simulate galaxy formation, the gas within galaxies is treated as a
single phase medium and is converted into stars on some characteristic
timescale, which is usually related to the local dynamical time.
Often, auxiliary criteria requiring that the flow be locally
converging or Jeans unstable are also invoked to restrict the
eligibility of gas fluid elements for star formation.  Based on an
assumed stellar initial mass function (IMF), it is then
straightforward to work out the expected rate of energy input due to
associated supernovae.  However, it is not clear how this ``feedback
energy'' should be deposited into the gas.  Simply augmenting the
thermal reservoir of the gas has little effect, because the cooling
time in the dense single-phase medium in these simulations is so short
that the energy is promptly radiated away.  Consequently, pure thermal
feedback in this manner fails to solve the overcooling problem.

Other approaches have been attempted to prevent this rapid loss of
feedback energy, thereby providing a back-reaction on star formation
\citep[see][for a recent comparison of the most common of these
methods]{Kay01}.  For example, \citet{Th2000} temporarily lower the
densities of SPH particles receiving feedback energy to prevent
immediate energy loss.  Alternatively, such particles have been
treated adiabatically for a brief period of time after acquiring
feedback energy \citep{Ge97}.  Although crude, these methods show some
success in regulating star formation.  For example, \citet{Th2001}
have recently found that the angular momentum problem in the formation
of disk galaxies can be alleviated through the combined effects of a
smoothed deposition of feedback energy with a delay of cooling by 30
Myr.

Another method for depositing the energy is through kinetic feedback,
where radial momentum kicks are imparted to particles surrounding
sites of star formation \citep[e.g.][]{Na93,Mi94}, but the fraction of
feedback energy that must be put into this channel depends on
numerical details of the simulations.  Also, the size of the affected
regions depends on spatial resolution, and is in general much larger
than the physical scale of individual supernova remnants.

\citet{SomLar99} experimented with yet more extreme versions of
feedback in order to investigate the effects of galactic outflows. For
example, in their ``blow-out'' simulations, they identified all dense,
cold gas at $z=2.4$ and instantly removed it and redistributed it
uniformly in spherical shells around the centres of the gas
clumps. While this is clearly a highly artificial model, it
demonstrated that outflows can have a large impact on galaxy
formation.  In particular, the loss of angular momentum by the gaseous
component was greatly reduced in their simulations, alleviating the
apparent discrepancy between observed and simulated values of the
specific angular momentum of galactic disks. \citet{Scan01c} used a
similar gas-removal technique to model galactic winds at high
redshift.

A disadvantage common to many of these recipes is their explicit or
implicit dependence on numerical details of the simulations.  For
example, if a criterion for star formation explicitly involves the
mass of a particle, it is unclear in what sense numerical convergence
for the star formation rate of a specific object can be expected if
the simulation is repeated with a different mass resolution.  Instead,
the free parameters describing star formation may need to be
recalibrated whenever the properties of the simulation are altered.
\cite{Spr99} tried to overcome this problem by developing a feedback
method based on an effective model for the star-forming interstellar
medium (ISM), where turbulent motions on unresolved scales were
postulated in an ad-hoc fashion to provide pressure support for the
gas, thereby regulating star formation.  In essence, feedback energy
was only slowly thermalised in this method, thereby delaying the
dissipation of feedback energy in a similar way to some of the models
noted above.

Because the Lagrangian nature of smoothed particle hydrodynamics makes
it relatively easy to obtain the high spatial dynamic range needed for
simulations of galaxy formation, it has so far been the most popular
numerical technique for studying this problem.  However, star
formation and feedback have also been incorporated into hydrodynamical
mesh codes \citep[e.g.][]{Ce93,Gne98}, but their limited spatial
resolution is a serious concern, because star formation occurs at
substantially higher overdensities than can be resolved with a fixed
grid in cosmological volumes.  This is quite different when
high-resolution Eulerian codes are applied to simulations of isolated
galaxies, as in the work of \citet{Mac99}, who carried out detailed
studies of starbursts in dwarf galaxies.  Eulerian codes have also
been employed to directly resolve the multi-phase structure of the
ISM, yet state-of-the-art simulations are either restricted to two
dimensions \citep{Wa99}, or can presently only resolve regions of size
$\sim$100 pc or so in three dimensions \citep{Wada01}.

\citet{Ye97} adopted an intermediate approach by developing a sub-grid
model of a two-phase ISM and implementing it in an Eulerian
code. Their spatial resolution was consequently relatively poor,
despite restricting themselves to rather small cosmological volumes,
but together with subsequent work \citep{Eli99b,Eli99a,Asc02} they
were able to demonstrate a number of interesting properties of this
scheme, which is physically better motivated than most of the models
currently employed in cosmological SPH simulations.  Because of this,
we will base our analysis in part on their approach and on a first
implementation of it in SPH by \citet{Hu99}, yet we will significantly
extend and modify the model.  Our specific aim is to formulate a
description of feedback which is physically motivated, numerically
well specified, and suitable for simulating galaxy formation within
large cosmological volumes.  In what follows, we show that our model
leads to self-regulation of star formation, and describe how it can be
normalised by analysing its properties.  Under the assumption that the
normalisation obtained from observations of disk galaxies in the
present Universe holds at all redshifts, we find that the resulting
cosmic star history leads to an overprediction of the luminosity
density of the universe.  We then argue that galactic winds are a
plausible mechanism for eliminating this discrepancy, and they
simultaneously help to account for other observational data, like the
metals found in the low-density intergalactic medium (IGM), and
describe a phenomenological strategy for including these winds in
cosmological simulations.

This paper is organised as follows. In Section 2, we summarise our new
hybrid multi-phase model for quiescent star formation in detail, and
analyse some of its basic properties.  In Section 3, we discuss
physical and observational constraints on the values of the model
parameters.  In Section 4, we extend the model by including galactic
winds.  We then describe our numerical implementation of the method in
a parallel TreeSPH code for cosmic structure formation in Section 5.
We present numerical results for the formation of disk galaxies in
isolation, and for small simulations of cosmic structure formation in
Section~6.  Finally, we give conclusions in Section~7.

\section{A multi-phase model for quiescent star formation}

We term our treatment of star formation and feedback a ``hybrid''
method because it does not attempt to explicitly resolve the spatial
multi-phase structure of the ISM on small scales, but rather makes the
assumption that crucial aspects of the global dynamical behaviour of
the ISM can be characterised by an effective ``sub-resolution'' model
that uses only spatially averaged properties to describe the medium.
In essence, by adopting a {\it statistical} formulation, we seek to
account for the impact of unresolved physics on scales that are
resolved.

In our hybrid approach, each SPH fluid element represents a region in
the ISM, whose properties are obtained from a suitable coarse-graining
procedure.  This is analogous to the N-body representation of
collisionless fluids, but here the structure of the unresolved matter
is more complex.  In particular, we picture the medium as a fluid
comprised of condensed clouds in pressure equilibrium with an ambient
hot gas.  The clouds supply the material available for star formation.
For these hybrid-particles, the equations of hydrodynamics are only
followed for the ambient gas. The cold clouds are subject to gravity,
add inertia, and participate in mass and energy exchange processes
with the ambient gas phase.  These processes are computed on a
particle by particle basis in terms of simple differential equations
using a specific model for the physics of the ISM.  Here, we attempt
to incorporate some of the key aspects of the theoretical picture of
the ISM outlined by \citet{McKee77}.

In the following, $\rho_h$ denotes the local density of the hot
ambient gas, $\rho_c$ is the density of cold clouds, $\rho_\star$ 
the density of stars, and $\rho=\rho_h+\rho_c$ is the total gas density.
Individual molecular clouds and stars cannot be resolved, thus
$\rho_c$ and $\rho_\star$ represent averages over small regions of the
ISM.  We implicitly assume that such a coarse-graining procedure has
been carried out, and we will formulate the interactions between the
phases assuming that the regions used to define the averages are of
constant volume.  In our model, the average thermal energy per unit
volume of the gas can then be written as $\epsilon = \rho_h u_h +
\rho_c u_c$, where $u_h$ and $u_c$ are the energy per unit mass of the
hot and cold components, respectively.

We model three basic processes that drive mass exchange between the
phases.  These are star formation, cloud evaporation due to
supernovae, and cloud growth due to cooling.  We discuss these
processes in turn, focusing first on self-regulated, ``quiescent''
star formation.  Later, we will extend the model to include additional
processes leading to the development of galactic winds.  For
simplicity of presentation, we will omit adiabatic terms in the
following definition of the model.

We assume that star formation converts cold clouds into stars on a
characteristic timescale $t_\star$, and that a mass
fraction $\beta$ of
these stars are short-lived and instantly die as supernovae.  This can
be described by: \be \frac{\dd\rho_\star}{\dd t} =
\frac{\rho_c}{t_\star} - \beta\frac{\rho_c}{t_\star} =
(1-\beta)\frac{\rho_c}{t_\star} .  \ee In principal, there is a time
delay between star formation events and associated supernovae equal to
the approximate lifetime of massive stars ($\sim 3\times 10^{7}$
years).  However, for the quiescent mode of star formation,
self-regulation is established so quickly that this time delay can be
neglected.

Star formation therefore depletes the reservoir of cold clouds at the
rate ${\rho_c}/{t_\star}$, and leads to an increase in the mass of the
ambient phase as $\beta{\rho_c}/{t_\star}$, because we assume that
ejecta from supernovae are returned as hot gas.  The parameter $\beta$
is the mass fraction of massive stars ($> 8 M_{\odot}$) formed for
each initial population of stars and hence depends on the adopted
stellar initial mass function (IMF). For a Salpeter IMF
(\citeyear{Sal55}) with slope $-1.35$ and upper and lower limits of
$40\,{\rm M}_\odot$ and $0.1\,{\rm M}_\odot$, respectively, it has the
value $\beta=0.106$.  We will typically adopt $\beta=0.1$, and note
that our results are not particularly sensitive to this choice.

In addition to returning gas (enriched with metals) to the ambient
phase of the ISM, supernovae also release energy.  The precise amount
of this energy depends on the IMF.  For the canonical value of
$10^{51} \;{\rm ergs}$ per supernova, we expect an average return of
$\epsilon_{\rm SN}=4\times 10^{48}\;{\rm ergs}\, {\rm M}_{\odot}^{-1}$
for each solar mass in stars formed for the IMF adopted here.  The
heating rate due to supernovae is hence \be \left.\frac{\dd}{\dd t}
(\rho_h u_h) \right|_{\rm SN} =\epsilon_{\rm SN} \frac{\dd
\rho_\star}{\dd t} = \beta u_{\rm SN} \frac{\rho_c}{t_\star} , \ee
where $u_{\rm SN} \equiv (1-\beta) \beta^{-1} \epsilon_{\rm SN}$ may
be expressed in terms of an equivalent ``supernova temperature''
$T_{\rm SN}= 2 \mu u_{\rm SN}/(3k) \simeq 10^{8}\; {\rm K}$.

In our formulation, we assume that the feedback energy from supernovae
directly heats the ambient hot phase.  In addition, we suppose that
cold clouds are evaporated inside the hot bubbles of exploding
supernovae, essentially by thermal conduction, thereby returning
material from condensed clouds to the ambient gas.  We take the total
mass of clouds that are evaporated to be proportional to the mass in
supernovae themselves, viz.  \be \left.\frac{\dd\rho_c}{\dd
t}\right|_{\rm EV} = A\beta\frac{\rho_c}{t_\star}. \ee The efficiency
$A$ of the evaporation process is expected to be a function of the
local environment. For simplicity, we will only take the expected
theoretical dependence on density, $A\propto \rho^{-4/5}$, into
account \citep{McKee77}.  We will not try to estimate the evaporation
efficiency from first principles, but rather treat its normalisation
as a parameter.  As we argue below, the range of permissible values
for $A$ is limited once we require a plausible temperature structure
for the ISM.

Finally, we invoke a process by which cold clouds come into existence
and grow.  We assume that a thermal instability operates in the region
of coexistence between the cold and hot phases, leading to mass
exchange between ambient gas and cold clouds
\citep{Fie65,Fie69,Beg90}.  In particular, we argue that radiative
energy loss by the hot gas leads to a corresponding growth of the
clouds, i.e.~the energy radiated cools gas from the temperature of the
hot phase to that of the cold phase and thus gives rise to a growth in
the mass of cold clouds. This mass flux is thus: \be
\left.\frac{\dd\rho_c}{\dd t}\right|_{\rm TI} =
-\left.\frac{\dd\rho_h}{\dd t}\right|_{\rm TI} = \frac{1}{u_h -
u_c}\Lambda_{\rm net}(\rho_h, u_h).  \ee Note that we assume that the
temperatures of and total volumes occupied by the hot and cold phases
remain constant during cloud growth. We compute the cooling function
$\Lambda_{\rm net}$ from the radiative processes appropriate for a
primordial plasma of hydrogen and helium essentially as described by
\citet{Ka96}.  The abundances of the various ionisation states of H
and He are computed explicitly, under the assumption of ionisation
equilibrium in the presence of an external UV background field, which
we here take to be a modified \citet{Ha96} spectrum with reionisation
taking place at redshift $z\simeq 6$ \citep[for details, see][]{Da99}.

The minimum temperature the gas can reach due to the radiative cooling
processes we include is about $10^4\,{\rm K}$, at which point the gas
becomes neutral, and further cooling would require molecular cooling
or that metals be present.  In fact, real molecular clouds will form
cores that are much colder than $10^4\,{\rm K}$.  However, in our
present statistical formulation we neglect the internal structure of
the clouds.  Their exact thermal energy content will be unimportant in
the current model as long as $u_h\gg u_c$, where we identify $u_h$
with the hot intercloud medium resulting from supernova remnants at
temperatures around $\sim 10^5-10^7\,{\rm K}$.  We will typically
assume a temperature of $T_c\simeq 1000\,{\rm K}$ for the cold clouds, but
note that the results do not depend on this choice for $T_c\ll
10^4\,{\rm K}$.

In the following, we will use a factor $f$ to differentiate between
situations where the thermal instability is assumed to be operating
($f=0$), and regions where it is ineffective and ordinary cooling
takes place ($f=1$).  In general, the gas can be expected to exhibit
thermally unstable behaviour if the cooling rate is a declining
function of temperature.  For a primordial mixture of gas, this is the
case roughly in the range $T\approx 10^5 - 10^6$ K, while for
metal-enriched gas the minimum of the cooling curve shifts to higher
temperatures, up to a few times $10^7$ K.  We thus require that the
temperature of the ambient gas is above $10^5\,{\rm K}$.  For the
onset of the thermal instability itself, we chose a simple density
threshold criterion, i.e. $f=0$ for $\rho>\rho_{\rm th}$, and $f=1$
otherwise.  This is motivated by the observed threshold behaviour of
star formation \citep{Ke89}.  Note that star formation will proceed
only once clouds can form; i.e.~when the gas density exceeds
$\rho_{\rm th}$.

Quantitatively, the rates at which the masses of the hot and cold
phases evolve can be written as: \be \frac{\dd\rho_c}{\dd t} = -
\frac{\rho_c}{t_\star} - A\beta\frac{\rho_c}{t_\star} + \frac{1-f}{u_h
- u_c}\Lambda_{\rm net}(\rho_h, u_h), \ee \be \frac{\dd\rho_h}{\dd t}
= \beta \frac{\rho_c}{t_\star} + A\beta\frac{\rho_c}{t_\star} -
\frac{1-f}{u_h - u_c}\Lambda_{\rm net}(\rho_h, u_h) . \label{eqevhot}
\ee In both of these equations, the first term on the right hand side
describes star formation and feedback, the second cloud evaporation,
and the third the growth of clouds due to radiative cooling of the
gas.

We now consider the energy budget of the gas, which we write as \be
\frac{\dd}{\dd t}\left( \rho_h u_h + \rho_c u_c\right) = -\Lambda_{\rm
net}(\rho_h, u_h) +\beta \frac{\rho_c}{t_\star}u_{\rm SN} -
(1-\beta)\frac{\rho_c}{t_\star}u_c ,
\label{eq1} \ee where $\Lambda_{\rm net}$ is the radiative cooling 
(or heating) function for the ambient hot medium. Only the ambient gas
is assumed to be subject to radiative processes. The second term 
describes the non-gravitational energy injected by exploding supernovae.
Finally, the third term describes the loss of energy in the gaseous
phase due to material that is locked up in collisionless stars.  The
loss term arises because we assume that the material that is converted
into stars is at the temperature of the cold clouds.

Within the framework of our model assumptions, we can split the energy
equation into two separate relations for the thermal budget of the hot
and cold components: 
\be \frac{\dd}{\dd t}\left(\rho_c u_c\right) =
-\frac{\rho_c}{t_\star} u_c - A\beta \frac{\rho_c}{t_\star}u_c +
\frac{(1-f)u_c}{u_h-u_c}\Lambda_{\rm net} \label{eq2}, \ee \be
\frac{\dd}{\dd t}\left(\rho_h u_h\right) = \beta\frac{\rho_c}{t_\star}
(u_{\rm SN} + u_c ) + A\beta \frac{\rho_c}{t_\star}u_c - \frac{u_h- f
u_c}{u_h-u_c}\Lambda_{\rm net} . \label{eq3} \ee In these equations,
the first term on the right hand side describes the effects of star
formation and supernovae, the second accounts for cloud evaporation,
and the third term the impact of thermal instability.

We assume that the cold clouds are at a fixed temperature, and so
$u_c$ will be treated as constant.  Consequently, the above equations
imply that the temperature of the hot phase will evolve according to
\be \rho_h \frac{\dd u_h}{\dd t}=\beta \frac{\rho_c}{t_\star}(u_{\rm
SN} + u_c - u_h) - A\beta \frac{\rho_c}{t_\star}(u_h - u_c) -
f\Lambda_{\rm net} \label{myeq1}.  \ee This is one of the basic
equations that is integrated in the simulation code, augmented of
course by terms coming from the external work done by pressure forces
and viscous effects.  In practice, we follow the evolution of the
entropy of the hot phase in our code, rather than the thermal energy,
for numerical reasons that are discussed by \citet{SprHe01}.

An inspection of equation~(\ref{myeq1}) reveals some interesting
properties of our model.  Suppose that the thermal instability is
operating ($f=0$), in which case the temperature of the hot phase
changes only due to star formation and feedback.  In fact, as long as
star formation is active, the temperature will evolve {\rm towards}
\be u_h= \frac{u_{\rm SN} }{ A+1 } + u_c .\label{myeq2}\ee Deviations
from this temperature decay on a timescale \be \tau_h = \frac{t_\star
\rho_h}{\beta(A+1) \rho_c}. \label{mytau} \ee Thus, provided star
formation is sufficiently rapid compared to adiabatic heating or
cooling due to gas motions, the temperature of the hot phase will be
maintained at the value set by equation (\ref{myeq2}), which is {\em
independent} of the star formation timescale $t_\star$.  Since usually
$A\gg 1$ and $u_{\rm SN}/A \gg u_c$ this temperature is in practice
just given by $u_h\simeq u_{\rm SN} / A$. For a supernova temperature
of $10^8\,{\rm K}$ and $A\sim100$, the self-regulated temperature of
the hot diffuse phase of the ISM will thus be maintained at $\sim
10^6\,{\rm K}$.

\begin{figure}
\bc
\resizebox{8.3cm}{!}{\includegraphics{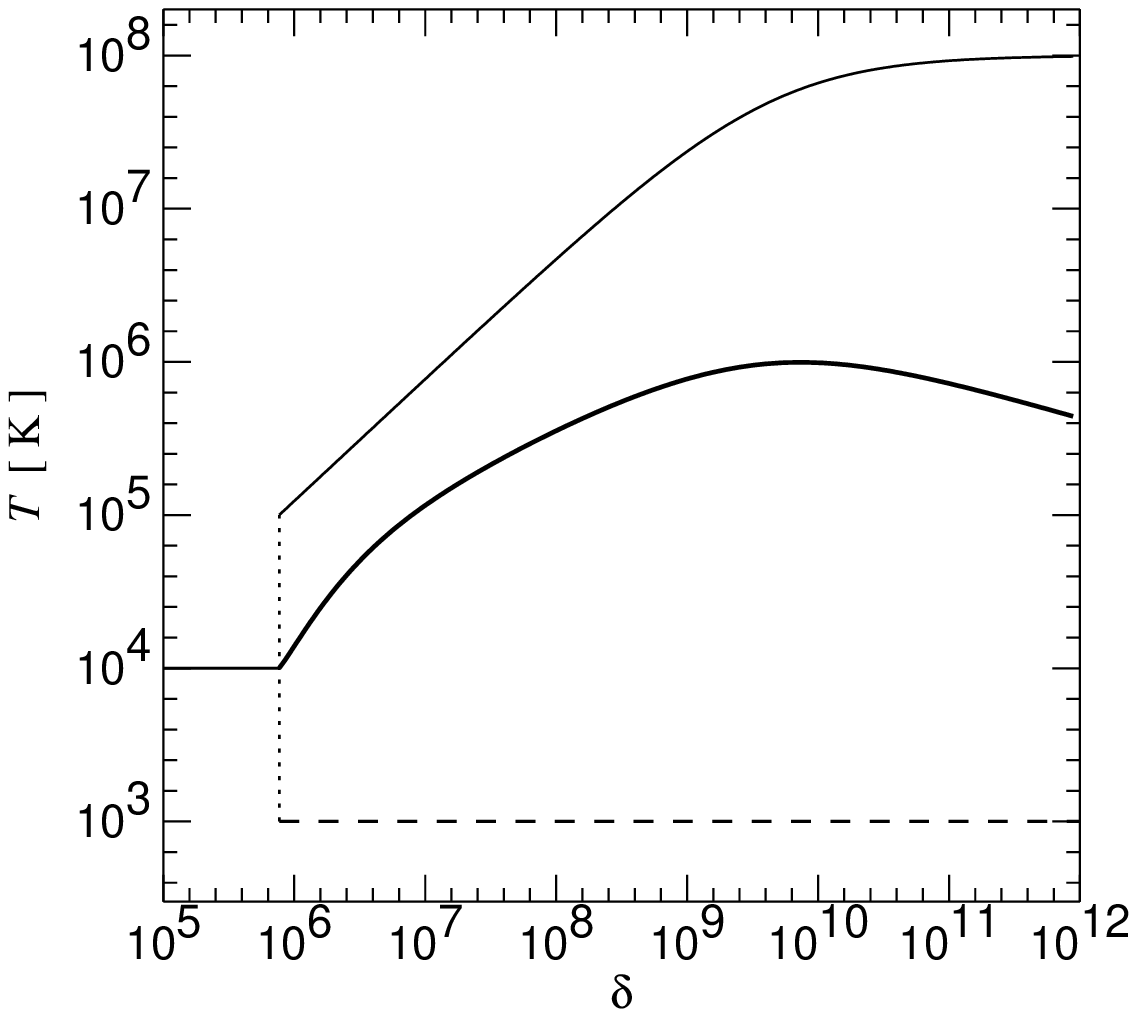}}\\%
\resizebox{8.3cm}{!}{\includegraphics{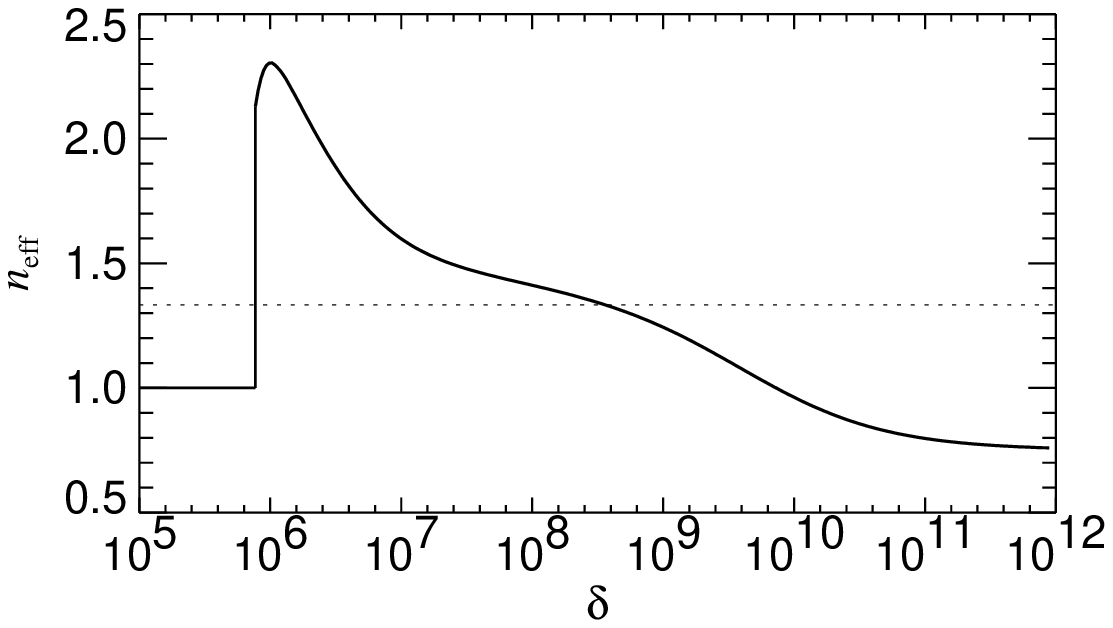}}\\%
\resizebox{8.3cm}{!}{\includegraphics{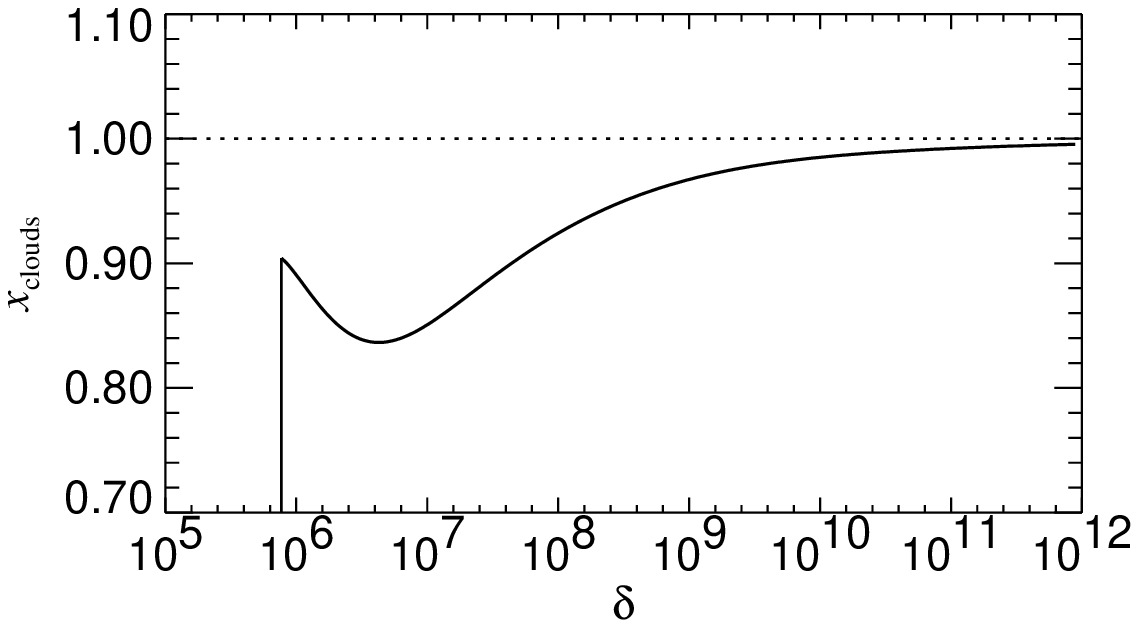}}\\%
\caption{Top panel shows the temperature structure of the multi-phase
medium as a function of baryonic overdensity (assuming $\Omega_b=
0.04$). Below a density of $\delta\simeq 8\times 10^5$, star formation
does not occur and the gas is treated as a single-phase fluid, with a
temperature that is maintained close to $10^4\,{\rm K}$ by cooling and
UV-heating.  When star formation sets in, a hot ambient medium
develops (thin solid line), and cold clouds are present at a fiducial
temperature of $10^3\,{\rm K}$ (dashed). The mass-weighted effective
temperature (thick solid line) is continuous at the onset of star
formation, and can be interpreted in terms of an effective pressure
$P_{\rm eff}$.  In the middle panel, we plot the local polytropic
index $n_{\rm eff}= \dd {\rm log} P_{\rm eff}/ \dd{\rm log} \rho$ of
the resulting effective equation of state. The bottom panel shows the
mass fraction in cold clouds as a function of overdensity.
\label{figeos}}
\ec
\end{figure}

A further interesting consequence of our feedback model is that it
leads to self-regulated star formation.  Owing to evaporation, star
formation reduces the density in cold clouds, lowering the star
formation rate. On the other hand, a higher density of hot gas leads
to an increase in the cooling rate, and hence to more rapid
replenishing of clouds, increasing the star formation rate.  In this
manner, a self-regulated cycle of star formation is established where,
in equilibrium, the growth of clouds is balanced by their evaporation
due to supernova feedback.

This condition can be seen in more detail by considering equation
(\ref{eq1}).  In the self-regulated regime, we expect the effective
pressure of the medium \be P_{\rm eff}=(\gamma -1)(\rho_h u_h + \rho_c
u_c ) \ee to be constant in time. This condition implies \be
\frac{\rho_c}{t_\star} = \frac{\Lambda_{\rm net}(\rho_h, u_h)}{\beta
u_{\rm SN} - (1-\beta)u_c} .  \ee Using $\Lambda_{\rm net}(\rho_h,
u_h)= [\rho_h/ \rho]^2 \Lambda_{\rm net}(\rho, u_h)$, we thus obtain
an expression for the expected ratio of masses in the cold and hot
phases, viz.  \be \frac{\rho_c}{\rho_h} = \frac{\rho_h}{\rho}\, y, \ee
where we have defined \be y\equiv \frac{t_\star \Lambda_{\rm
net}(\rho, u_h)} {\rho \left[\beta u_{\rm SN} - (1-\beta)u_c\right]} .
\label{eqydef} \ee Note that $y$ is a function only of $\rho$ in the
self-regulated regime, provided that $t_\star$ and $A$ depend only on
density. We can then express the mass fraction \be x\equiv
\frac{\rho_c}{\rho} \ee of cold clouds as \be
x=1+\frac{1}{2y}-\sqrt{\frac{1}{y}+\frac{1}{4y^2}} \, .
\label{eqxdef} \ee The effective pressure of the gas will then take on
the value \be P_{\rm eff}=(\gamma -1)\rho \left [ (1-x) u_h + x
u_c\right]\, ,
\label{peff} \ee where the term in square brackets is the effective
mass-weighted ``temperature'' $u_{\rm eff}$ of the medium.  

In Figure~\ref{figeos}, we show the density dependence of this
effective temperature, and of the temperatures of the hot and cold
components in our multi-phase model of the ISM. Also shown is the mass
fraction of gas in cold clouds, and the local logarithmic slope
$n_{\rm eff}$ of the effective equation of state $P_{\rm eff}(\rho)$
of the star-forming medium.  It is clear that the functional
dependence of $P_{\rm eff}$ on density is particularly important for
the dynamical stability of star-forming regions. For $n_{\rm
eff}>4/3$, the effective pressure can provide enough vertical
thickening to stabilise gaseous disks against rapid break-up into
clumps due to dynamical instabilities.

\section{Selecting parameters}

Following \citet{McKee77}, we express the density dependence of the
supernova evaporation parameter $A$ as \be A(\rho)= A_0
\left(\frac{\rho}{\rho_{\rm th}}\right)^{-4/5}, \ee where $\rho_{\rm
th}$ and $A_0$ are parameters of our model.  To specify the star
formation timescale, $t_\star$, we make the common assumption that
this quantity is proportional to the local dynamical time of the gas.
This plausible choice results in a Schmidt-type law for the dependence
of the star formation rate on density, as observed.  We thus set \be
t_\star(\rho)= t^\star_0 \left(\frac{\rho}{\rho_{\rm
th}}\right)^{-1/2}, \ee where $t^\star_0$ is an additional parameter
of the model.

Before we study the properties of our model in detail, we discuss how
its free parameters can be constrained.  To this point, we have
introduced two parameters that depend on the IMF, $\beta$ and $u_{\rm
SN}$, and three parameters that determine the regulation of the
multi-phase medium due to star formation; these are $\rho_{\rm th}$,
$A_0$, and $t^\star_0$.

For the purposes of this work, we neglect uncertainties in the IMF and
treat $\beta$ and $u_{\rm SN}$ as known constants.  However, the other
three parameters are crucially important to the behaviour of the
model.  We proceed by first requiring that at the onset of thermal
instability, the equilibrium temperature of the hot medium is such
that the thermal instability in fact becomes operative.  This means
that the cooling function should start to fall at this temperature,
which happens at $\simeq 10^5 {\rm K}$. We thus require that \be
T_{\rm SN}/A_0 = 10^5\; {\rm K}, \ee which fixes $A_0$ to a value of
$A_0 \simeq 1000$.

Next, we argue that the effective pressure of the gas should be a
continuous function of density at the onset of the regime of
self-regulated star formation. The gas just below the threshold will
have cooled down to $10^4\, {\rm K}$, where it becomes neutral.
Further cooling could happen only due to less efficient molecular
cooling processes that we ignore. (For now, we also neglect cooling
due to metals, which could however become significant once the gas
becomes chemically enriched from star formation.)  Hence we impose the
condition $u_{\rm eff}(\rho_{\rm th})= u_4$, where $u_4$ is the
specific energy corresponding to a temperature of $10^4\, {\rm K}$.
Based on the equations for the self-regulated regime, we thus have \be
\rho_{\rm th}= \frac{x_{\rm th}}{\left(1-x_{\rm th}\right)^2}\,
\frac{\beta u_{\rm SN}-(1-\beta)u_c}{t_\star^0 \Lambda(u_{\rm
SN}/A_0)}, \ee which sets the density threshold $\rho_{\rm th}$ for a
given a value of $t^\star_0$.  Here, $x_{\rm th} = 1 +
(A_0+1)(u_c-u_4)/u_{\rm SN}\simeq 1 - A_0 u_4/u_{\rm SN}$ is the mass
fraction in clouds at the threshold, and $\Lambda$ is defined 
by $\Lambda(\rho,u) = \Lambda_{\rm net}(\rho,u)/\rho^2$, such that
$\Lambda$ looses its dependence on density for temperatures high
compared to $10^4\, {\rm K}$.

We have thus reduced the description of our model to one free
parameter, the star formation timescale $t^\star_0$.  It is clear that
this parameter sets the overall gas consumption timescale, which is,
in principle, well-constrained by observations of the efficiency of
star formation.

Observationally, there appears to be a tight correlation between
disk-averaged measurements of the star formation rate per unit area
and the gas surface density, a ``global Schmidt-law'', given by
\citet{Ke98} as \be \Sigma_{\rm SFR}= (2.5 \pm 0.7) \times 10^{-4}
\left( \frac{\Sigma_{\rm gas}}{ { \rm M_{\odot} pc^{-2}}}\right)^{1.4
\pm 0.15} \frac { {\rm M_{\odot}} }{ {\rm yr\,kpc^2}}
\label{KK}.
\ee This correlation is valid from typical disk-averaged gas densities
of $10 \,{\rm M_{\odot} pc^{-2}}$ in normal spirals to gas densities
as high as $10^5\,{\rm M_{\odot} pc^{-2}} $ in the central regions of
starbursting galaxies.  Roughly the same law also appears to hold
locally for azimuthally averaged gas densities and star formation
rates, although perhaps with a slightly smaller amplitude than the one
given by equation (\ref{KK}).  In addition, there is a clear threshold
behaviour for star formation, with $\Sigma_{\rm SFR}$ very rapidly
falling at densities below $\sim 10\,{\rm M_{\odot} pc^{-2}}$
\citep{Ke89,Ke98,Mar01}. It has been suggested that the threshold
might be related to the onset of gravitational instabilities in the
disk, but presently the evidence for this argument remains
inconclusive.

\begin{figure}
\bc \resizebox{8cm}{!}{\includegraphics{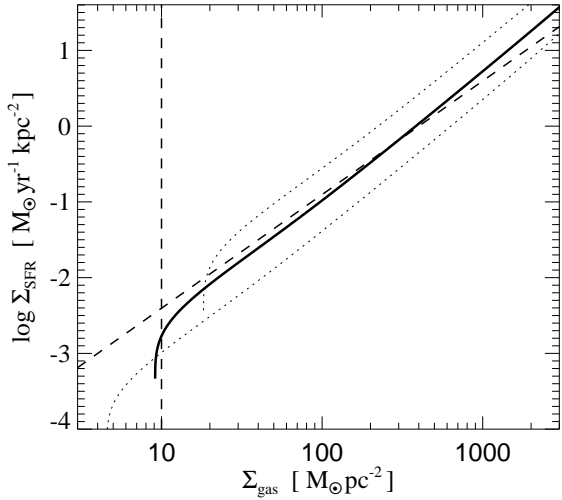}}%
\caption{Star formation rate per unit area versus gas surface density
in a star-forming disk of gas. The dashed inclined line shows the
Kennicutt law of equation~(\ref{KKons}), with the vertical line drawn
to indicate the observational cut-off in the star formation rate.  The
solid and dotted lines have been computed numerically by solving the
equations of hydrodynamic equilibrium together with our multi-phase
model for self-gravitating sheets of gas. The solid line shows the
result for $t^\star_0= 2.1\,{\rm Gyr}$, the lower dotted line is
obtained for $t^\star_0= 8.4,{\rm Gyr}$, and the upper dotted line for
$t^\star_0= 0.53\,{\rm Gyr}$. \label{figkenntheory}} \ec
\end{figure}

By modelling individual galaxies, we will directly try to fit the
Schmidt-law of equation~(\ref{KK}), using however a slope of 1.5,
which prompts us to lower the normalising coefficient in front by a
factor of 2. This correction results when changing the best-fit slope
to 1.5 and simultaneously requiring the star formation rate to remain
unchanged at the intermediate density value of $10^3\, {\rm M_{\odot}
pc^{-2}}$. The resulting local Schmidt-law then provides a good fit to
the azimuthal data of 21 spirals presented in Fig.~3 of \citet{Ke98}.
The observational constraint we thus try to match can also be
expressed as a gas consumption timescale, taking the form \be t_{\rm
SFR} = \frac{\Sigma_{\rm gas}}{\Sigma_{\rm SFR}}= 3.2\,{\rm Gyr}\,
\left(\frac{\Sigma_{\rm gas}}{10\,{\rm
M_{\odot}\,pc^{-2}}}\right)^{-0.5}.
\label{KKons}
\ee At the threshold for the onset of star formation, this thus
indicates a timescale of about 3.2~Gyr, becoming shorter towards
higher densities. This compares well with the cited median gas
consumption timescales of 2.1 Gyr \citep{Ke98} and 2.4 Gyr
\citep{Row99}.

In Figure~\ref{figkenntheory}, we show the relation between star
formation density and gas surface density predicted by our multi-phase
model for various choices of $t^\star_0$. The curves have been
computed by solving the equations of hydrostatic equilibrium for
self-gravitating layers of gas with varying surface density. Clearly,
the amplitude of the star formation rate is very sensitive to the
value of $t^\star_0$, with $t^\star_0= 2.1\,{\rm Gyr}$ providing a
good fit to the Kennicutt law.

Once the amplitude of star formation has been matched by adjusting
$t^\star_0$, the slope and the cut-off obtained can be used as
additional checks on the applicability of our model. The slope is
matched quite well, even though this is not entirely trivial because
it requires that the vertical structure that develops under the action
of $P_{\rm eff}$ for the self-gravitating, star-forming sheet of gas
leads to star formation rates per unit area which are compatible with
the observed slope of the Schmidt-law. Interestingly, the cut-off
induced by the best-fit value of $t^\star_0$ also lies approximately
in the right location.  It is presently unclear whether this has any
profound significance, or whether it is just a fortunate coincidence
in the present simple model. Recall that the cut-off in the model is
induced by an imposed physical density threshold $\rho_{\rm th}$ for
the onset of cloud formation, and that this density is tied to
the value for the star formation timescale.
 
\begin{figure}
\bc
\resizebox{8cm}{!}{\includegraphics{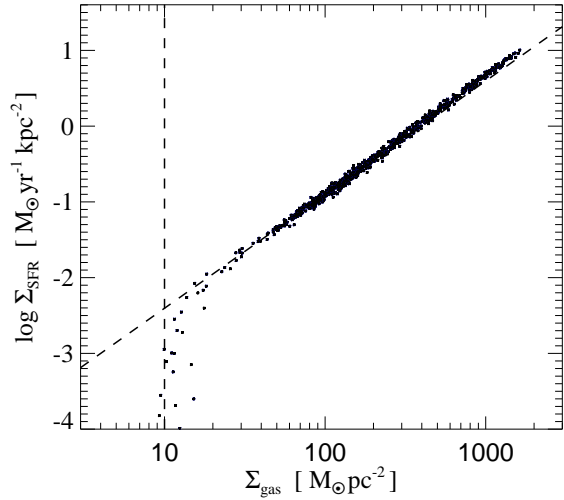}}%
\caption{Star formation rate per unit area versus gas surface density
in a self-consistent simulation of a disk galaxy which quiescently forms
stars.  The symbols show azimuthally averaged measurements obtained
for our fiducial choice of $t^\star_0= 2.1\,{\rm Gyr}$. The dashed
inclined line gives the Kennicutt law of equation (\ref{KKons}), and
the vertical line marks the observed cut-off of star formation.
\label{figkennmeasured}}  \ec
\end{figure}

Finally, we examine how well full three-dimensional simulations of
spiral galaxies obey the Kennicutt law that we used to set the star
formation timescale. In Figure~\ref{figkennmeasured}, we show
azimuthally averaged measurements obtained for our fiducial choice of
$t^\star_0= 2.1\,{\rm Gyr}$ in a compound galaxy consisting of a dark
halo, and a star-forming gaseous disk. There is good agreement with
the corresponding analytic curve in Fig.~\ref{figkenntheory},
validating the numerical implementation of the multi-phase model in
our simulation code.

\section{Winds and starbursts}

\subsection{Winds}

As summarised above, our multi-phase model leads to the
establishment of a physically
motivated and numerically well controlled regulation cycle for star
formation in gas that has cooled and collapsed to high baryonic
overdensities.  Gas contained in dark matter haloes can thus cool and
settle into rotationally supported disks where the baryons are
gradually converted into stars, at a rate consistent with observations
of local disk galaxies.  In this model, the thickness and the star
formation rates of gaseous disks are regulated by supernova feedback,
which essentially provides finite pressure support to the star forming
ISM, thereby preventing it from collapsing gravitationally to
exceedingly high densities, and also allowing gaseous disks to form
that are reasonably stable against axisymmetric perturbations.

However, it is clear that the model we have outlined so far will not
be able to account for the rich phenomenology associated with
starbursts and galactic outflows, which are observed at both low
\citep[e.g.][]{Heck95,Bland95,Lehn96,Dahlem97,Heck00} and high
redshifts \citep[e.g.][]{Frye01,Pett00,Pett01}. This is because our
multi-phase model by itself offers no obvious route for baryons to
climb out of galactic potential wells {\em after} having collapsed
into them due to cooling.  Note that for the hybrid model of quiescent
star formation we explicitly assume that the cold clouds and the hot
surrounding medium remain tightly coupled at all times.  The high
entropy gas of supernova remnants is thus trapped in potential wells
by being tied into a rapid cycle of cloud formation and
evaporation. In principal, tidal stripping of enriched gas in galaxy
interactions \citep{Gne97} could lead to a transport of enriched gas
back into the low-density IGM.  However, high-resolution simulations
of galaxy collisions \citep{Ba88,Ba92b,He92,He93d} have shown that
such dynamical removal of gas from the inner regions of galaxies
appears to be rather inefficient, especially for the deep potential
wells expected for haloes in CDM universes \citep{Sp98}.

On the other hand, it is becoming increasingly clear that galactic
winds and outflows may play a crucial role not only in chemically
enriching and possibly heating the IGM
\citep{Nath97,Mad01,Agui01c,Agui01a,Agui01b}, in polluting the IGM
with dust \citep{Agui99b,Agui99a}, and in enriching the intracluster
and intragroup medium, but may also be an important mechanism in
regulating star formation on galactic scales
\citep{Scan00a,Scan01b,Scan01a}.  Since winds can reheat and transport
collapsed material from the center of a galaxy back to its extended
dark matter halo and even beyond, they can help to reduce the overall
cosmic star formation rate to a level consistent with observational
constraints.  Because radiative cooling is very efficient at high
redshifts and in small haloes \citep{Whi78,Wh91}, numerical
simulations of galaxy formation typically either overproduce stars
compared with the luminosity density of the universe, or harbour too
much cold gas in galaxies.  The self-regulated model we present above
will also suffer from this problem, because it does not drastically
alter the total amount of gas that cools.  It is plausible, however,
that galactic winds may solve this ``overcooling'' problem, provided
that they can expel sufficient quantities of gas from the centres of
low-mass galaxies. Removal of such low-angular momentum material may
also help to resolve the problem of disk sizes being too small in CDM
theories \citep{Na94,Nav00,Bin01}.  Note that semi-analytic models of
galaxy formation must also invoke feedback processes that reheat cold
gas and return to the extended galactic halo or eject it altogether.

We are thus motivated to extend our feedback model to account for
galactic winds driven by star formation.  Winds have been investigated
in a number of theoretical studies \citep[][among
others]{Mac99,Agui01c,Agui01a,Agui01b,Ef2000,Mad01,Scan01c}, but the
mechanism by which galactic outflows originate is not yet well
understood.  In the star-forming multi-phase medium, it is plausible
that not all of the hot gas in supernova remnants will remain confined
to the disk by being quickly used up to evaporate cold
clouds. Instead, supernova bubbles close to the surface of a
star-forming region may break out of a disk and vent their hot gas
into galactic haloes. As a result, a galactic-scale wind associated
with star formation may develop. Note that this process does not
necessarily require a prominent starburst, but could be a common
phenomenon even with quiescent star formation \citep{Ef2000}.  In the
latter case, winds may often not be strong enough to escape from all
but the smallest galaxies, but they can nevertheless represent a
crucial process for transporting metals and energy released by
supernovae into the extended galactic halo.

Unfortunately, our understanding of the small-scale hydrodynamics that
is actually responsible for driving winds is crude, although
high-resolution simulations of starbursts \citep{Mac99} provide some
clues about the physics.  Here, we will not attempt to formulate a
detailed theory for the production of winds, but will rather use a
phenomenological approach that can easily be combined with our model
for star formation and feedback.  To this end, we will parameterise
the winds in analogy to \citet{Agui01a}, who studied wind propagation
and metal enrichment of the IGM using detailed analytic models applied
to a sequence of simulation time slices.  Their phenomenological model
of winds was constructed to be physically and energetically plausible,
to be simple, and to be constrained by observational data to the
extent possible, an approach we will also follow.

We start by assuming that the disk mass-loss rate that goes into a
wind, $\dot M_{\rm w}$, is proportional to the star formation rate
itself, viz.  \be \dot M_{\rm w}=\eta \dot M_\star \, , \ee where
$\eta$ is a coefficient of order unity.  In fact, \citet{Mar99}
presents observational evidence that the disk mass-loss rates of local
galaxies are of order the star formation rates themselves, with $\dot
M_{\rm w} / \dot M_{\star} \sim 1-5$, and without evidence for any
dependence on galactic rotation speed.  Note that $\dot M_{\rm w}$
just describes the rate at which gas is lost from a disk and fed into
a wind.  Whether or not this material will be able to escape from the
halo will then depend on a number of factors: the velocity to which
the gas is accelerated, the amount of intervening and entrained gas,
and the depth of the potential well of the halo.  If the wind is slow
and the halo is of sufficient mass, the gas ejected from the disk will
remain bound to the halo, and may later fall back to the star-forming
region, giving rise to a galactic ``fountain.''  On the other hand,
even a slow wind may escape dwarf galaxies and, at sufficiently high
redshift, can potentially pollute a large volume without overly
perturbing the thermal structure of the low-density IGM.  Note that
for values of $\eta$ above unity, a wind is expected to greatly
suppress star formation in galaxies that are of low enough mass to
allow the wind to escape.  We will adopt $\eta=2$ in our fiducial wind
model, consistent with the observations of \citet{Mar99}.

We further assume that the wind carries a fixed fraction $\chi$ of the
supernova energy. Equating the kinetic energy in the wind with the
energy input by supernovae, \be \frac{1}{2} \dot M_{\rm w}{v_{\rm
w}^2} = \chi \epsilon_{\rm SN} \dot M_\star, \ee we obtain the wind's
velocity when it leaves the disk as \be v_{\rm w}=\sqrt{\frac{2 \beta
\chi u_{\rm SN}}{\eta(1-\beta)}}. \label{eqwind} \ee We will treat
$\chi$ as a further parameter of the wind model.  While it cannot be
larger than $1$ in the present formalism, it is nevertheless to be
expected that the wind energy constitutes a sizable fraction of the
overall supernova energy input. We will assume $\chi=0.25$ in most of
the test simulations discussed in this paper. Note that it should in
principle be possible to use observations of the cosmic star formation
history or the metal enrichment of the IGM to constrain this
parameter.  We do not explicitly model a distribution of wind
velocities in this study.  Of course, some variation of the local wind
speed will be automatically produced by the dynamics of local gas
interactions.

\subsection{Starbursts}

We have shown that our hybrid model leads to an efficient
self-regulation of star formation, which can be understood in terms of
an effective equation of state. In general, the model is thus best
viewed as describing quiescent star formation, where the star
formation rate will in general only gradually accelerate when the gas
density is increased, in accordance with the empirical Schmidt law.
The winds we have introduced as an extension of this model will not
change this picture.  We specifically postulate that a wind leaves a
star-forming region without significantly perturbing it dynamically,
and our numerical implementation of wind formation is designed to
ensure this behaviour.  Winds thus merely reduce the amount of gas
available for star formation.  Depending on the depth of the potential
well, the gas may or may not come back to the galactic disk at a later
time and become available for star formation again.

It is interesting to note, however, that we expect the quiescent mode
of star formation to eventually become ``explosive'' (i.e. very rapid)
for sufficiently high gas densities.  Physically, it is plausible that
self-regulation should break down at high densities.  In the
self-regulated regime, cold clouds are constantly being formed and
evaporated.  If clouds are not replenished by cooling, they will be
consumed on a timescale $t_c= {t_\star}/({\beta A}) =
({\rho}/{\rho_{\rm th}})^{3/10}{t^{\star}_{0}}/({\beta A_0})$. This
timescale $t_c$ describes the rate at which clouds are reprocessed. At
the onset of star formation, it is about a factor 100 shorter than the
star formation timescale itself, on the order of a few times $10^7$
years.  However, at higher densities, clouds are reprocessed more
slowly, and eventually $t_c$ will become larger than the maximum
lifetime of individual clouds, which is estimated to be as high as
$10^8$ yr for giant molecular clouds. At this point, it will become
difficult to maintain tight self-regulation because the clouds will
not survive long enough to await evaporation. Instead they may all
collapse and form stars on the timescale of their lifetime, i.e.~the
timescale of star formation will suddenly become shorter in this
regime and deviate from the scaling we have assumed so far.

Since it is unclear how this transition to accelerated star formation
proceeds in detail, we refrain from modelling it explicitly in this
study.  However, we remark that already the effective pressure model
discussed so far shows a possibility for run-away star formation, for
purely dynamical reasons.  Recall that the ISM is stabilised against
gravitational collapse by the effective pressure provided in the
multi-phase model. For sufficiently high densities, the corresponding
equation of state eventually becomes soft. More specifically, there is
a certain overdensity, where the local polytropic index falls below a
slope of $4/3$. It is well known that barotropic gas spheres with an
index below this value are unstable.  We thus expect that once we
assemble a sufficiently large amount of cold gas, the effective
pressure will be insufficient to stabilise the cloud against further
gravitational collapse to much higher densities, leading to a run-away
conversion of the gas into stars.

\section{Numerical implementation}

The hybrid multi-phase model outlined in Section~2 can be represented
numerically in a number of different ways. \citet{Ye97} used an
Eulerian hydro code, where the properties of the flow are discretised
on a mesh. The mesh size provides a natural scale for the
coarse-graining procedure inherent in our model. One may also
discretise the {\em mass}, as is done in the Lagrangian SPH
technique. \citet{Hu99} presented a first implementation of the
approach of \citet{Ye97} using SPH.

We also employ SPH to implement our model, in the form of a modified
and extended version of the parallel TreeSPH code {\small GADGET}
\citep{SprGadget2000}.  Throughout, we will use the novel formulation
of SPH that was recently derived by \citet{SprHe01}, which manifestly
conserves energy and entropy when appropriate, and greatly reduces
numerical inaccuracies that otherwise can arise in poorly resolved
cooling flows.

On a technical level, we have actually implemented two different
methods for representing our star formation model in the code. In an
``explicit'' treatment, which we will describe first, each SPH
particle can have three different mass components, describing
``ordinary'' ambient gas, cold clouds, and collisionless stars.  The
various processes of cloud formation, cloud evaporation and star
formation are then explicitly followed in terms of mass exchange
between these components of the hybrid particles.

However, because the self-regulation timescale is typically short
compared to the dynamical and star formation timescales, the mass
fraction contained in cold clouds at a given density can be obtained
with good accuracy by just assuming the equilibrium value expected for
self-regulation, where the latter can be worked out analytically.  We
can thus replace the explicit treatment with a simplified method based
on these equilibrium values.  Star particles can then be formed
stochastically from the reservoir of clouds, thereby avoiding an
artificial coupling of gas and collisionless stellar material.  This
idea forms the basis of our second method for implementing the
multi-phase model, to be described after the explicit treatment.

\subsection{Explicit implementation of the multi-phase model}

In our explicit variant for implementing the multi-phase model, we
describe each SPH hybrid-particle by a total mass $m$, a mass
$m_\star$ in stars, a mass $m_c$ in cold clouds, and an internal
energy $u_h$ in the ambient gas. The mass in the ``ordinary'' gas
phase is therefore given by $m_h = m - m_\star - m_c$. Note that we
here build up stellar mass gradually in each star-forming particle,
unlike in the effective treatment discussed in the next section, which
avoids the implied temporary coupling of gas and collisionless stellar
material.  Also note that most SPH particles in a cosmological
simulation will never reach the high densities required for star
formation, and will therefore have $m=m_h$, with $u_h$ describing
their normal thermal reservoir.

We compute the hydrodynamical forces in the usual manner, but take the
pressure to be \be P=(\gamma -1) \rho (m_h u_u + m_c u_c)/(m_h+m_c),
\ee and compute kernel-estimated densities with the gas mass $m -
m_\star$ of each particle only. The stellar masses $m_\star$ add
inertia to the particles and contribute to the source function of the
gravitational field.  Star formation is then treated in the code on a
particle by particle basis in the following manner:
\begin{enumerate}
\item For each timestep $\Delta t$, we first work out the mass of
clouds that turns into stars and produces supernovae: \be \Delta
m_{\rm sf}= m_c \frac{\Delta t}{t_\star}\, .\ee
\item If the thermal instability is active ($f=0$), we then compute
the new mass in the hot phase.  To this end, we consider the
Lagrangian analogue of equation (\ref{eqevhot}): \be \Delta m_h =
\beta(A+1) \Delta m_{\rm sf} - \frac{\Lambda_{\rm net}}{u_h -
u_c}\frac{m_h}{\rho_h}\Delta t , \ee and we solve this equation
implicitly for the new mass $m_h'$ in the following way: \be m_h' =
m_h + \beta(A+1) \Delta m_{\rm sf} - \frac{\Lambda_{\rm net}(\rho_h
m_h'/m_h, u_h)}{u_h - u_c}\frac{m_h}{\rho_h}\Delta t \ee This
guarantees stability even when $m_h$ changes rapidly. In order to
improve the robustness of the integration, we however do not allow
$m_h'$ to fall below half the value of $m_h$ in a single step.  If the
thermal instability is not active, we simply have $m_h' = m_h +
\beta(A+1) \Delta m_{\rm sf}$.
\item Once $m_h'$ is determined, we can compute the mass that is
actually transferred from the hot to the cold phase due to the thermal
instability alone: \be \Delta m_{\rm TI}= m_h +\beta(A+1)\Delta m_{\rm
sf} - m_h' \, .\ee Note that $\Delta m_{\rm TI}=0$ if the thermal
instability is not operating.  The mass evaporated by supernovae is
\be \Delta m_{\rm evap}= \beta A \Delta m_{\rm sf}\, .  \ee
\item The new masses of cold and hot phase can now be obtained as: \be
m_c' = m_c - \Delta m_{\rm sf} + \Delta m_{\rm TI} - \Delta m_{\rm
evap} \ee \be m_h'= m_h + \beta\Delta m_{\rm sf} - \Delta m_{\rm TI} +
\Delta m_{\rm evap} \ee And the new stellar mass is: \be m_\star'=
m_\star + (1-\beta)\Delta m_{\rm sf} \ee Note that the total mass is
conserved, $ m_h' + m_c' + m_\star' = m_h + m_c + m_\star$, by 
construction.
\item If the thermal instability is operating, the rate of change of
internal energy of the ambient medium is given by the normal adiabatic
term plus the contribution \be \frac{\dd u_h}{\dd t}= \frac{\beta
\Delta m_{\rm sf}\left[ u_{\rm SN}+(A+1)(u_c- u_h)\right]}{m_h \Delta
t} \ee from star formation and feedback.  Otherwise, if ordinary
cooling is active, the rate of change of internal energy additionally
gets a contribution from the normal cooling function, with the latter
being solved implicitly for the new temperature at the end of the
timestep \citep{SprGadget2000}.
\end{enumerate}

Note that the coupling of the three phases can cause difficulties in
certain situations. For example, if a hybrid particle is dynamically
stripped from a star forming region and moves to an environment of
lower density, the particle will at first have most of its gas mass
still bound in cold clouds. These clouds will quickly be evaporated by
residual star formation of the particle, provided its density has not
fallen so rapidly that the star formation timescale has become very
long. If this should happen, cloud material may escape to very low
density, and survive there for a long time. In order to prevent this,
we assume that all clouds are evaporated instantly if the density of a
particle falls below $\rho_{\rm th}$, with the required energy
extracted from the reservoir $u_h$. Star formation will thus be
strictly confined to the regions $\rho > \rho_{\rm th}$, and all gas
of lower density is just ``ordinary'' gas.

Somewhat more problematic is the tight coupling of the stellar
material that has formed to the mass that still resides in gas.  This
is in principle unphysical, even though it may be a reasonable
approximation in many cases, because the stars and the star-forming
medium tend to move together, except in highly dynamic situations, as,
for example, in major mergers.  Nevertheless, it is clear that the
explicit treatment described above requires a mechanism that
eventually decouples the formed stellar material from the gas,
otherwise one could obtain at late times large numbers of ``gas''
particles whose masses are dominated by their stellar components.  In
order to achieve such a decoupling, we turn a hybrid particle whose
stellar mass has grown above its gas mass, i.e.~for $m_\star \ge 0.5
m$, into a star particle of mass $m_\star$. The rest of the mass,
consisting of clouds and ambient gas, is distributed kernel-weighted
among the SPH neighbours of the particle that is being converted. We
distribute clouds and ambient mass separately, as well as the thermal
energy in the ambient medium.

This mechanism maintains a roughly constant mass resolution, and it
keeps the total particle number exactly constant. However, it is clear
that subtle dynamical effects due to the temporary coupling of gaseous
and stellar material may remain.  This is one of the reasons why we
prefer a simplified, ``effective'' implementation of the multi-phase
model which we describe next.

\subsection{Simplified treatment assuming self-regulation}

According to equation~(\ref{myeq1}), the ambient hot medium evolves
towards an equilibrium temperature on a timescale given by
equation~(\ref{mytau}), provided that the density is sufficiently high
for star formation to occur.  This time is short compared to the star
formation timescale; i.e.~conditions for self-regulation are achieved
quickly.  In order to simplify the computation of the gasdynamics, it
should, therefore, be a good approximation to assume that the
conditions of self-regulated star formation always hold. In this case,
the temperature of the ambient hot medium is simply given by
equation~(\ref{myeq2}), and the mass fraction in clouds follows from
equations~(\ref{eqydef}) and (\ref{eqxdef}). This then also determines
the star formation rate and the effective pressure (\ref{peff}) of
each particle.  In this way, we avoid treating the mass exchange
processes explicitly, simplifying the code, making it faster, and
reducing its storage requirements.

Additional advantages follow if the build-up of the stellar component
is not described ``smoothly'' as in the explicit approach, but instead
probabilistically with expectation value consistent with the star
formation rate.  The star formation rate of an SPH particle of current
mass $m$ is given by $ \dot M_\star = (1-\beta)\, x\,{m}/{t_\star}$.
Given a timestep $\Delta t$, we spawn a new star particle of mass
$m_\star= m_0/N_g$ once a random number drawn uniformly from the
interval $[0,1]$ falls below \be p_\star= \frac{m}{m_\star} \left
\{1-\exp\left[- \frac{(1-\beta)\,x \Delta
t}{t_\star}\right]\right\}. \ee The initial phase-space variables of
the new particle are copied from the gas particle, which in turn is
reduced in mass by $m_\star$. Here, $m_0$ is the initial gas mass of
each SPH particle, and $N_g$ is an integer which gives the number of
``generations'' of stars each gas particle may form.  If a gas
particle has already spawned $N_g-1$ stars, it will simply be turned
into a star particle should it become the site of another star
formation event.

Note that all the star particles will have identical masses in this
approach.  For choices of $N_g$ in the range between 1 and about 4,
one achieves good mass resolution for the stellar component, while
simultaneously maintaining a reasonably constant mass resolution in
the gaseous phase. For $N_g>1$ the total number of particles increases
due to star formation, but there is a firm upper bound to this number,
and if only a small fraction of the baryons are turned into stars, as
in cosmological simulations, the relative increase is moderate.

More important, there is no artificial dynamical coupling between the
gas particles and the stars in this approach, because all the stellar
mass is contained in independent star particles at all times, unlike
in the explicit treatment, where each component is tied to hybrid
gas-star particles.  Another advantage is that the distribution of
formation times of the star particles directly reflects the underlying
star formation history, without any bias, thereby simplifying the
analysis and interpretation of simulation outputs. The equal masses of
star particles also keep possible mass segregation effects due to
collisional relaxation to a minimum, and in the limit of very large
$N_g$ one naturally approaches the limit of continuous star formation.

\subsection{Metal enrichment}

Stellar winds and supernova explosions of massive stars enrich
surrounding gas with metals (i.e. all elements heavier than helium).
We assume that for each mass element $\Delta M_\star$ locked up in
long-lived stars, the mass in metals returned is $ \Delta M_{\rm met}
= p \Delta M_\star$, where $p$ is the {\em yield}.  Provided that
these metals are well mixed with the local gas of mass $M_g$, and
provided no gas enters or leaves the system (i.e. a closed-box) the
metallicity $Z\equiv M_{\rm met}/M_g$ will increase by $\Delta Z = p
\Delta M_\star/M_g$ due to the formation of the stars.  This already
takes into account that some of the metals in the gas are also lost to
the forming stars.

Here, we assume that each gas element locally behaves as a closed box,
with metals being instantaneously mixed between clouds and ambient
gas.  The metallicity of a patch of the ISM will therefore increase
during one timestep by \be \Delta Z= (1-\beta)\,p\,x\, \frac{\Delta
t}{t_\star}, \label{eqnmet} \ee where $x$ is the local fraction
of gas contained in cold clouds.  For gas below the threshold density
for star formation, we have $x=0$ and the metallicity remains
constant.

It is straightforward to follow the growth of the metallicity of each
gas particle using equation~(\ref{eqnmet}), both for the explicit
treatment of the multi-phase structure as described in section 5.1, or
for the simplified ``effective'' method described in section 5.2.
However, depending on the approach used, it is more problematic to
define the metallicity of the star particles.  In the explicit
treatment, the stellar component is built up gradually, with stars
being formed from gas of varying metallicity.  The independent star
particles that are eventually produced will then have a mass-weighted
average of the metallicities of the gas elements from which they were
formed.  This complicates the interpretation of the metallicity
distribution of these star particles, because it will not directly
reflect the metallicity distribution obtained if star formation were
instead always to immediately generate independent star particles.

In the probabilistic method of generating star particles, this problem
does not arise.  The metallicity of a newly spawned star particle is
just given by the current metallicity of the star-forming gas
particle. In this method, the expectation value of the total mass in
metals found in gas and stars at any given time will be equal to the
yield times the total mass in stars. Note that the effective model
also avoids any metal diffusion, which can occur at a small level in
the explicit treatment where remaining gas and metals are distributed
among neighbouring particles once a gas particle is dissolved, with
its hybrid stellar component being turned into a collisionless
particle.

Although there is no efficient mechanism for metal diffusion in our
model, metals can of course be transported along with the gas flow.
In particular, winds may deposit metals far from where they were
produced, leading, for example, to enrichment of the haloes of
galaxies, or of the IGM.  However, currently we do not yet include
metal-line cooling in our code, hence the metallicity acts
fundamentally only as a tracer variable without having any dynamical
consequences.  In future work, we plan to investigate effects that
arise when this part of the model is refined.

\subsection{Wind formation}

As discussed in Section~4, we assume that star formation can be
accompanied by mass-loss from star-forming regions, giving rise to
winds.  There is strong observational evidence that such winds exist,
but the detailed physics underlying this phenomenon is unclear.  But,
even if there were a firm theoretical basis for understanding these
winds, we currently lack the spatial resolution in our simulations to
directly model the relevant physical processes.  We are thus led to
adopt a phenomenological treatment of how such winds are generated. At
a technical level, we proceed in a similar way to the probabilistic
method for spawning star particles.

During a timestep $\Delta t$, a gas particle is added to the wind if a
uniformly distributed random number out of the interval $[0,1]$ falls
below \be p_w= 1-\exp\left[- \frac{\eta(1-\beta)\,x \Delta
t}{t_\star}\right]. \ee If this is the case, we modify the particle's
velocity $\vec{v}$ according to \be \vec{v}'=\vec{v} + v_w \,\vec{n}\,
, \ee where $v_w$ is the wind velocity given by equation
(\ref{eqwind}). For the unit vector $\vec{n}$, we either select a
random direction on the unit sphere (isotropic wind), or we choose a
random orientation along the direction $\vec{v}\times \nabla \phi$,
where $\phi$ is the gravitational potential (``axial'' wind). If the
latter scheme is chosen, the wind particles are preferentially ejected
along the rotation axis of a spinning object. This mimics the expected
preference of wind ejection to occur in a direction orthogonal to
rotationally flattened, disk-like galaxies.  However, even for an
isotropic wind the presence of a dense disk will lead to a bipolar
wind pattern, simply because the wind can propagate much more easily
in the direction orthogonal to the disk.  Note that because of the
random orientation of the momentum kick along the direction $\vec{n}$,
the expectation values of momentum and angular momentum remain
unchanged. While the momentum is not strictly balanced when a single
particle is put into the wind, it is hence statistically conserved.
Also, the expected average increase of the specific kinetic energy of
a particle that enters the wind is given by $v_w^2/2$, as desired.

\begin{figure}
\bc
\resizebox{8cm}{!}{\includegraphics{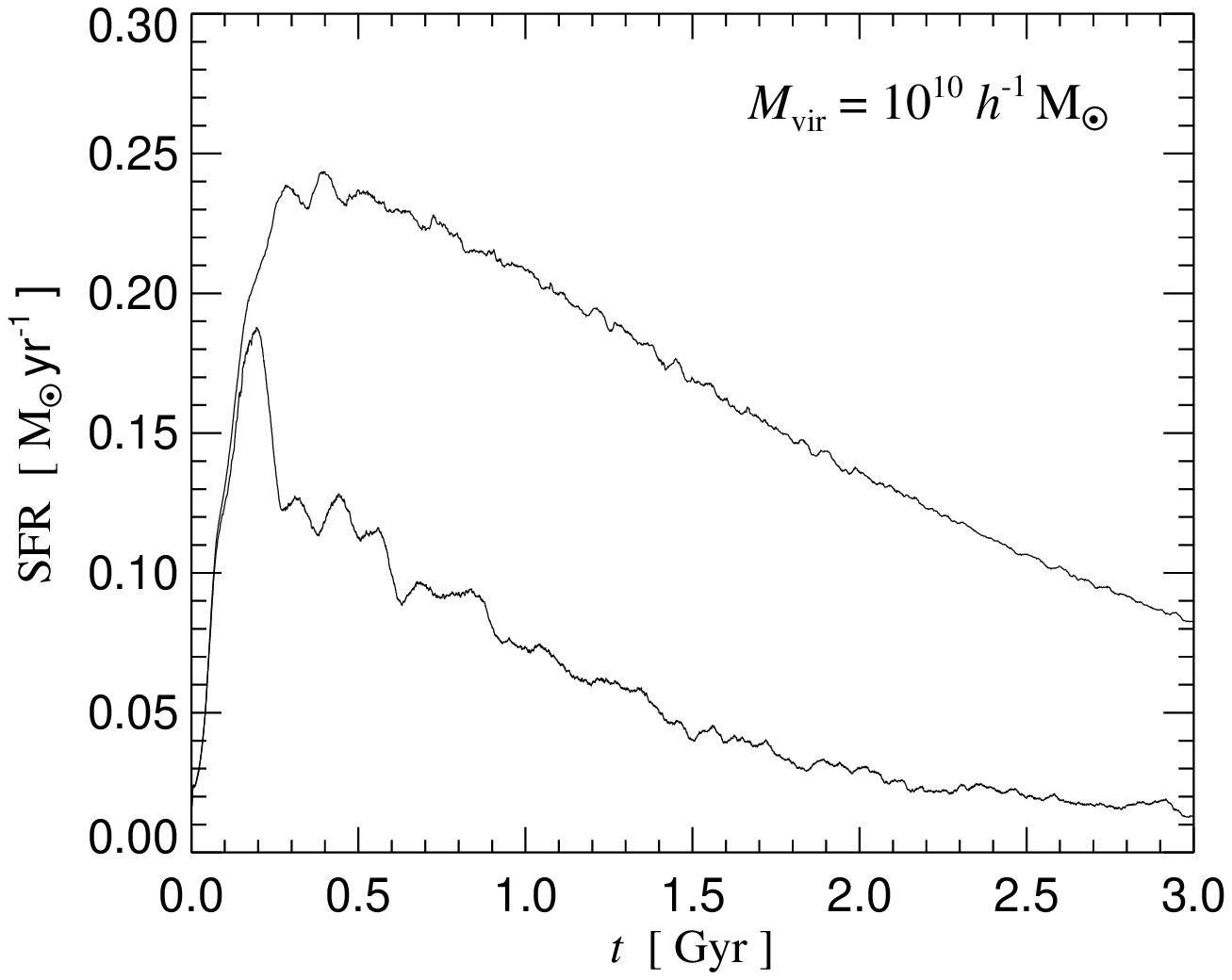}}\\%
\resizebox{8cm}{!}{\includegraphics{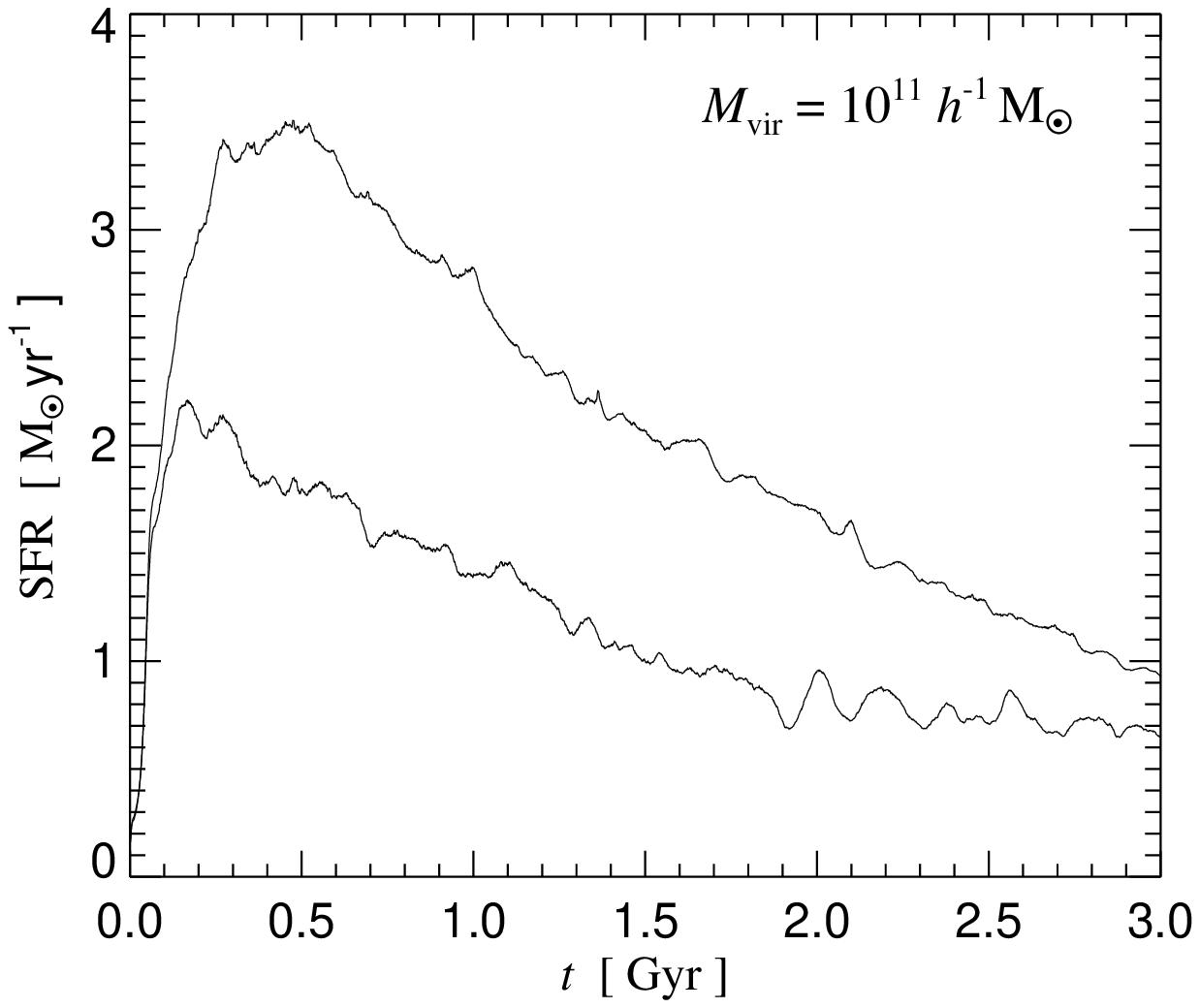}}\\%
\resizebox{8cm}{!}{\includegraphics{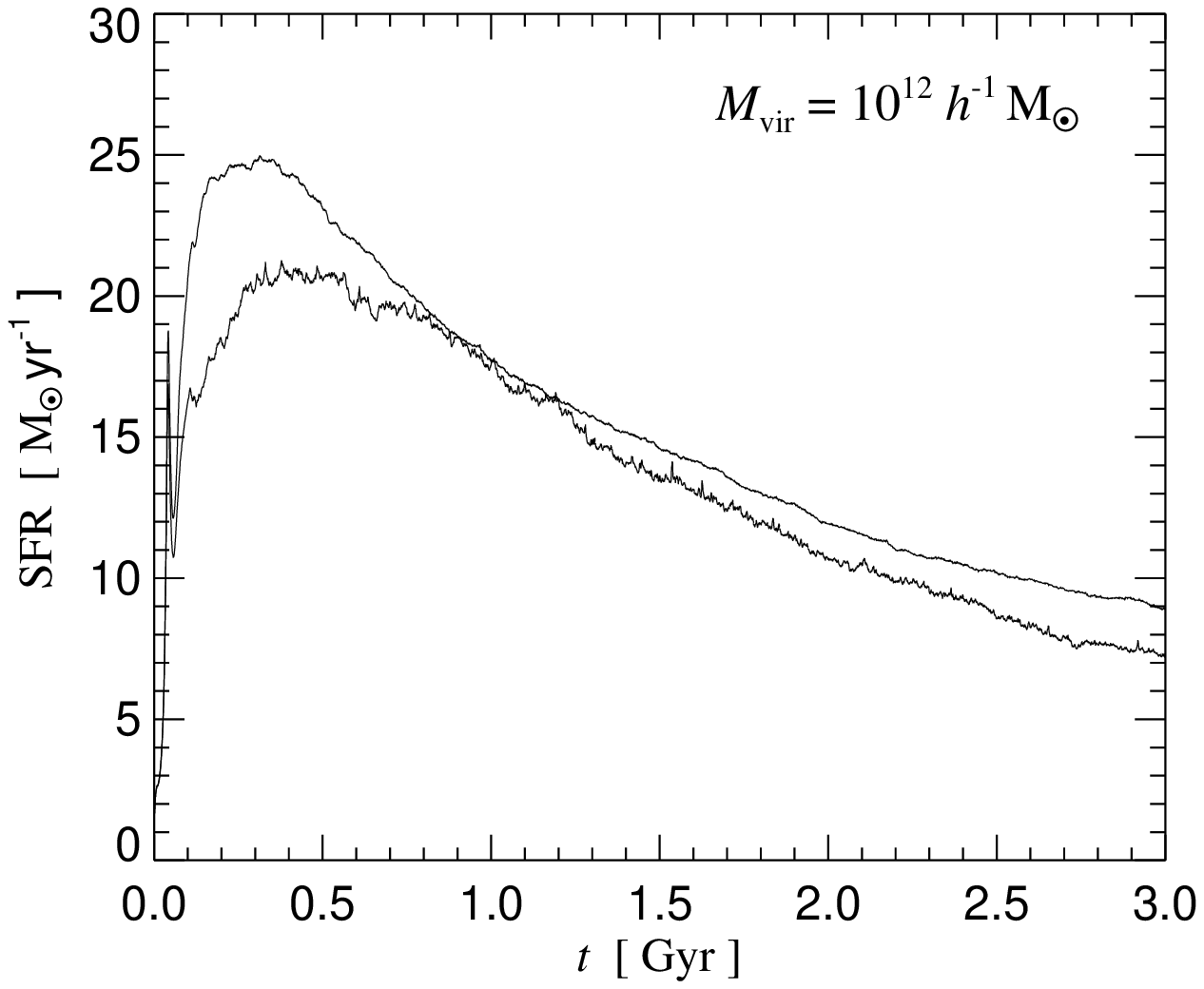}}%
\caption{Star formation rates in simulations of disks growing within
haloes of different masses, with or without the inclusion of a
wind. From top to bottom, the panels show results for haloes of
differing total mass: $M_{\rm vir}= 10^{10} \, h^{-1}{\rm M}_\odot$,
$10^{11} \, h^{-1}{\rm M}_\odot$, and $10^{12} \, h^{-1}{\rm
M}_\odot$, respectively. The inclusion of winds always leads to a net
suppression of star formation and hence to a lower curve, but the
importance of this effect is a strong function of halo
mass.\label{figdisksfr}} \ec
\end{figure}

An interesting problem arises because of the finite thickness of a
star-forming region. A real wind plausibly originates from a region
close to the surface of a star-forming disk, allowing it to leave
without greatly impacting ongoing star formation.  In the simple model
developed here, we do not try to restrict wind formation to a surface
layer. Instead, we allow all the SPH particles in the star-forming
region to enter the wind if chosen by the probabilistic
criterion. However, wind particles from inner parts of star-forming
disks would normally be stopped again by other particles in the dense
disk, resulting in rapid dissipation and thermalisation of the kinetic
energy.  In addition, the momentum input would lead to a strong
perturbation and, in extreme cases, to a disruption of the disk. This
is in disagreement with our assumption that the wind can escape from
the dense, star-forming phase without directly affecting it. Only
outside the disk, would gasdynamical interactions within the halo stop
the wind.  To allow such behaviour, we ``decouple'' a spawned wind
particle for a brief time from hydrodynamic interactions, i.e.~it
neither exerts nor receives hydrodynamic forces during this
period. The particle is however included in the computation of the gas
density, and for gravity. The full hydrodynamic interactions are
enabled again once the density of the particle has fallen to
$0.1\,\rho_{\rm th}$, or once a time of $\Delta t= 50\,{\rm Myr}$ has
elapsed, whichever happens earlier. This allows the wind particle to
travel ``freely'' up to a few kpc until it has left the dense
star-forming phase. In this way, we can mimic quite well the strong,
but quiescent mass-loss from a star-forming region. We have checked
that the strength of the wind and the efficiency of wind escape are
insensitive to the detailed values chosen for the parameters of this
decoupling prescription.

\section{Tests and results}

The feedback model discussed above has widespread implications for
cosmological simulations, influencing nearly all predicted properties
of galaxy populations.  However, it is beyond the scope of this paper
to fully investigate all these consequences in detail.  Rather, we
focus here on the methodology of the modelling technique.  The
simulations we present below were selected as test cases to allow us
to validate our approach, and to give us insight into key effects of
the feedback model in a cosmological context.

More specifically, we focus on two small sets of simulations, one
designed to study the problem of disk formation in a highly idealised
situation, and the other selected as a full-blown, yet rather small
cosmological simulation of structure formation.  Results of modelling
with higher dynamic range will be presented in due course.

\subsection{Isolated spiral galaxies}

Currently, simulations of cosmic structure formation still fail to
match observed properties of spiral galaxies, although some progress
has been achieved recently using improved feedback prescriptions
\citep[e.g.][]{Somm02}.  One cause of the difficulty has been that
collapsed gas loses too much angular momentum to the dark matter, so
that disks are too compact when compared with observations.  On the
other hand, the spin distribution of dark matter haloes is well
understood both analytically and numerically.  Simple analytic models
indicate that disk sizes in such haloes should match the observed
distribution, provided that baryons conserve most of their specific
angular momentum during collapse \citep{Mo98,vdB01b,vdB01a,vdB02}.

\begin{figure*}
\bc
\resizebox{8.0cm}{!}{\includegraphics{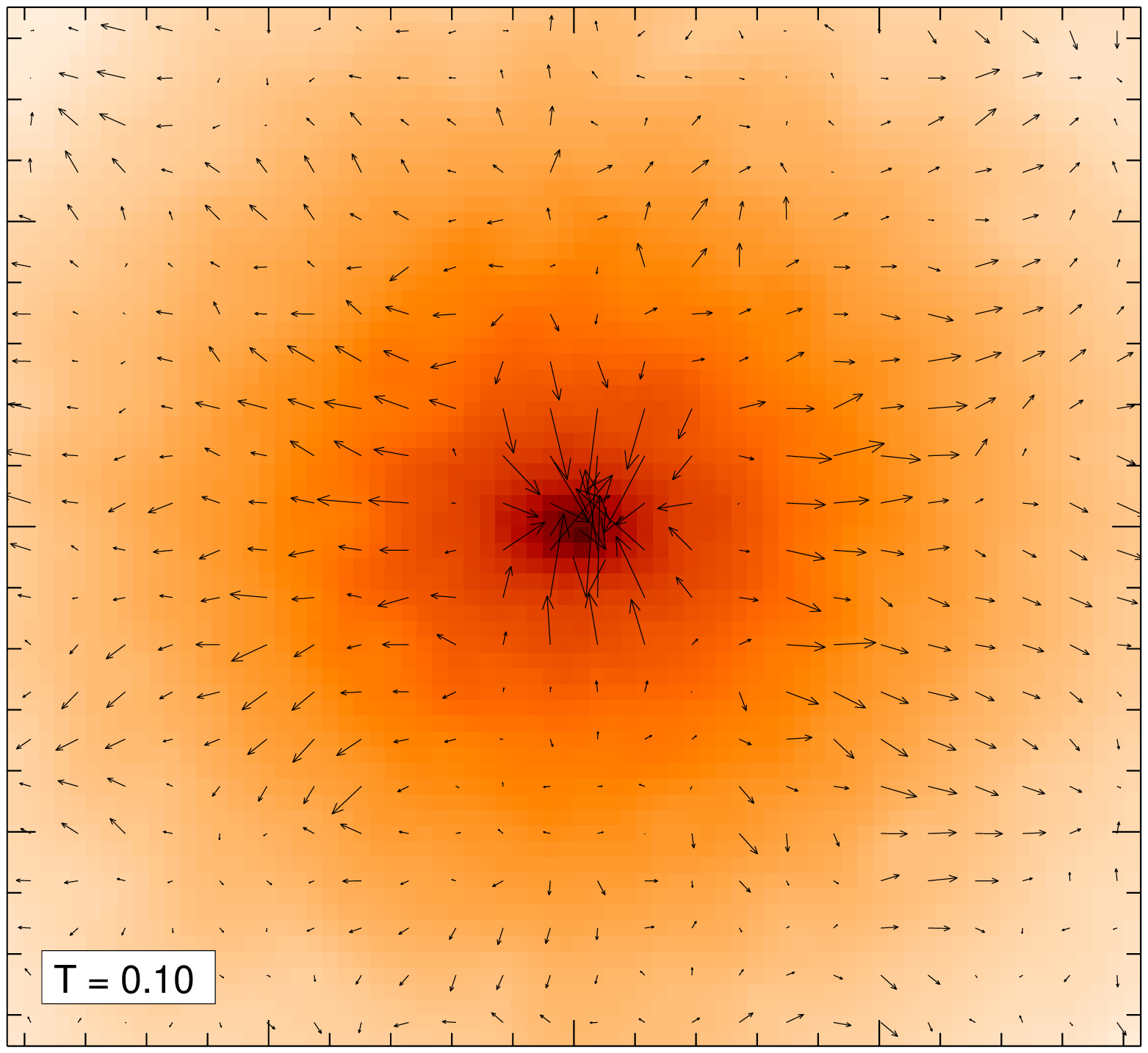}}%
\resizebox{8.0cm}{!}{\includegraphics{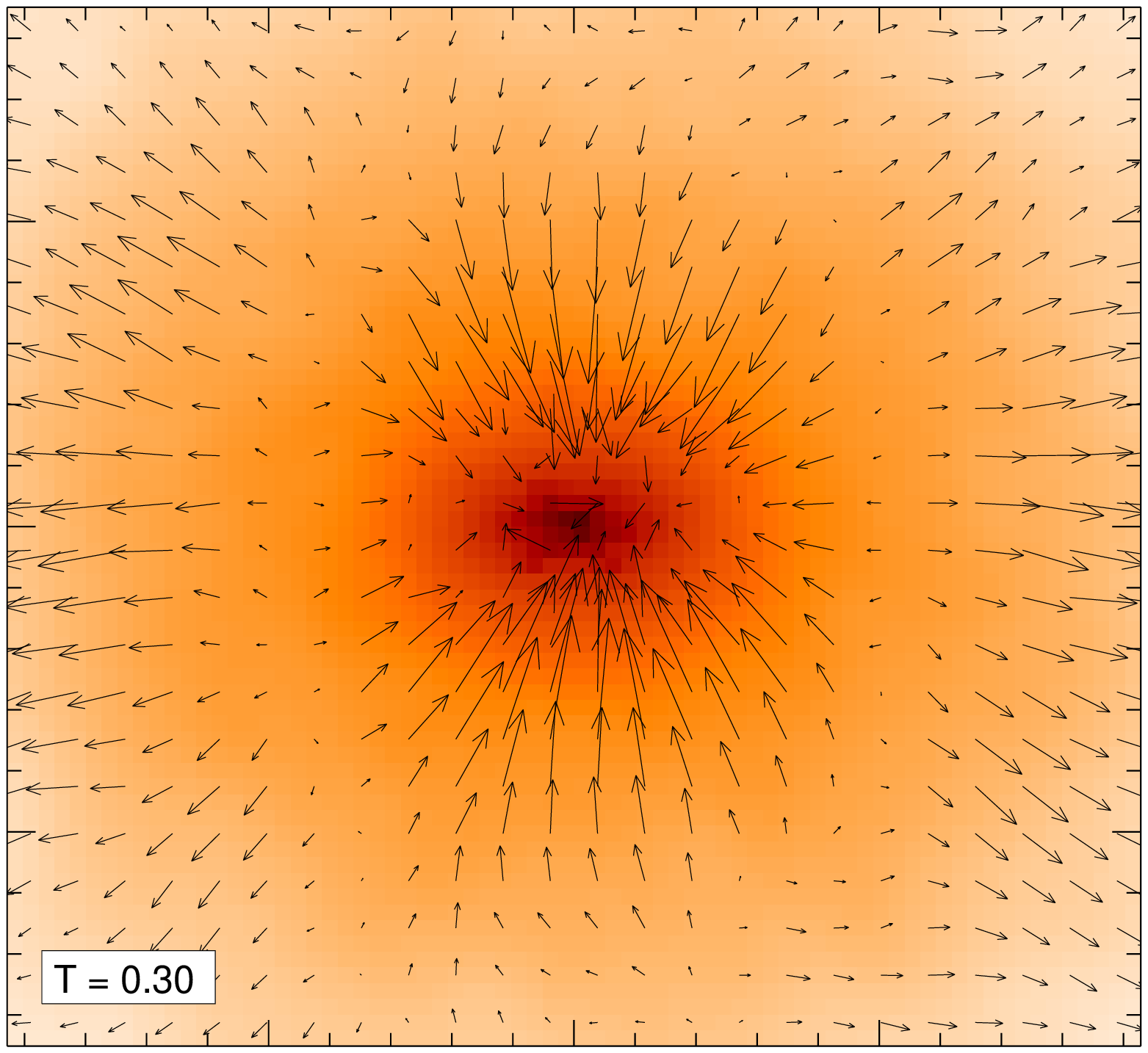}}\\%
\resizebox{8.0cm}{!}{\includegraphics{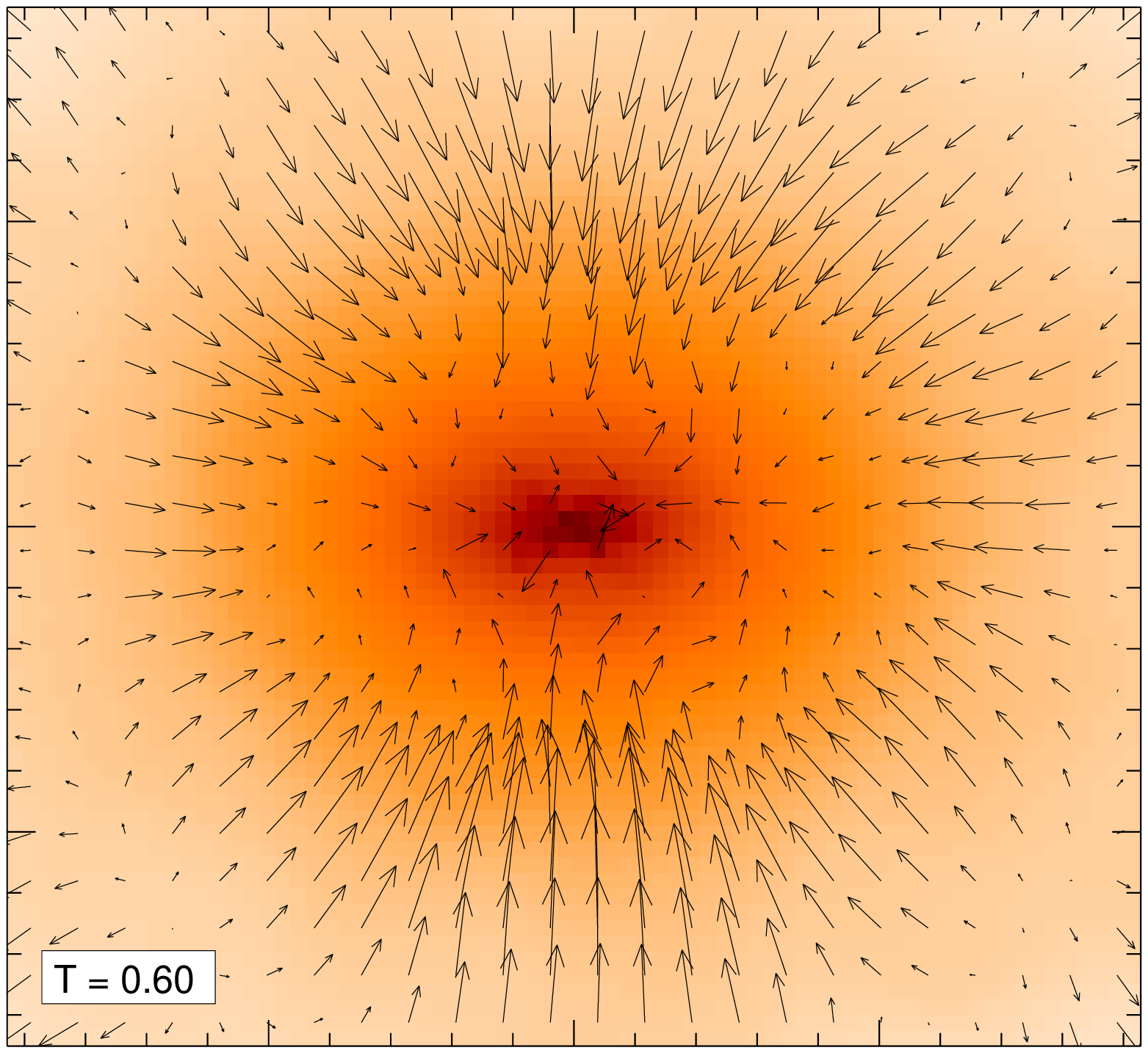}}%
\resizebox{8.0cm}{!}{\includegraphics{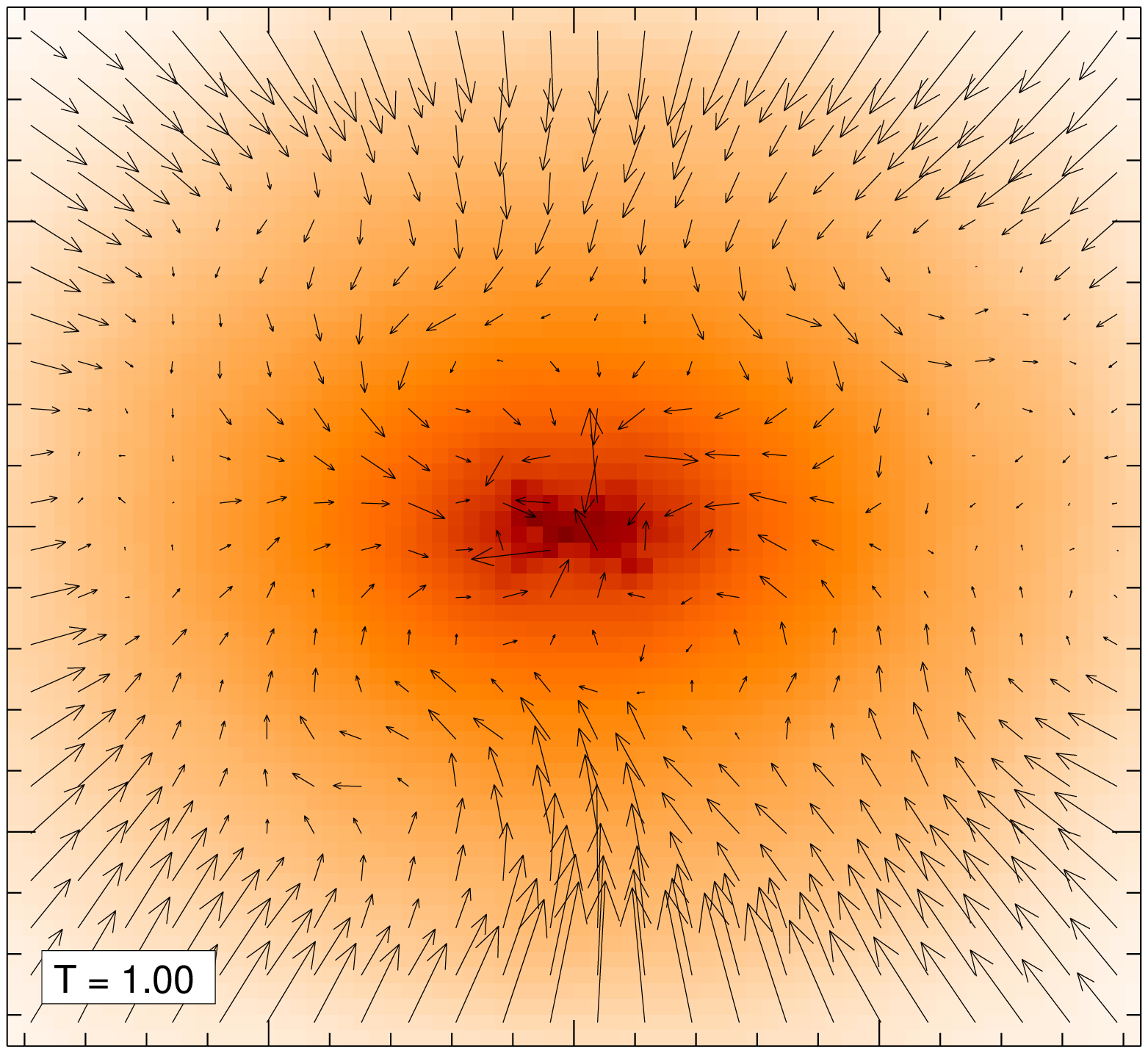}}\\%
\resizebox{8.0cm}{!}{\includegraphics{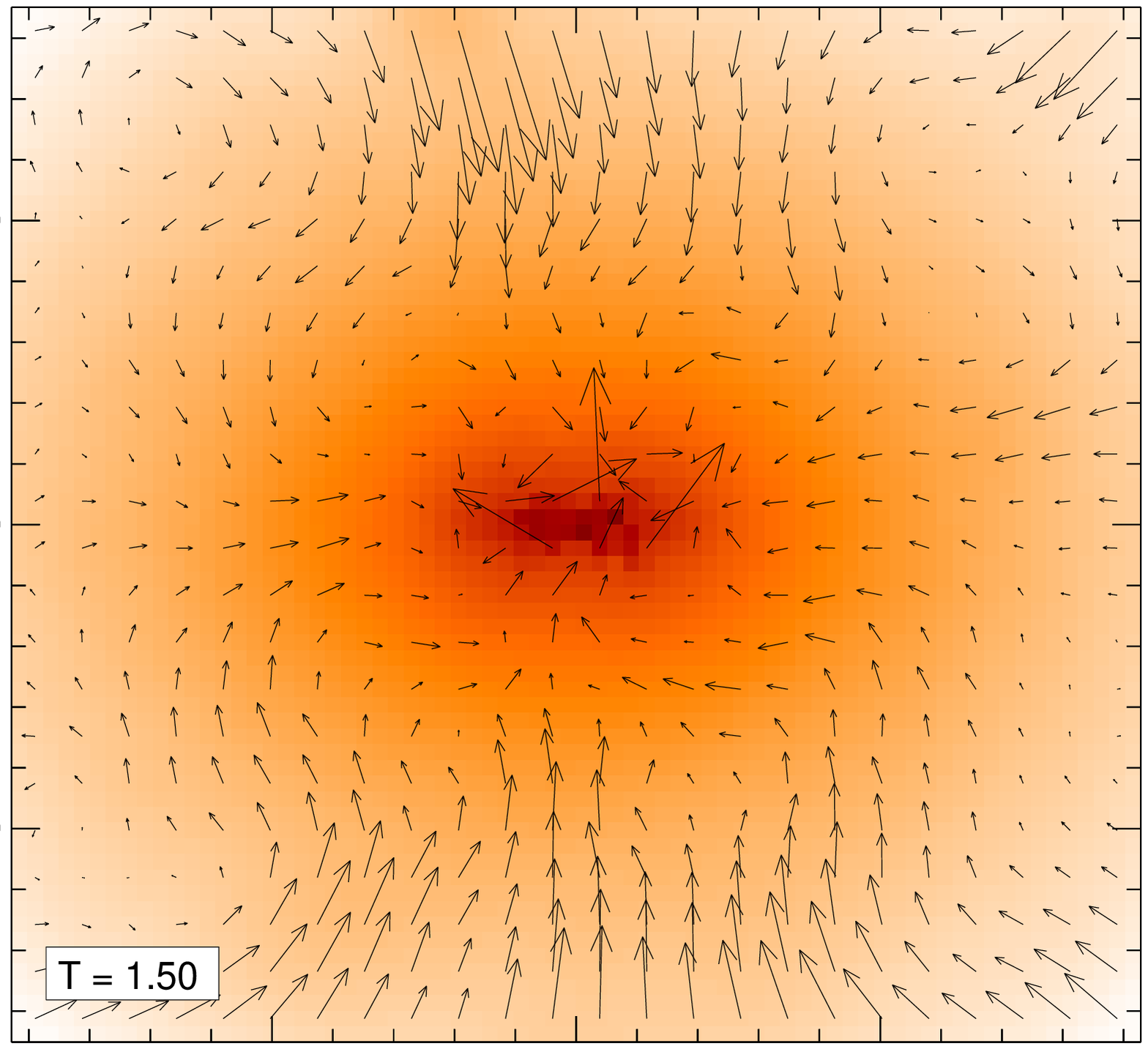}}%
\resizebox{8.0cm}{!}{\includegraphics{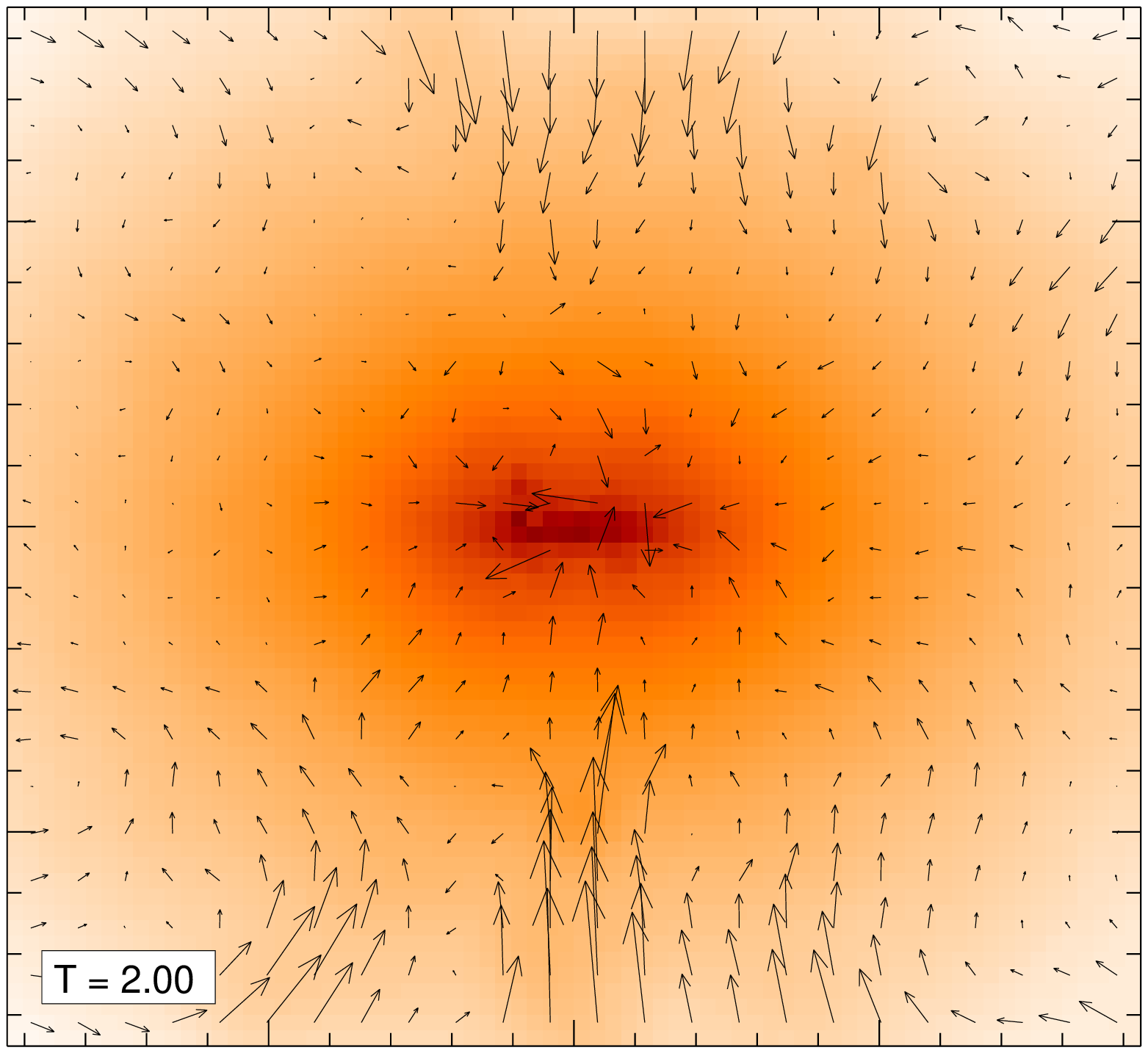}}%
\vspace{0.7cm}\\%
\caption{Time evolution of the gas flow in a halo of total mass
$M_{\rm vir}= 10^{12} \, h^{-1}{\rm M}_\odot$.  The velocity field is
represented by arrows and the logarithm of the gas density is
indicated as a colour-scale.  Labels give the elapsed time in
$h^{-1}{\rm Gyr}$ since the start of the simulation.  A wind of speed
$242\,{\rm km}\,{\rm s}^{-1}$ is included in this model, but it is
stopped already very close to the disk. \label{figVfield100}} \ec
\end{figure*}

\begin{figure*}
\bc
\resizebox{8cm}{!}{\includegraphics{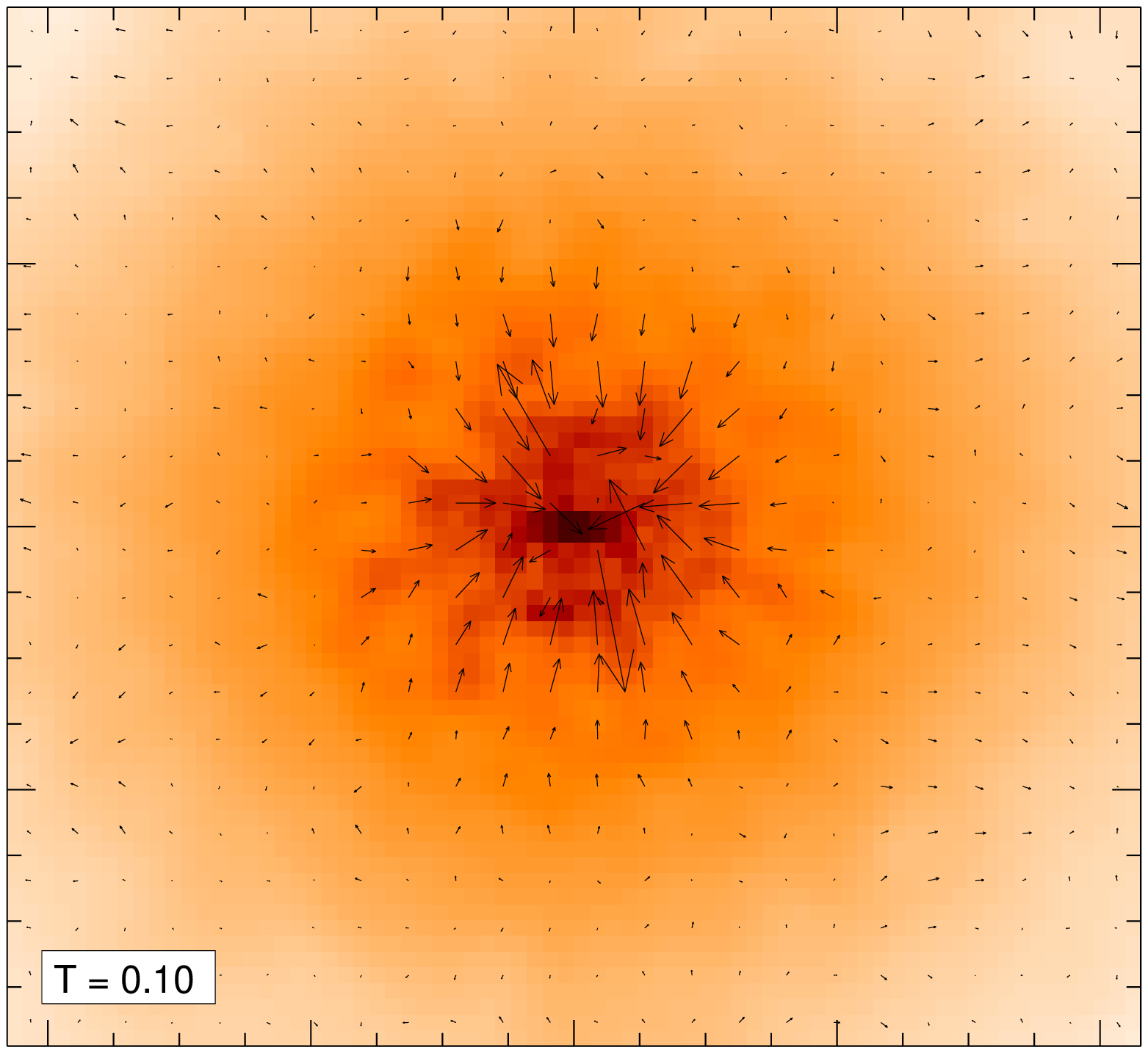}}%
\resizebox{8cm}{!}{\includegraphics{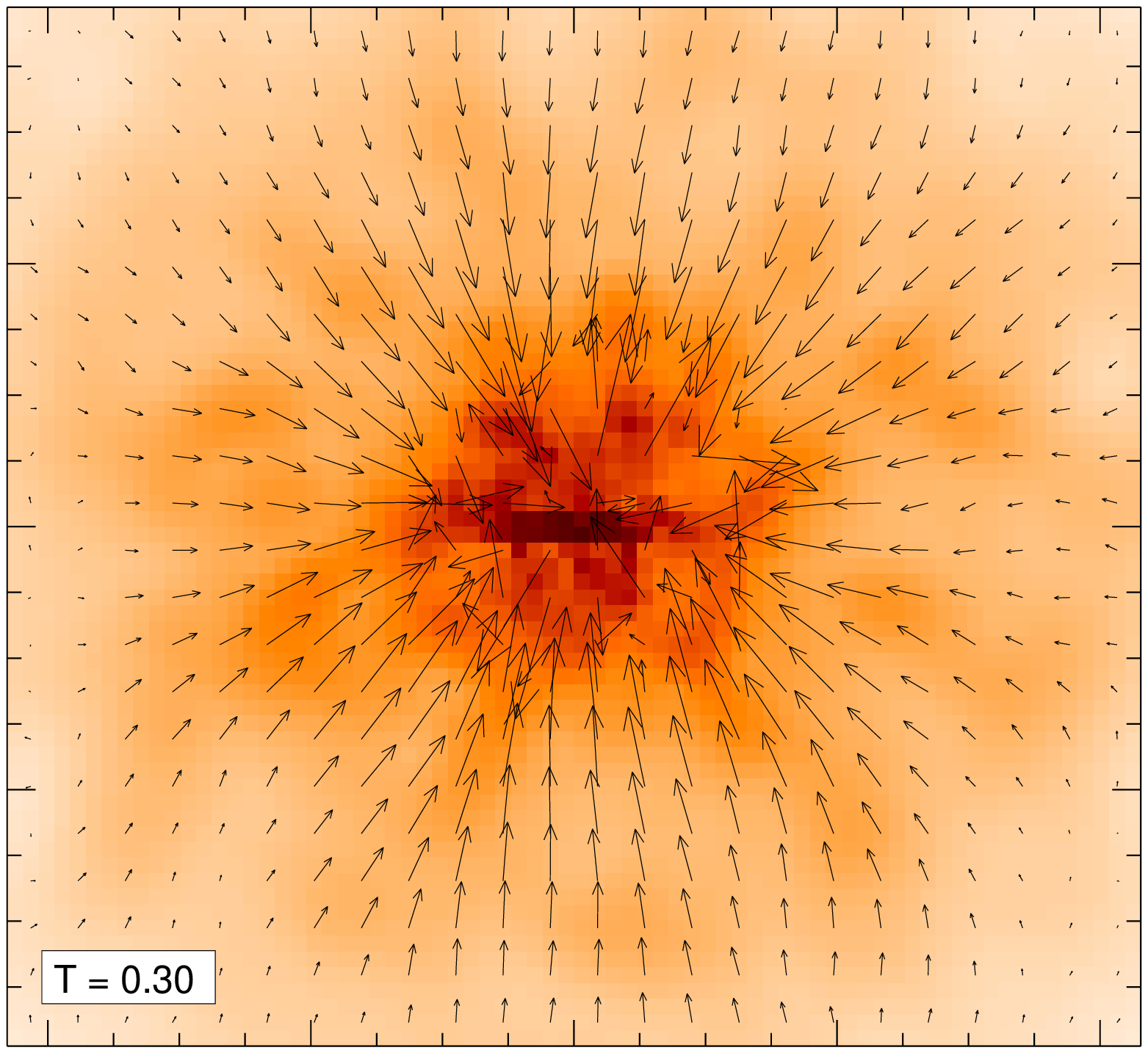}}\\%
\resizebox{8cm}{!}{\includegraphics{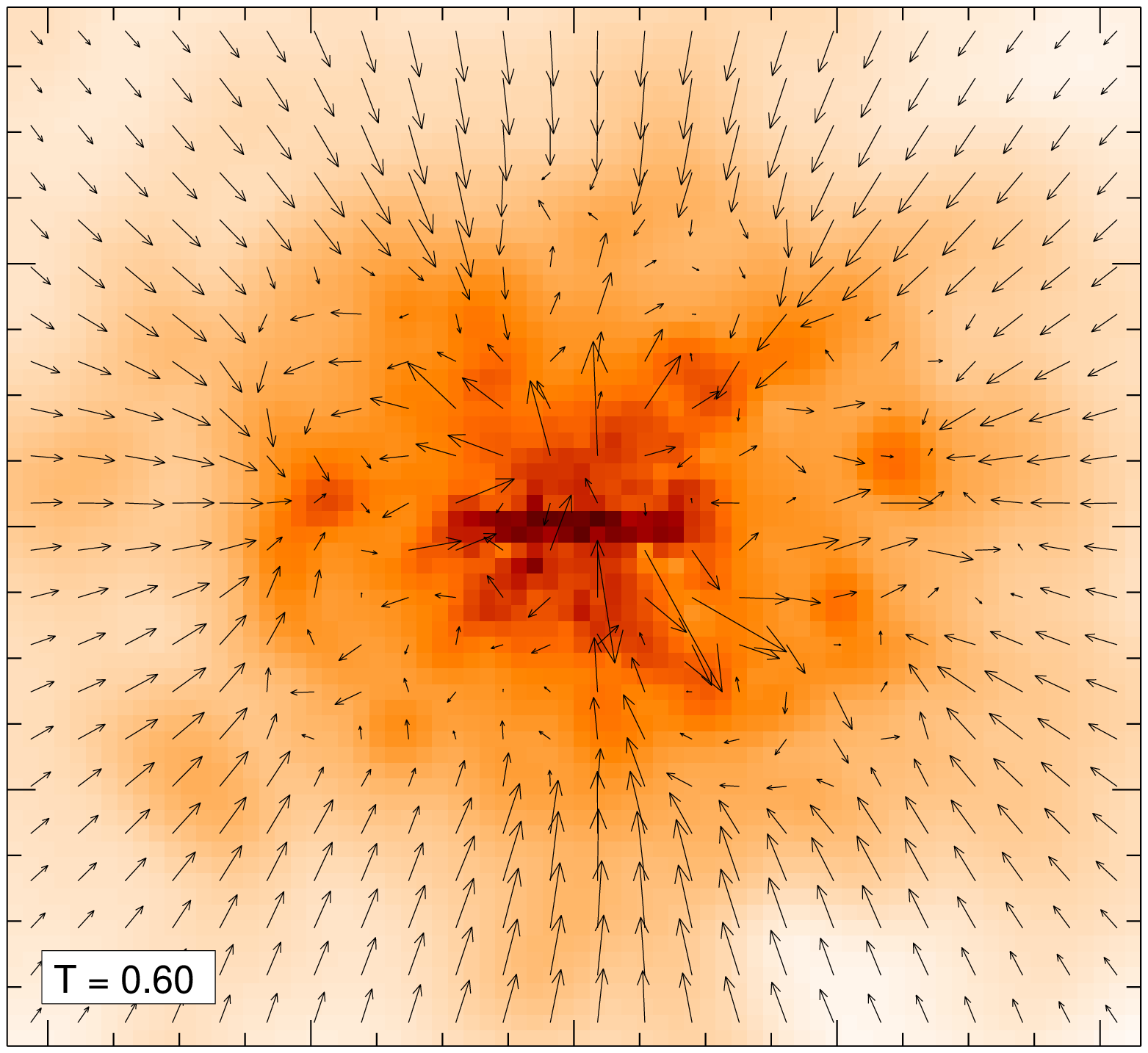}}%
\resizebox{8cm}{!}{\includegraphics{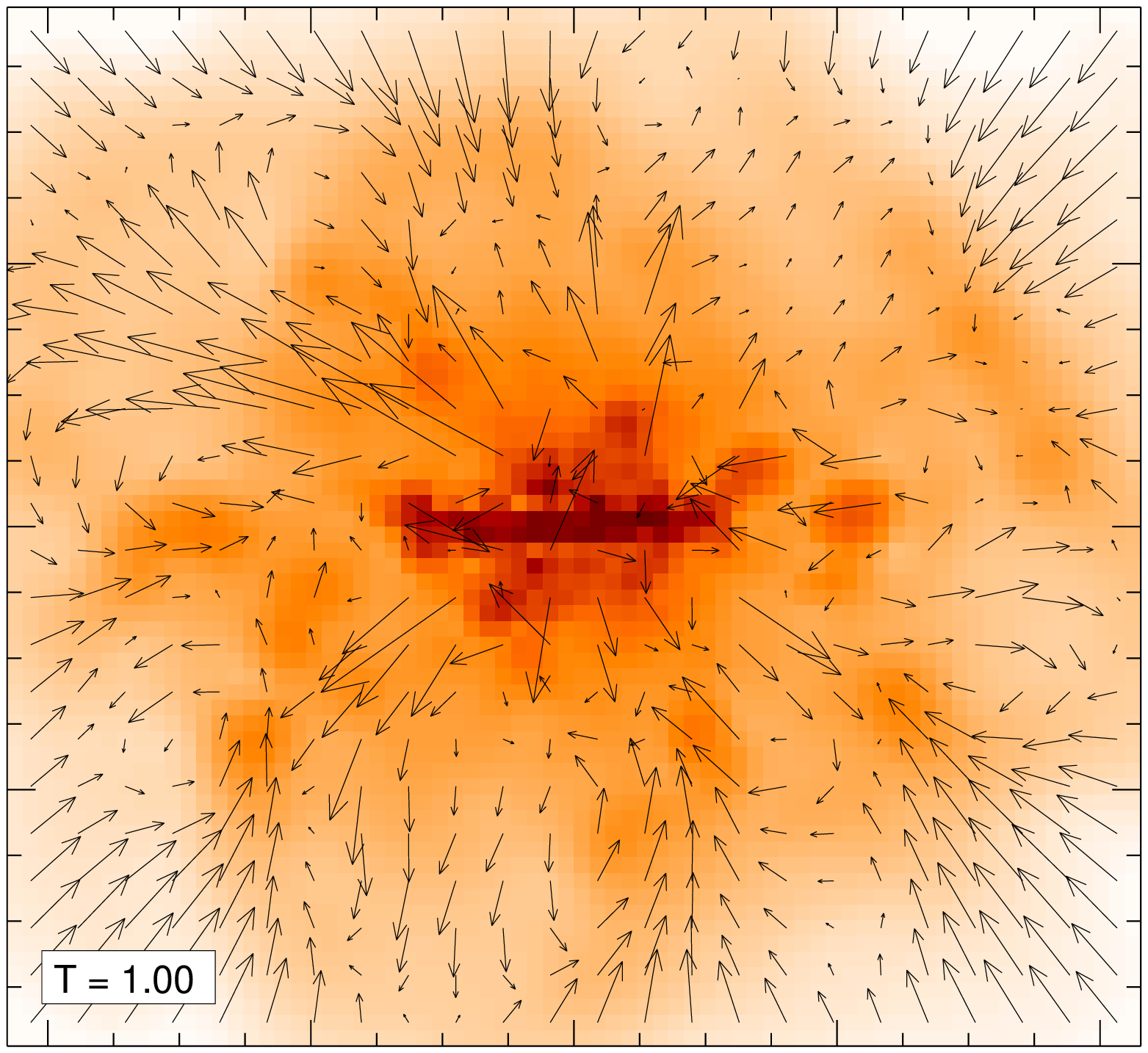}}\\%
\resizebox{8cm}{!}{\includegraphics{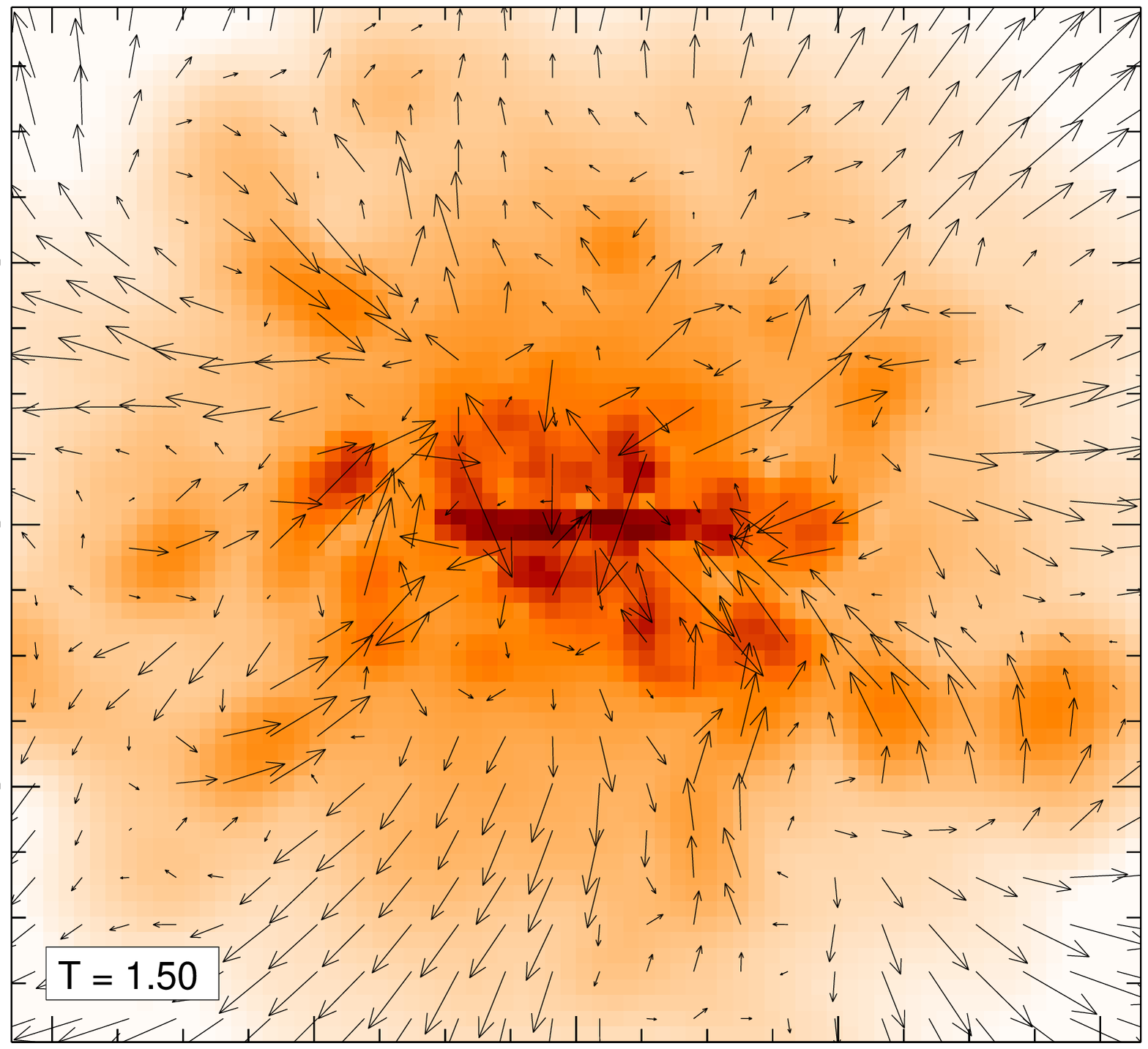}}%
\resizebox{8cm}{!}{\includegraphics{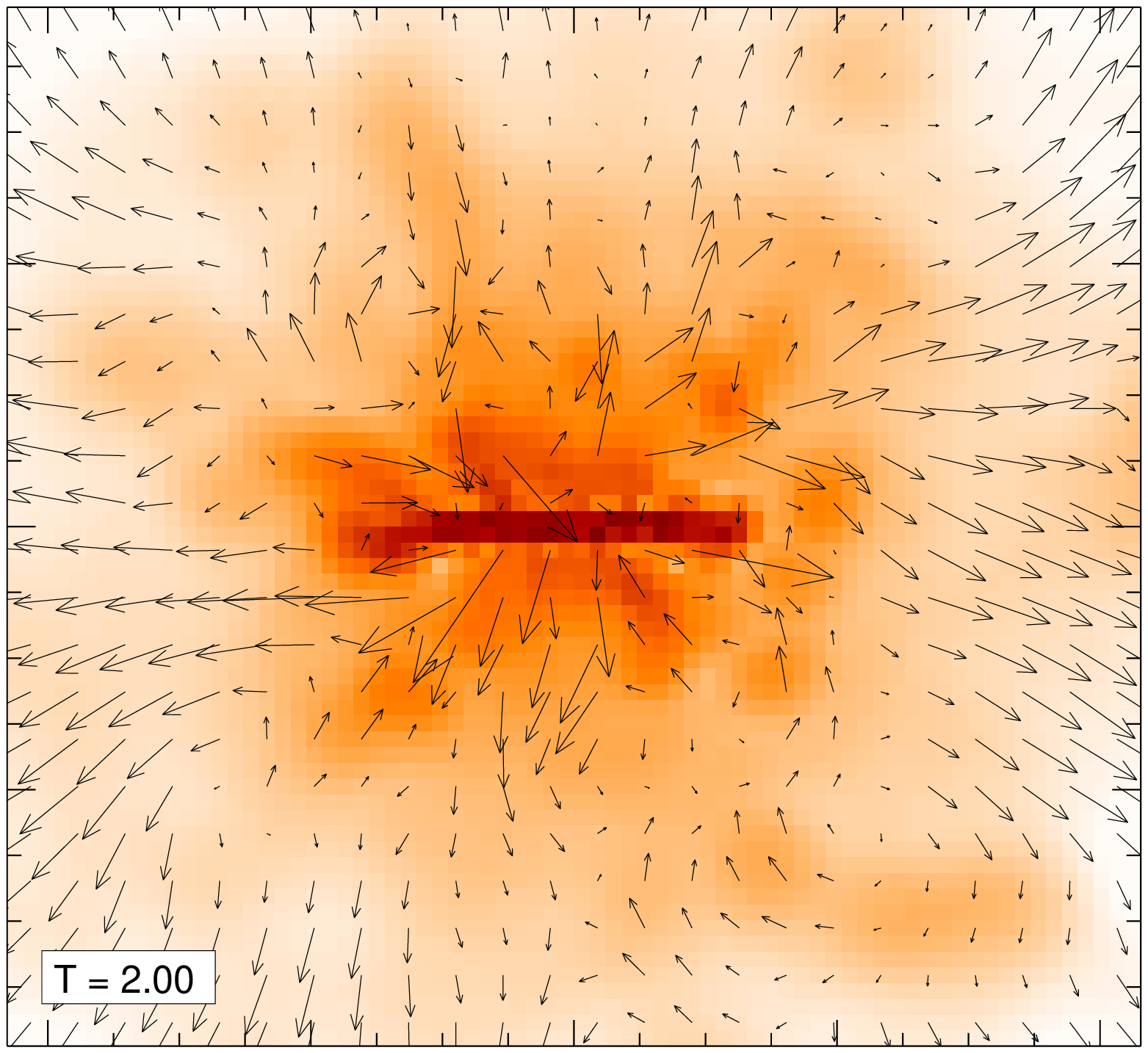}}%
\vspace{0.7cm}\\%
\caption{Time evolution of the gas flow in a halo of total mass
$M_{\rm vir}= 10^{11} \, h^{-1}{\rm M}_\odot$.  The velocity field is
represented by arrows and the logarithm of the gas density is
indicated as a colour-scale.  Labels in each panel give the elapsed
time in $h^{-1}{\rm Gyr}$ since the start of the simulation.  A wind
of speed $242\,{\rm km}\,{\rm s}^{-1}$ is included in this model,
somewhat slower than the escape speed of $v_{\rm esc}\simeq 280\,{\rm
km}\,{\rm s}^{-1}$ from this halo. As a result, gas escapes to
substantial heights in the halo, but falls back later onto the disk in
a galactic ``fountain''.\label{figVfield10}} \ec
\end{figure*}

\begin{figure*}
\bc
\vspace*{-0.3cm}\resizebox{8cm}{!}{\includegraphics{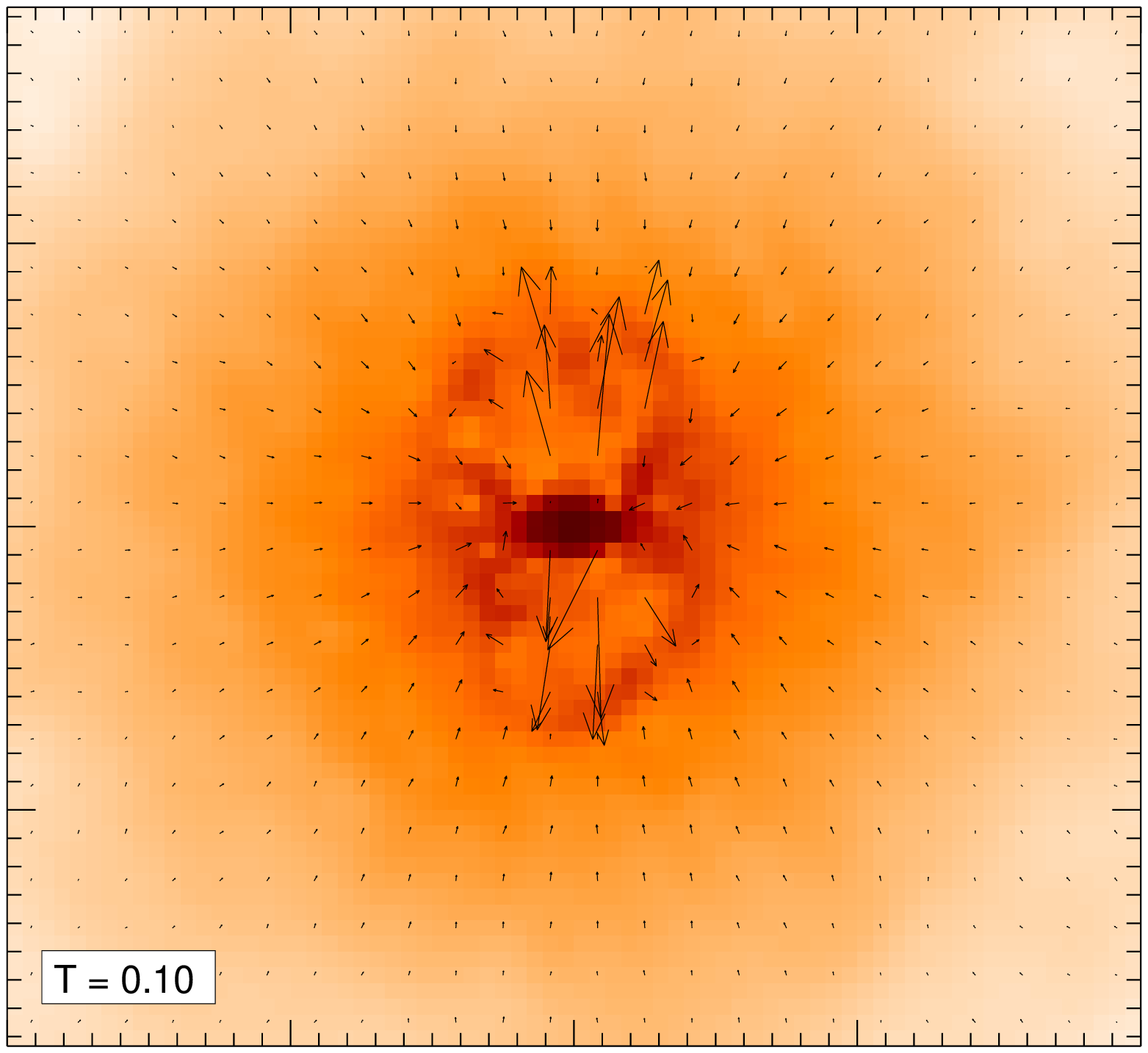}}%
\resizebox{8cm}{!}{\includegraphics{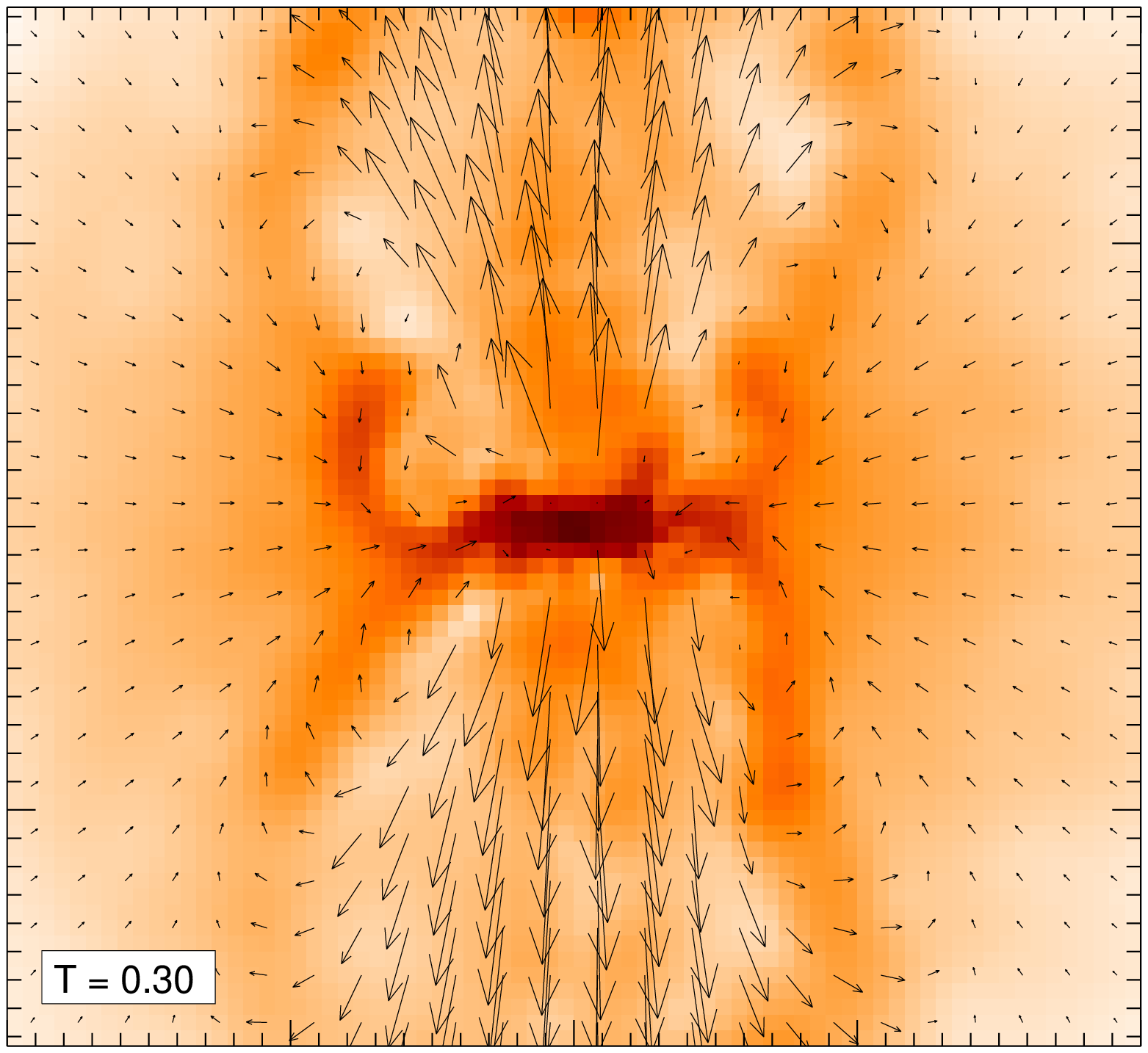}}\\%
\resizebox{8cm}{!}{\includegraphics{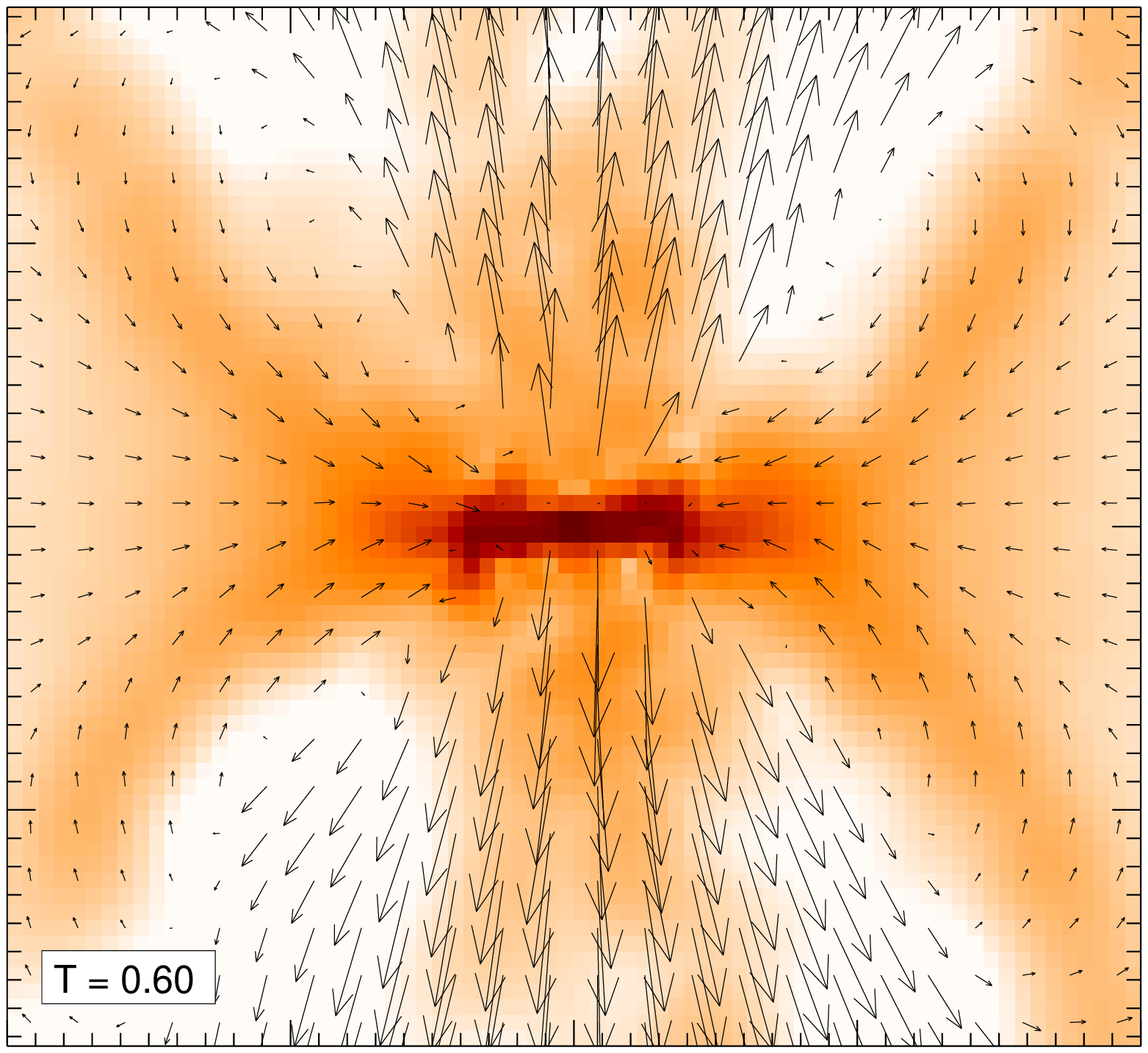}}%
\resizebox{8cm}{!}{\includegraphics{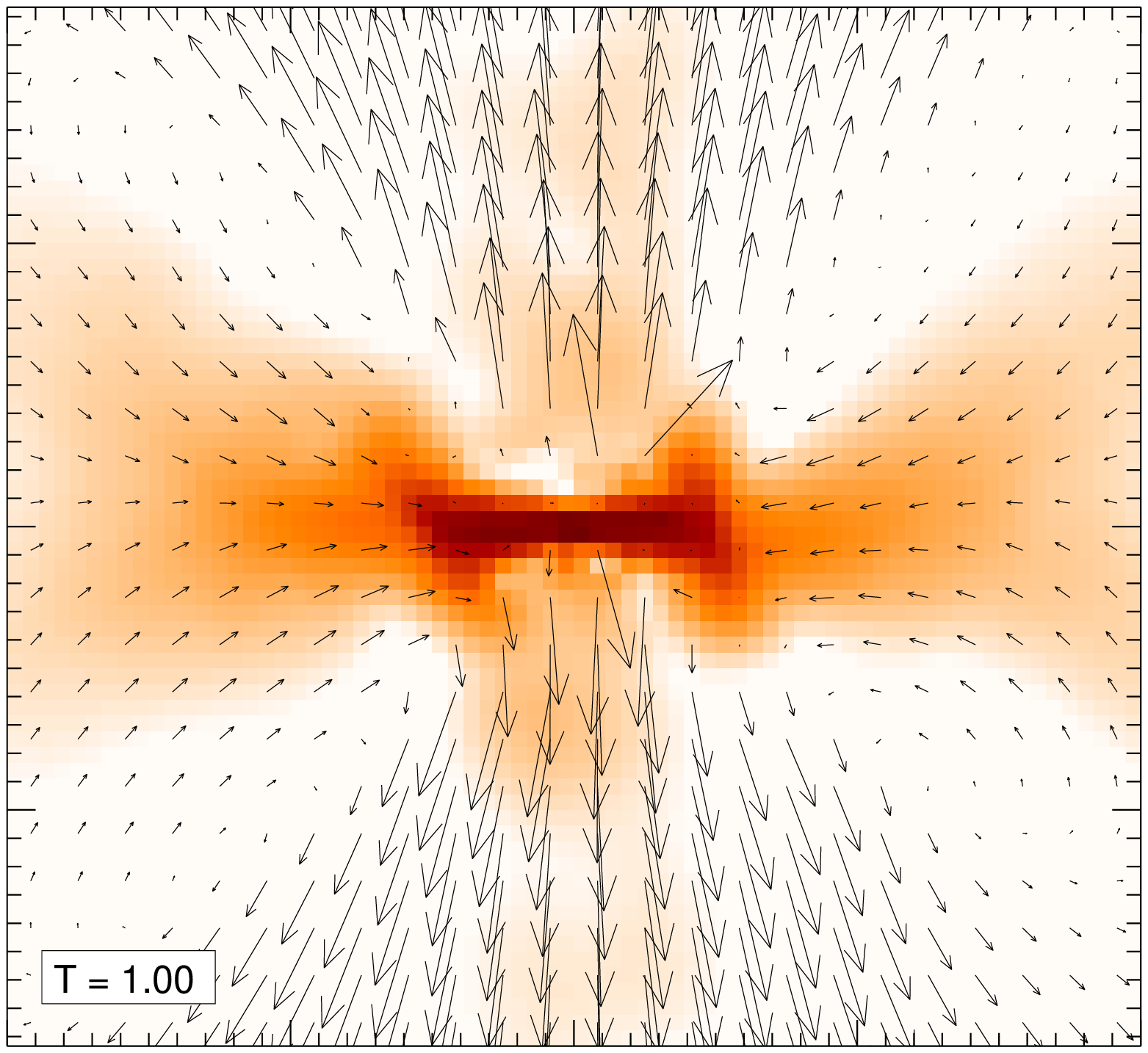}}\\%
\resizebox{8cm}{!}{\includegraphics{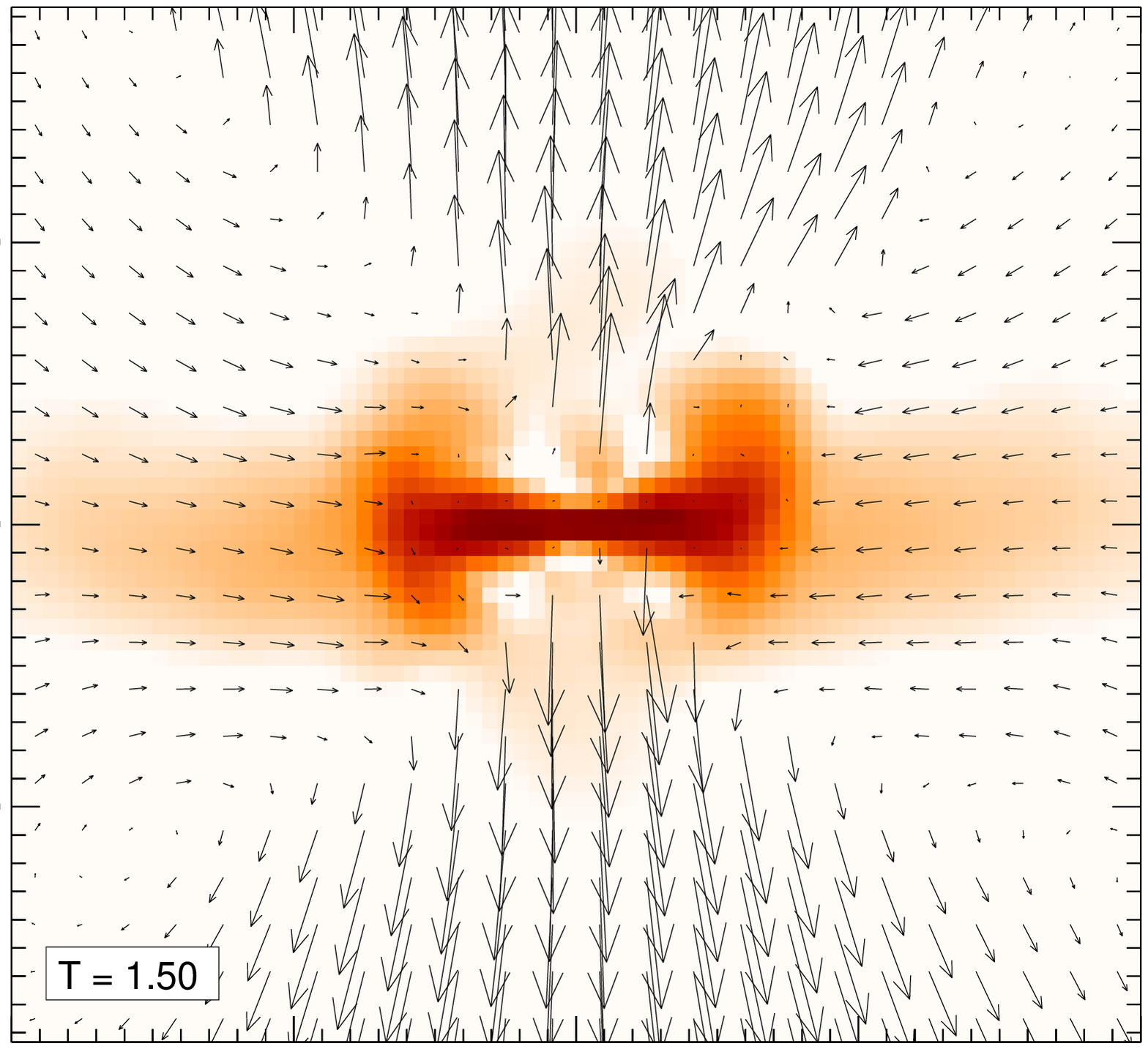}}%
\resizebox{8cm}{!}{\includegraphics{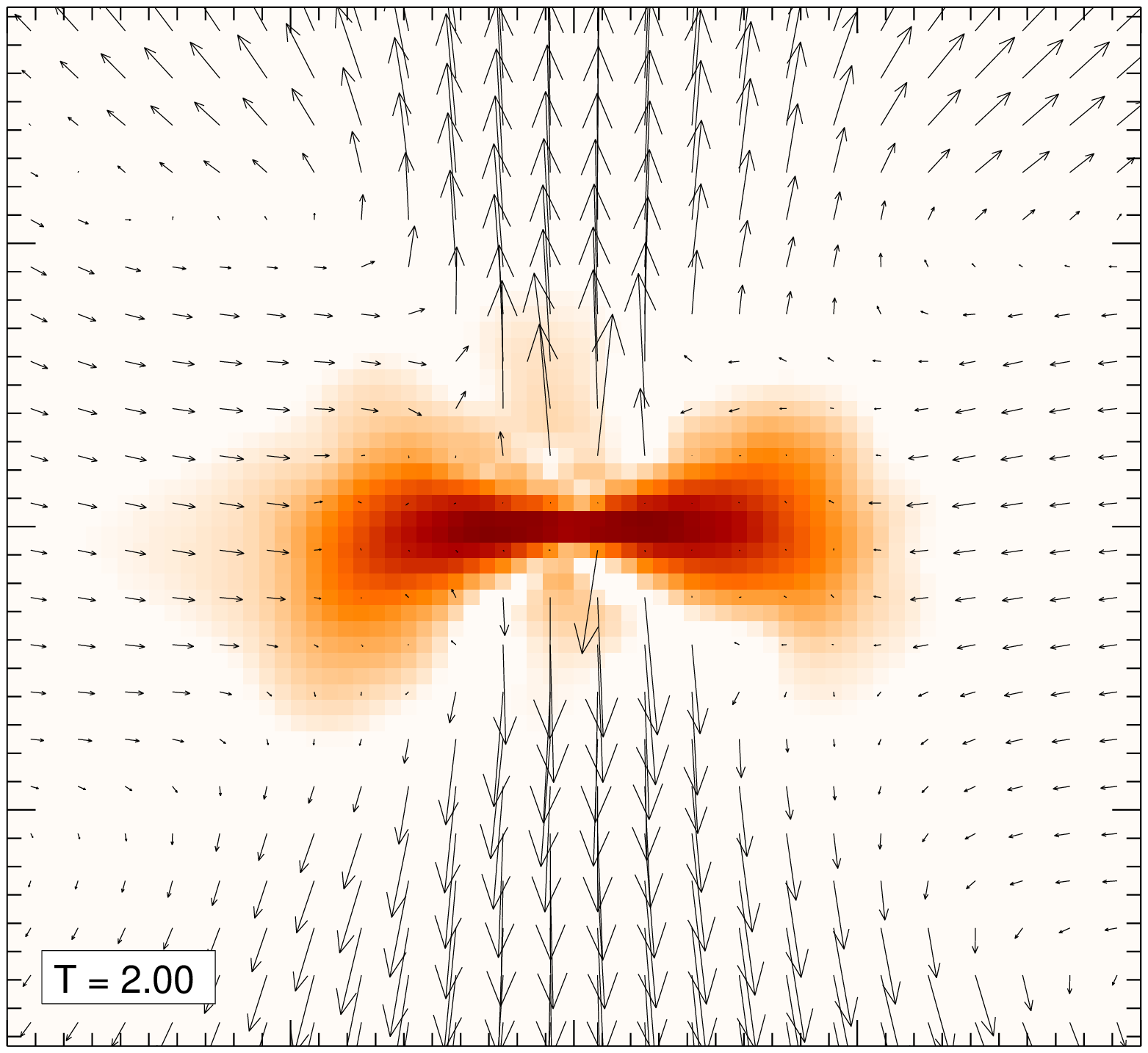}}%
\vspace{0.5cm}\\%
\caption{Time evolution of the gas flow in a halo of total mass
$M_{\rm vir}= 10^{10} \, h^{-1}{\rm M}_\odot$.  The velocity field is
represented by arrows and the logarithm of the gas density is
indicated as a colour-scale.  Labels in each panel give the elapsed
time in $h^{-1}{\rm Gyr}$ since the start of the simulation.  A wind
of speed $242\,{\rm km}\,{\rm s}^{-1}$ is included in this model,
considerably higher than the escape speed of $v_{\rm esc}\simeq
130\,{\rm km}\,{\rm s}^{-1}$ from this halo. As a result, a galactic
super-wind develops which blows out of the galaxy, entraining a
significant fraction of the gas from the halo. \label{figVfield1}} \ec
\end{figure*}

\begin{figure*}
\bc
\vspace*{-0.3cm}\resizebox{8cm}{!}{\includegraphics{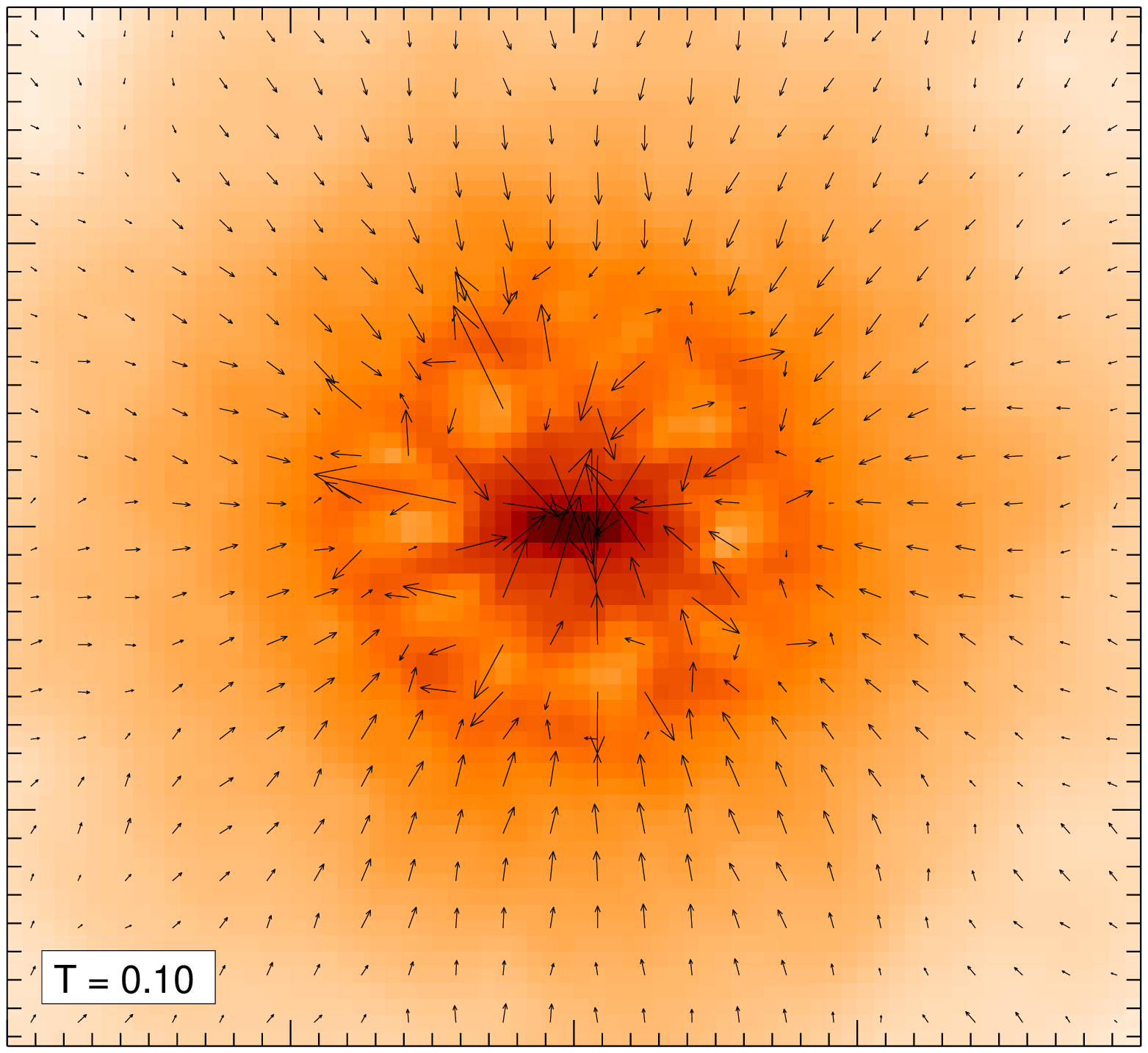}}%
\resizebox{8cm}{!}{\includegraphics{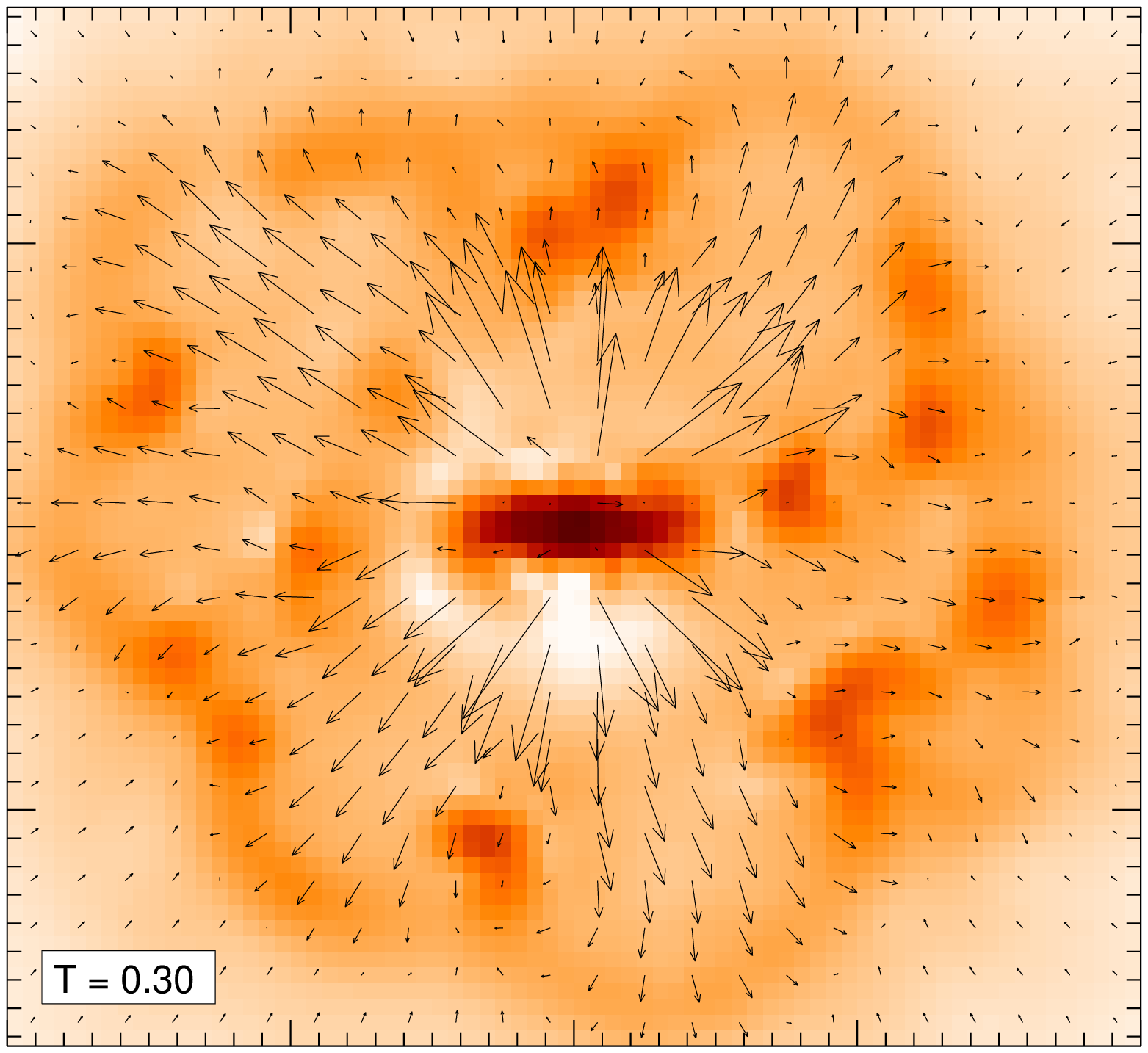}}\\%
\resizebox{8cm}{!}{\includegraphics{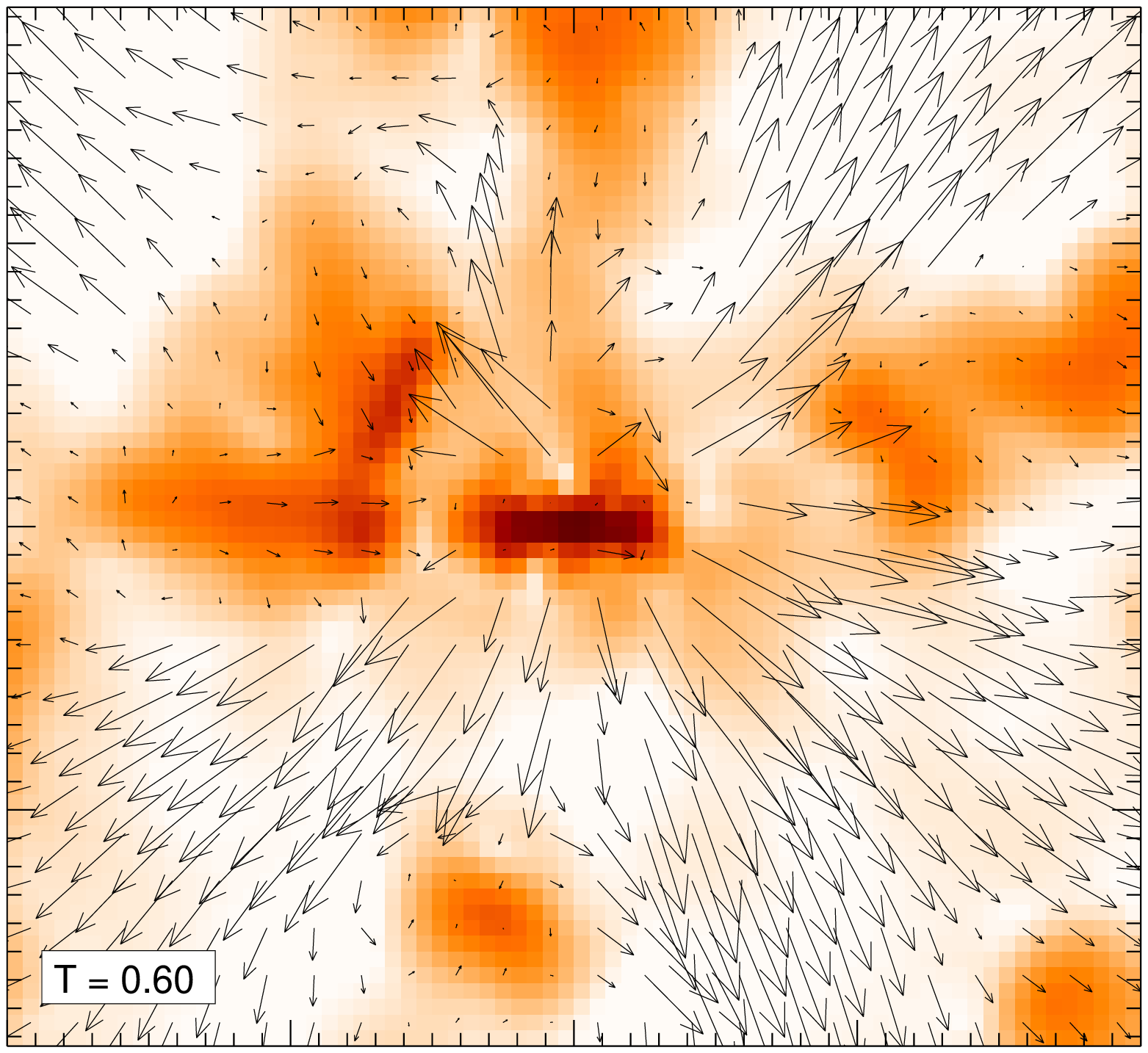}}%
\resizebox{8cm}{!}{\includegraphics{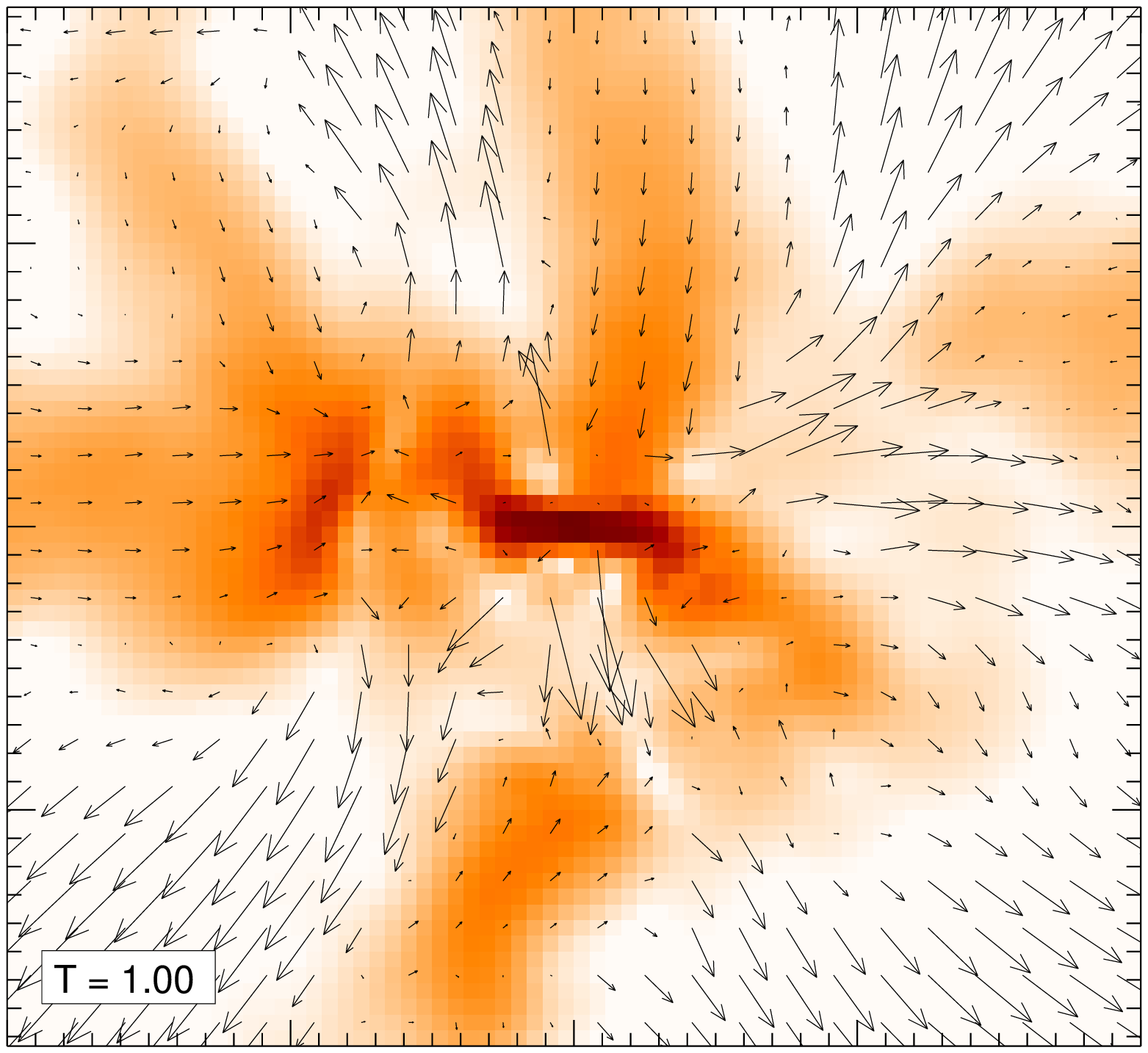}}\\%
\resizebox{8cm}{!}{\includegraphics{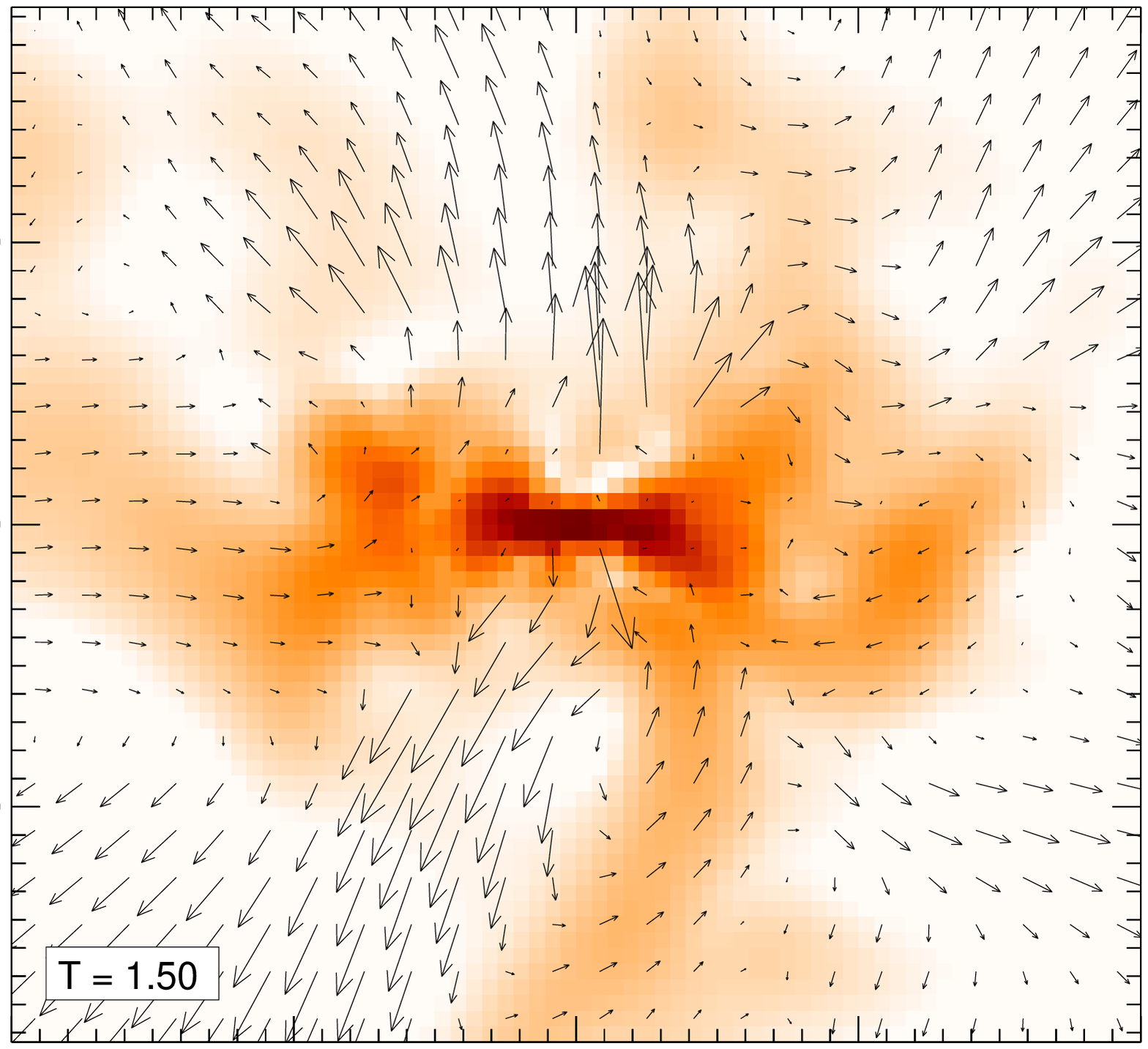}}%
\resizebox{8cm}{!}{\includegraphics{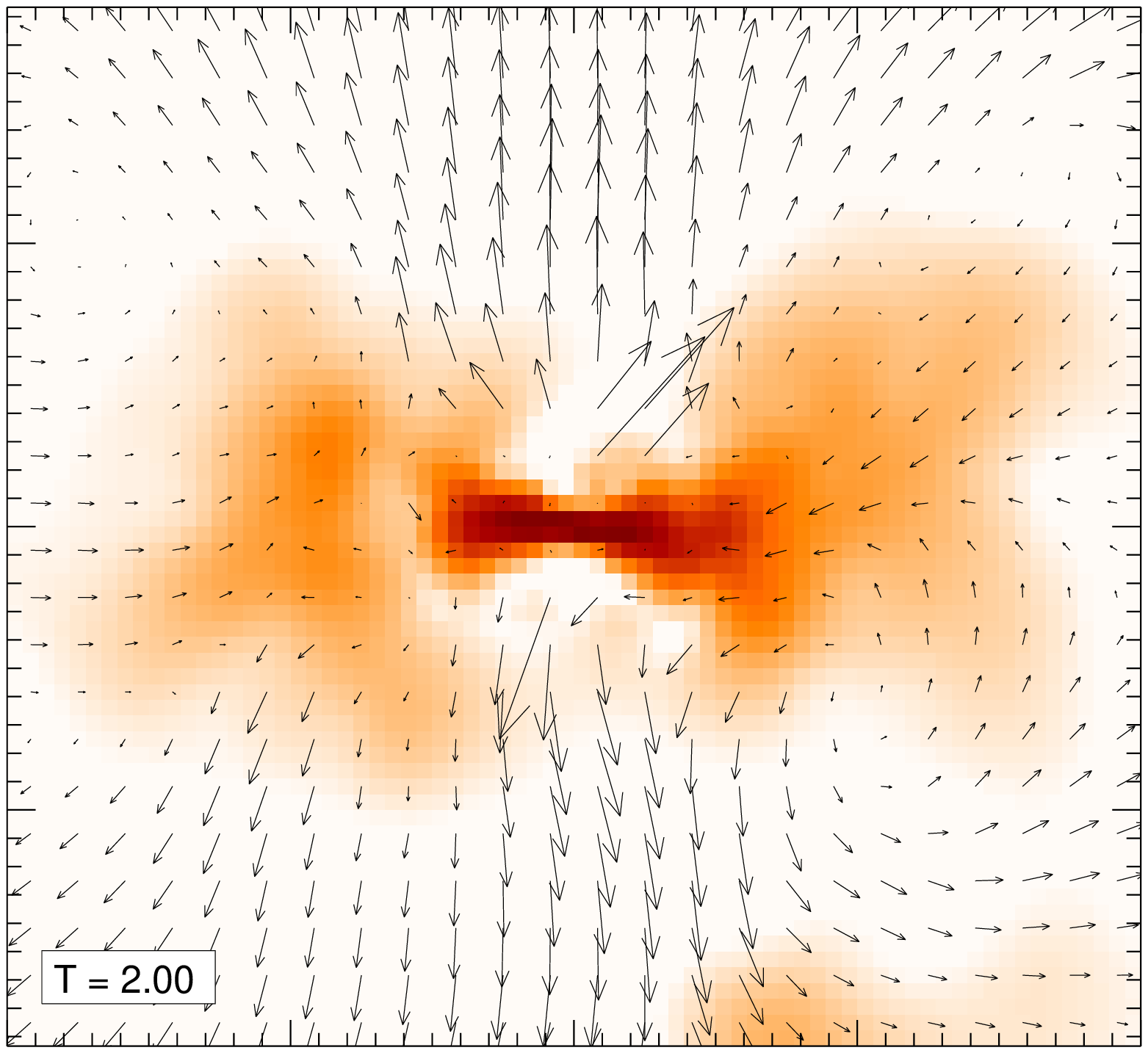}}%
\vspace{0.5cm}\\%
\caption{The same as Figure~\ref{figVfield1}, except that the
simulation included isotropic winds instead of explicitly modelling an
outflow that is preferentially oriented along the rotation axis of the
disk.  A bipolar outflow pattern orthogonal to the disk still develops
for the isotropic wind due to the dense disk that forms in the
$xy$-plane, and due to the non-isotropic inflow pattern.  Overall, the
gas flow is considerably less ordered in this case, but it is
qualitatively still quite similar to the model with axial winds.
\label{figVfield1iso}}
\ec
\end{figure*}

\begin{figure*}
\bc
\hspace*{-0.3cm}\resizebox{10.5cm}{!}{\includegraphics{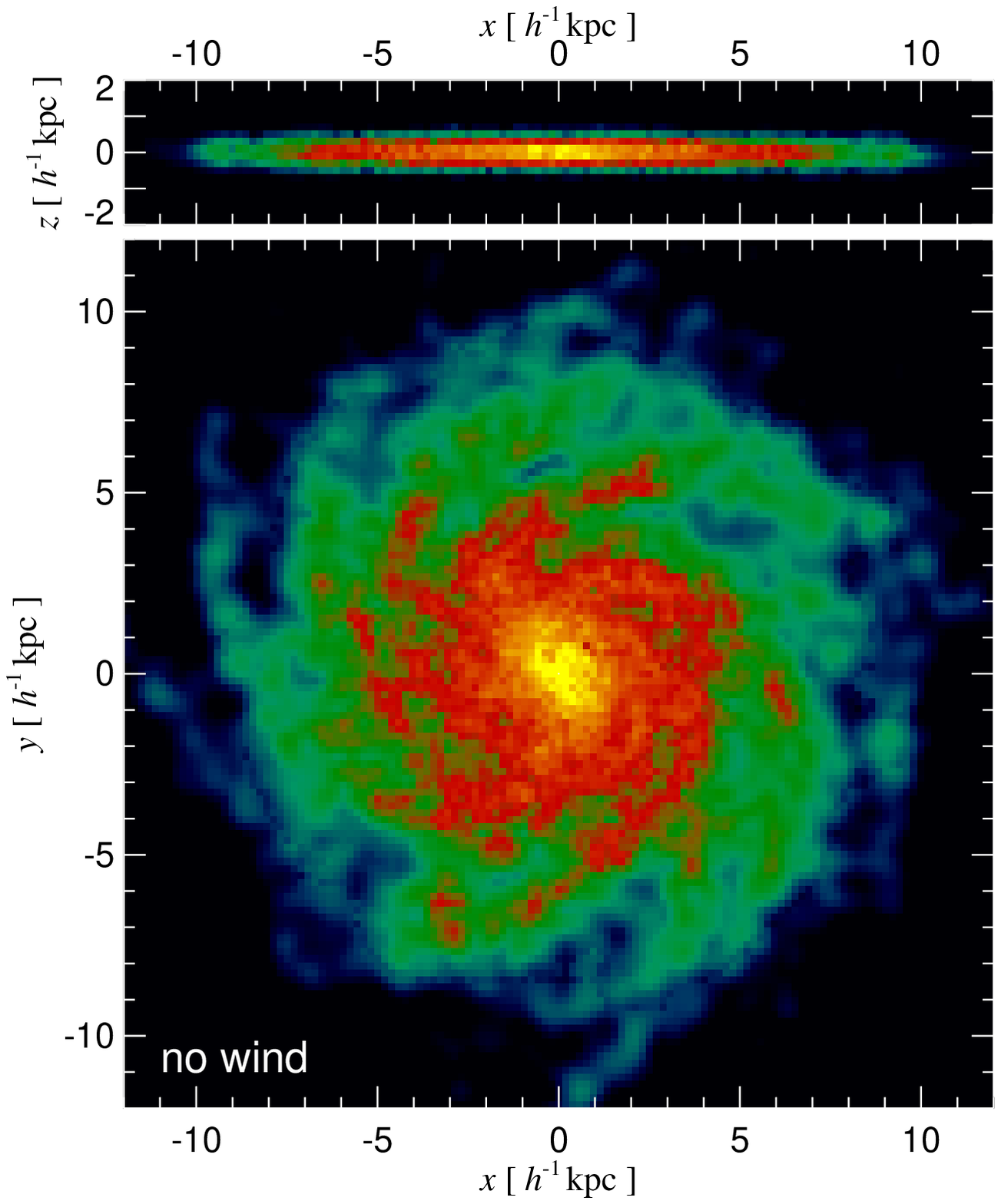}}%
\hspace*{-2.4cm}\resizebox{10.5cm}{!}{\includegraphics{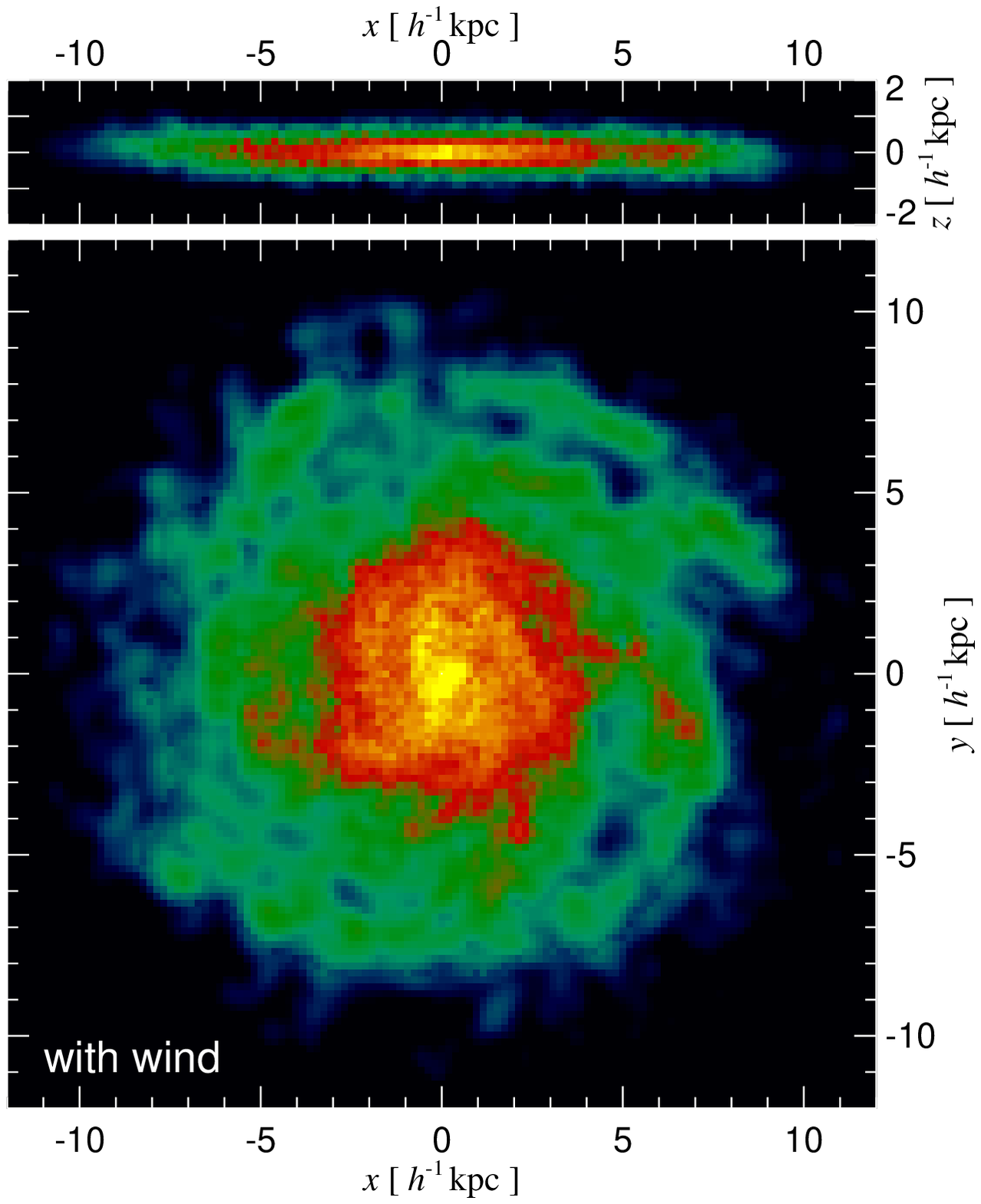}}\\%
\caption{Stellar disk that formed after $3\,{\rm Gyr}$ in a halo of
mass $M_{\rm vir}= 10^{12} \, h^{-1}{\rm M}_\odot$, seen both face on
and edge on. On the left, the simulation does not include winds, on
the right it does. The colour-contours in the pictures have been set
such that they contain roughly the same fraction of the total light in
each case. \label{figlumdens100}} \ec
\end{figure*}

In our tests here, we will not be concerned with these issues
directly.  Instead, we examine the inside-out formation of a disk
galaxy under idealised settings which allows us to cleanly study the
properties of our self-regulated model for star formation, with or
without winds.  To this end, we consider a series of simulations which
effectively follow the \citet{Fall80} picture of the formation of a
disk galaxy.  We set up an NFW-halo \citep{NFW,NFW2} in isolation,
with the dark matter and gas initially in virial equilibrium. When
evolved in time using adiabatic gas physics, our initial conditions
are nearly stable, showing only negligible secular evolution over a
Hubble time.  However, when the gas is allowed to cool radiatively,
the gas in the center of the halo quickly loses its pressure support
and settles into a rotationally supported disk that grows from inside
out, with a size that depends on the initial angular momentum of the
gas. Note that due to the axisymmetric set-up we have chosen, angular
momentum transport between the gas and the dark matter is essentially
absent in these simulations.

For definiteness, we have chosen a halo of mass $M_{\rm vir}=
10^{12}\,h^{-1}{\rm M}_\odot$, with 10\% of the mass being in baryonic
form.  The initial angular momentum $J$ of the halo can be described
in terms of the spin parameter $\lambda=J|E|^{1/2}/(G M_{\rm
vir}^{5/2})$, for which we adopt a value $\lambda=0.1$ to produce a
large disk. The relative distribution of angular momentum within the
halo was chosen under the assumption that the specific angular momenta
$j(r)$ of spherical shells are all aligned, and that their magnitude
is given by the fitting formula \be j(r)= {j_{\rm max}}
\left(\frac{M(r)}{M_{\rm vir}}\right)^s \ee of \citet{Bull01}. Within
each shell, we distribute the angular momentum as if the shell were in
solid body rotation. For $s=1$, the choice we adopt here, this implies
that in the $xy$-plane, the initial azimuthal streaming velocity is
proportional to the circular velocity squared. We have also simulated
haloes with masses of $10^{11}\,h^{-1}{\rm M}_\odot$, and
$10^{10}\,h^{-1}{\rm M}_\odot$. They are simply scaled down versions
of the $10^{12}\,h^{-1}{\rm M}_\odot$ halo, differing by factors of 10
in mass, and by $10^{1/3}\simeq2.15$ in length and velocity scale.

In order to simplify these test simulations further, we have
represented the dark matter halo in most of our runs by a static
gravitational potential.  This neglects the contraction of the halo
due to the infall of the gas, but for the purposes of the present
analysis this effect is not important. Simulations that include a live
dark halo give very similar results, but are more expensive,
especially when one wants to reduce the additional particle noise from
the dark halo to a negligible level.

For each of the three halo masses, we run simulations with and without
a wind. Typically we used 40000 self-gravitating SPH particles in each
case, except for a number of resolution tests which we will discuss
separately below. The parameters of our multi-phase model were set to
our set of fiducial values, i.e.~$t_0^\star =2.1\,{\rm Gyr}$, and for
the wind sector, $\eta=2$ and $\chi=0.25$.

In Figure~\ref{figdisksfr}, we show the star formation rates in these
simulations as a function of time. In all cases, the rapid initial
development of the gas disk is accompanied by an abrupt rise of the
star formation rate to a maximum soon after the start of the
simulations.  Thereafter, the star formation rate declines nearly
exponentially for some time, but the decline becomes increasingly
slower at later times. In general, the models with winds yield a lower
star formation rate at all times, but the magnitude of the suppression
is a strong function of halo mass. The $10^{12} \, h^{-1}{\rm
M}_\odot$ halo forms stars nearly unaffected by a wind, but the
$10^{11} \, h^{-1}{\rm M}_\odot$ halo and particularly the $10^{10} \,
h^{-1}{\rm M}_\odot$ halo suffer a strong reduction in their star
formation rates.

The cause for these differences between the effects of the wind in
haloes of different mass becomes readily apparent when one examines
the flow of the gas in more detail. In Figures~\ref{figVfield100},
\ref{figVfield10}, and \ref{figVfield1}, we show the time evolution of
the velocity field in the $xz$-plane, which is a slice orthogonal to
the gas disk, with the rotation axis along the $z$-axis. We indicate
the gas velocity field with arrows, overlaid on a colour-scale map
showing the gas density in the slice.

For the massive $10^{12} \, h^{-1}{\rm M}_\odot$ halo, the wind is
never able to propagate very far from the star-forming disk. Instead,
any ejected material is quickly stopped by the ram pressure of ambient
gas, and, even more important, by the strength of the gravitational
force field.  Note that the wind speed of $242\,{\rm km}\,{\rm
s}^{-1}$ is comparable to the circular velocity of this halo, but it
is still much smaller than the halo's escape velocity.

However, the evolution is already somewhat different in the $10^{11}
\, h^{-1}{\rm M}_\odot$ halo, which is shown in
Figure~\ref{figVfield10}.  In this case, the wind is still not
energetic enough to unbind material from the halo, which has an escape
velocity of $v_{\rm esc}\simeq 280\,{\rm km}\,{\rm s}^{-1}$.  However,
ejected gas is able to rise to substantial heights in the halo and
to displace infalling material, before it eventually falls back to the
disk, forming a ``galactic fountain'' \citep{Shap76}.  This process
slows down the processing of gas into stars, and it mixes metals into
the halo, which thus becomes gradually enriched.

Finally, if the potential well is shallow enough, the wind can escape
from the galaxy entirely. This is the case for the $10^{10} \,
h^{-1}{\rm M}_\odot$ halo, as shown in Figure~\ref{figVfield1}. The
wind velocity is now substantially larger than the escape velocity of
$v_{\rm esc}\simeq 130\,{\rm km}\,{\rm s}^{-1}$, allowing the wind to
``blow out'' into the intergalactic medium, where it deposits both
metals and its residual kinetic energy. The wind can also entrain some
of the halo's gas, thereby further reducing the infall rate onto the
star-forming disk.  Note that our wind model has been specifically
constructed to be ``quiescent'', i.e.~to leave the star-forming ISM
intact. The wind does not blow away the ISM entirely, but it instead
leads to quiescent mass-loss with strength determined by the star
formation rate. This removal of baryons, together with the
interception of infalling material, leads to a strong reduction of the
star formation rate in low-mass haloes.

\begin{figure*}
\bc
\resizebox{8cm}{!}{\includegraphics{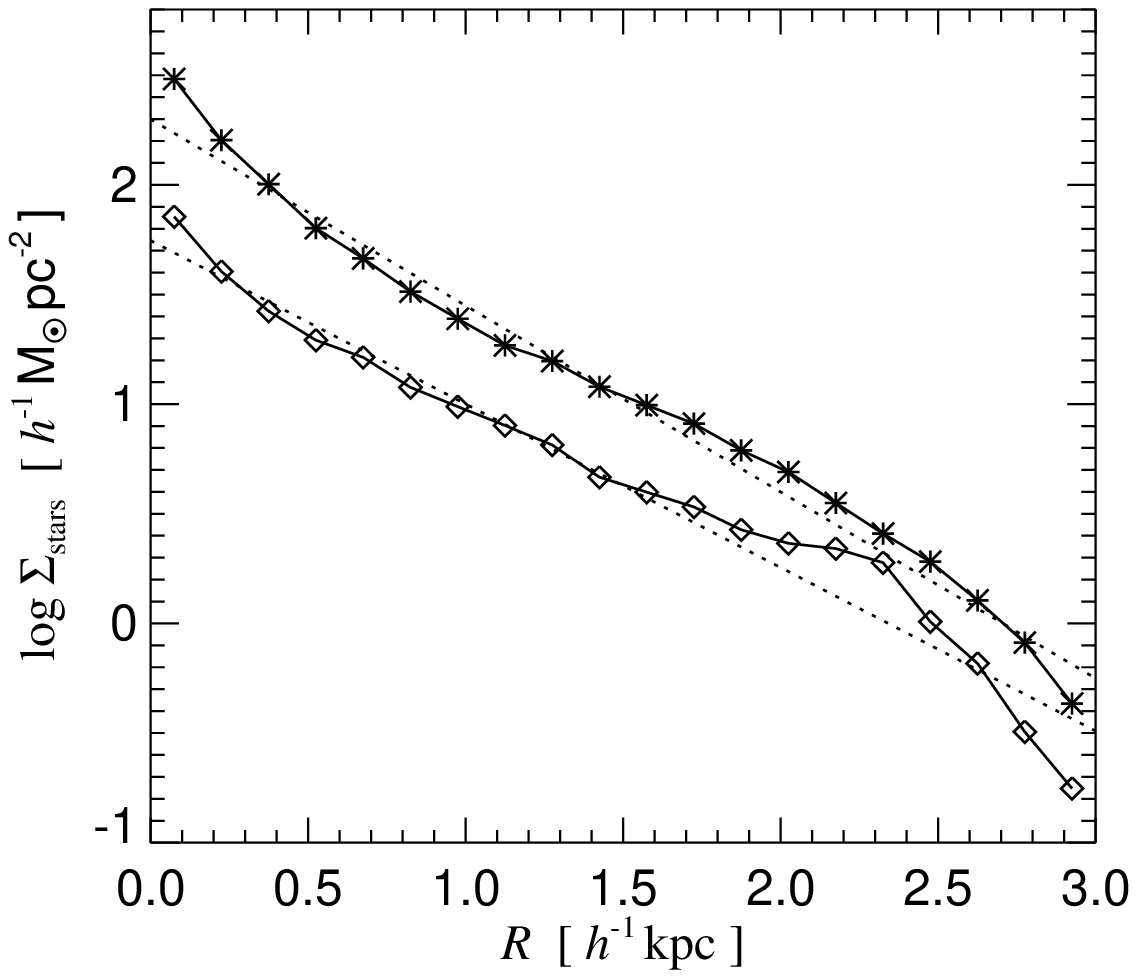}}%
\resizebox{8cm}{!}{\includegraphics{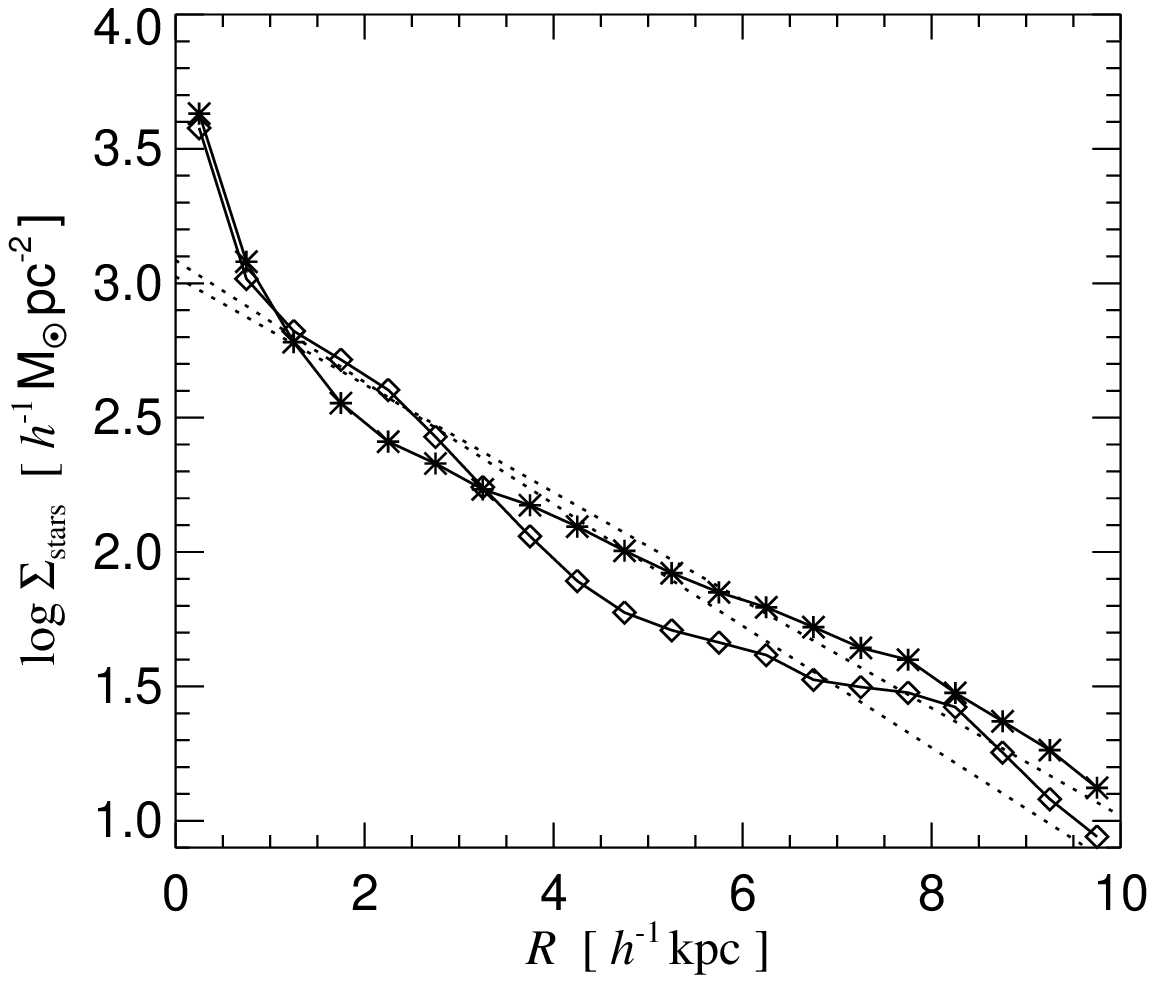}}\\%
\caption{Radial surface brightness profiles for disks that formed in
isolated haloes after $3\,{\rm Gyr}$.  On the left, we show results
for a halo of total mass $10^{10} \, h^{-1}{\rm M}_\odot$, both for a
run without winds (stars) and one with winds (diamonds). The dotted
lines are exponential profiles, fitted under the constraint that they
give the same total light as the entire simulated disk. The
exponential scale lengths of the disks are $0.51$ and
$0.58\,h^{-1}{\rm kpc}$, respectively.  On the right, the same results
are shown for a halo of total mass $10^{12} \, h^{-1}{\rm M}_\odot$,
with the exponential fits giving scale lengths $2.1$ and
$1.9\,h^{-1}{\rm kpc}$, respectively.
\label{figdiskprofile}}
\ec
\end{figure*}

\begin{figure*}
\bc
\hspace*{-0.3cm}\resizebox{10.5cm}{!}{\includegraphics{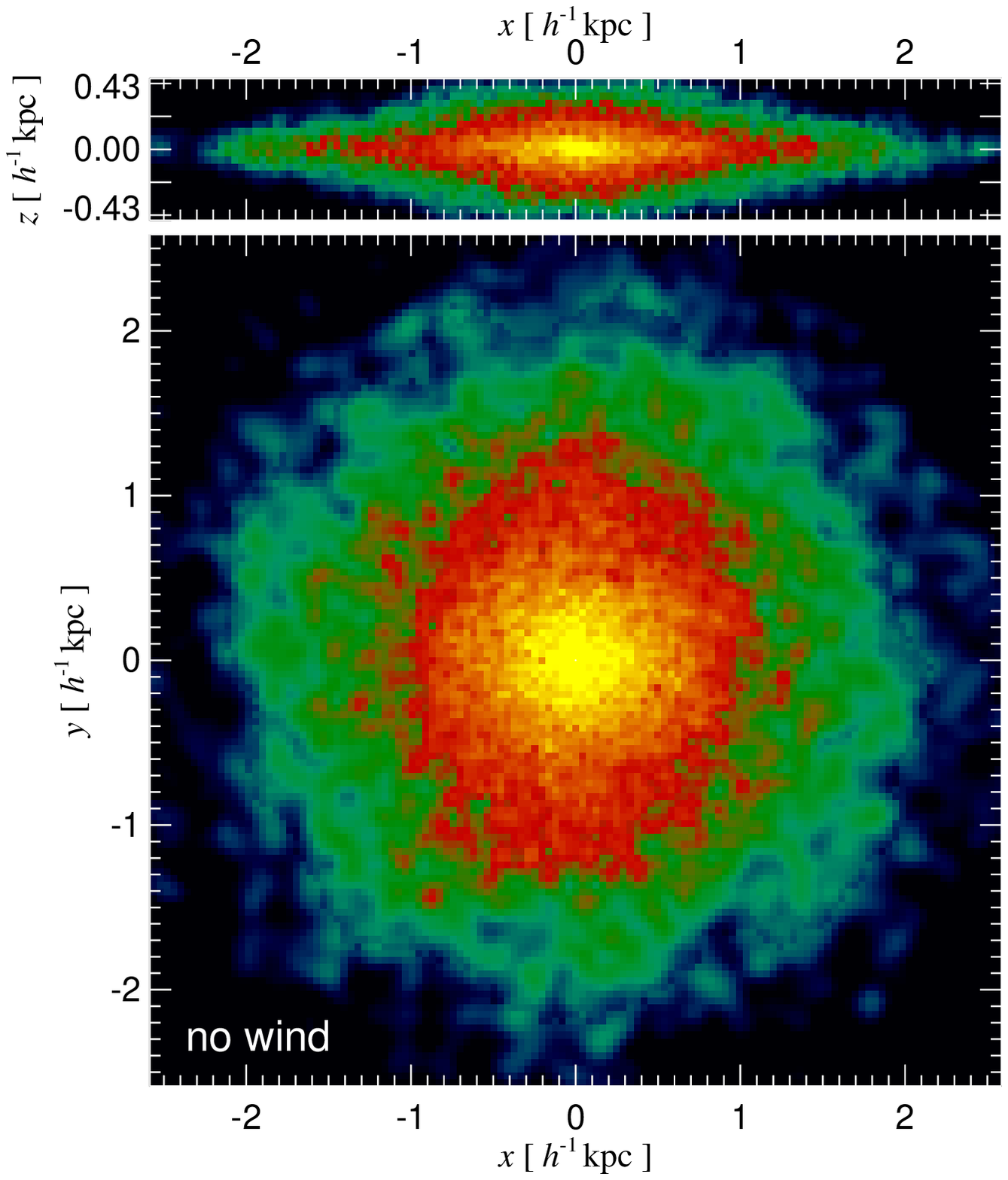}}%
\hspace*{-2.4cm}\resizebox{10.5cm}{!}{\includegraphics{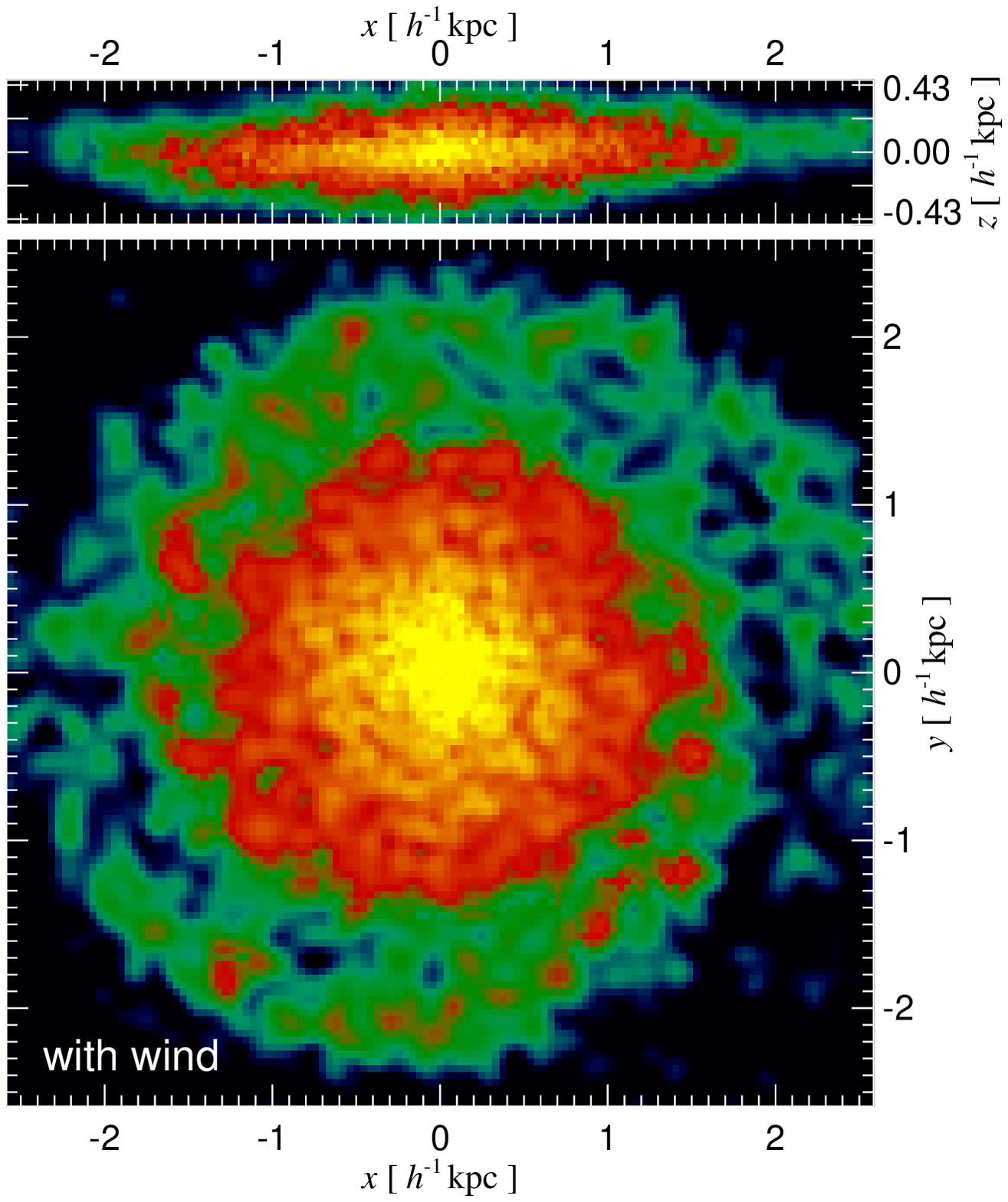}}\\%
\caption{The stellar disk that formed after $3\,{\rm Gyr}$ in a halo
of mass $M_{\rm vir}= 10^{10} \, h^{-1}{\rm M}_\odot$, seen both face
on and edge on. On the left, the simulation does not include winds, on
the right it does. The colour-contours in the pictures have been set
such that they contain roughly the same fraction of the total light in
each case. Note that the the morphology of the two disks appears
similar, but the model that includes the wind has only 36\% of the
luminosity of the run that does not. \label{figlumdens1}} \ec
\end{figure*}

In the three wind runs discussed above, ``axial'' wind formation has
been used.  In the small-mass galaxy shown in Figure~\ref{figVfield1},
this leads to a highly collimated, bipolar outflow.  In
Figure~\ref{figVfield1iso}, we show the same model, but computed with
isotropic winds. In this case, a bipolar outflow pattern orthogonal to
the disk still develops, but due to the dense disk that forms in the
$xy$-plane and the non-isotropic infall pattern of the gas.  Overall,
this results in a gas flow that is qualitatively still quite similar
to the model with axial winds, and the resulting star formation rate
is also very similar. In practice, even in this extreme case of a thin
centrifugally supported disk, it makes little difference whether
isotropic winds or the axial model are used.

It is also interesting to study the stellar disks that develop in
these simulations. In Figure~\ref{figlumdens100}, we show the
luminosity density of the stellar disk that formed in the two $10^{12}
\, h^{-1}{\rm M}_\odot$ simulations after a time of {\rm 3\,Gyr}. As
expected, a thin stellar disk has grown. Interestingly, the disk
morphology appears nearly unaffected by whether or not a wind is
included.  In both cases, the disks are quite smooth, as are the
gaseous layers that form the stars. This is a result of the
pressurisation of the gas in our multi-phase model.  Note that if the
gas was treated as a single-phase medium, it would essentially behave
as an isothermal gas at temperature $10^4\,{\rm K}$.  The disk would
then be rather unstable to gravitational perturbations and would be
bound to quickly break up into many lumps.

The radial surface brightness profiles of the disks are shown in
Figure~\ref{figdiskprofile}.  In the main body of the galaxy, the
profile is reasonably well fitted by an exponential, but there is a
clear deviation towards higher luminosities in the centre.  This is
probably a consequence of the initial angular momentum distribution
that we adopted. For similar assumptions about the distribution of
initial angular momentum, analytic computations of disk formation
predict a similar effect \citep{vdB01a}.

Figure~\ref{figdiskprofile} also shows how the influence of the wind
depends on halo mass.  While the stellar profile for the $10^{12}\,
h^{-1}{\rm M}_\odot$ halo is nearly unaffected by the winds, the small
$10^{10}\, h^{-1}{\rm M}_\odot$ halo suffers a reduction in its
surface brightness by nearly a factor of 3. However, the radial scale
lengths are essentially unaffected, as are the disk morphologies. The
latter can be seen in more detail in Figure~\ref{figlumdens1}, where
we show the projected stellar mass densities for the low-mass halo.
In comparison to Fig.~\ref{figlumdens100}, it is clear that the ratio
of disk scale height to radial scale length is larger for the smaller
halo. This is a direct consequence of the effective equation of state
of the multi-phase medium.  The effective pressure stabilises the gas
disks vertically against gravity, but in doing so it imposes a certain
scale, i.e.~the scale height depends only weakly on surface mass
density and does not fall below $\sim 150\,{\rm pc}$ in our model.

\begin{figure}
\bc
\resizebox{8cm}{!}{\includegraphics{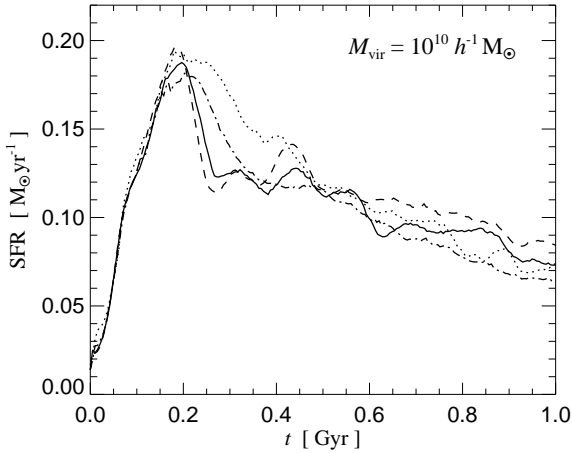}}%
\caption{Convergence study of the star formation rate in four runs
that differ in total by a factor 64 in mass resolution. The
simulations include winds and describe the formation of a disk in a
$10^{10}\,h^{-1}{\rm M}_\odot$ halo. We show results for simulations
with 40000 (solid), 160000 (dashed), 10000 (dot-dashed), and 2500
(dotted) particles.  The gravitational softening used in the
simulations was varied in proportion to the cube root of the mass
resolution, ranging from 0.03 to $0.12\,h^{-1}{\rm kpc}$.
\label{figconv}}
\ec
\end{figure}

Finally, we briefly examine the sensitivity of our numerical results
to the mass resolution employed. This is a particularly important
test, not only to assess numerical convergence in general, but also
for establishing the suitability of the method for cosmological
simulations of structure formation.  In such calculations, haloes of
different mass are invariably resolved with a different number of
particles. If the numerical results for star formation and feedback
depend sensitively on mass resolution, differential resolution effects
would be introduced that could severely compromise an interpretation
of the results.  Note that this also implies that the free parameters
of any feedback model that behaves well in this respect should be
independent of resolution.

By construction, we expect the model described in this paper to meet
these requirements, primarily because the feedback and star formation
schemes are formulated in terms of an effective subresolution model
which does not require an explicit reference to the particle
formalism. We test this contention by repeating the wind-simulation of
the $10^{10}\,h^{-1}{\rm M}_\odot$ halo at a number of different mass
resolutions, both with smaller and larger particle numbers than the
40000 used above. In particular, we selected mass resolutions of 2500,
10000, and 160000 particles, thereby obtaining a series of runs that
stretches a dynamic range of 64 in mass resolution.  We also varied
the 3D spatial resolution (i.e. the gravitational softening length) in
proportion to the cube root of the mass resolution, as it is typically
done in cosmological simulations of structure formation.  In
Figure~\ref{figconv}, we compare the star formation rates measured as
a function of time in these simulations. Reassuringly, the agreement
is rather good between all four runs, confirming that the model
outlined above is numerically well posed. The formulation in terms of
an effective sub-resolution model has a well-specified continuum
limit, and we do not expect the results of this test to change even
when the resolution is increased indefinitely. This means, however,
that much higher resolution would not begin to eventually resolve the
``true'' spatial structure of the ISM -- in order to do this
faithfully, the effective model would have to be dropped.

\subsection{Cosmic star formation history}

We now employ our multi-phase model in full cosmological simulations
of structure formation.  In this paper, we restrict ourselves to
highlighting some of the main consequences of the model with respect
to the overall star formation rate and the metal enrichment of the
IGM.  For these purposes, it is sufficient to consider moderate-sized
simulations of a relatively small volume. In particular, we work with
$2\times 50^3$ dark matter and SPH particles in a box of length
$11.3\,h^{-1}{\rm Mpc}$ per side. This gives a mass resolution of
$m_{\rm gas}=1.28\times 10^8 \, h^{-1}{\rm M}_\odot$ in the gas and
$m_{\rm dm}=8.33\times 10^8 \, h^{-1}{\rm M}_\odot$ in the dark
matter, for a canonical $\Lambda$CDM cosmology with parameters
$\Omega_0=0.3$, $\Omega_\Lambda=0.7$, $\Omega_b=0.04$, $h=0.67$, and
$\sigma_8=0.9$. We have run several of these simulations for different
parameter choices, and in addition one run where we stepped up the
mass resolution by a factor of eight to $2\times 100^3$ particles.

\begin{figure}
\bc
\resizebox{8cm}{!}{\includegraphics{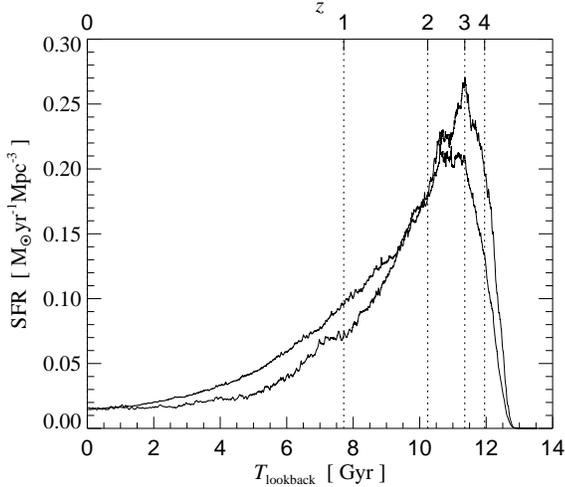}}%
\caption{Comoving star formation history of two small cosmological
simulations, using the hybrid multi-phase model without winds for star
formation timescales $t_0^\star = 1\, {\rm Gyr}$ and $t_0^\star =
4.2\, {\rm Gyr}$, respectively. For a smaller value of $t_0^\star$,
the peak of the star formation rate moves to higher redshift, and the
decline towards the present epoch proceeds faster. However, the
integrated mass of stars that forms up to $z=0$ is insensitive to
$t_0^\star$: the short-timescale model turns 18.2\% of the baryons
into stars, the long-timescale model 18.4\%. \label{figsfrtime}} \ec
\end{figure}

Note that the simulation volume in these runs is much too small to be
representative of the universe as a whole, and in fact, the largest
modes in the box become non-linear by $z=0$.  Nevertheless, the
simulations suffice to identify the systematic effects that arise from
our feedback schemes, and from the winds, in particular.  However, it
should be kept in mind that simulations of larger size with comparable
or better mass resolution are required to arrive at quantitatively
accurate results for cosmological expectation values such as the mean
luminosity density.

In Figure~\ref{figsfrtime}, we show the cosmic star formation history
for two of our wind-less $2\times 50^3$ simulations, differing only in
the value adopted for the star formation timescale $t^\star_0$. While
a larger value of this parameter shifts star formation towards later
times, the integrated star formation history is insensitive to the
value of $t^\star_0$. Ultimately, if only the quiescent model of star
formation is considered, the mass of stars that forms is approximately
given by the amount of baryons that can cool, because the gas
consumption timescale is significantly shorter than the age of the
universe. It is, however, well known that cooling alone is quite
efficient
\citep[e.g.][]{Whi78,Wh91,Ka93,Pea99,Lew00,Dave2001,Bal01,SprHe01},
leading to a collapse fraction well in excess of observational bounds
on the amount of baryons in stars and cold gas. For example,
\citet{Fuk98} use optical data to constrain the fraction of baryons in
the form of cold gas or stars to be in the range of 6.2\% to
16.7\%. \citet{Bal01} derive an even lower value of about 5\%, based
on the $K$-band luminosity function \citep{Cole2001}, which should in
principle provide a more robust estimate due to the relatively weak
dependence of the $K$-band mass-to-light ratio on star formation
history. However, some uncertainty arises from the adopted IMF; as
\citet{Cole2001} discuss, an estimate of the mass fraction in stars
higher by about a factor of 2 is obtained if a Salpeter IMF
(\citeyear{Sal55}) instead of a Kennicutt (\citeyear{Ke83}) IMF is
assumed. Note that there are also tight limits on the amount of gas in
neutral and molecular form, implying that they together can only
account for at most 10\% of the mass bound in stars. Hence, most of
the condensed gas should in fact be bound in stars, and there is not
much room to ``hide'' gas that has cooled.

\begin{figure}
\bc
\resizebox{8cm}{!}{\includegraphics{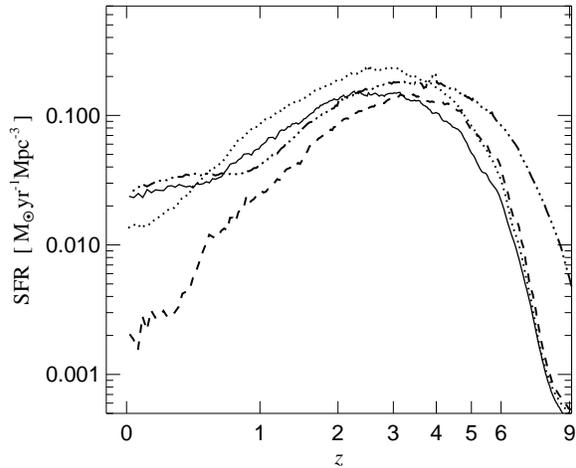}}\\%
\caption{Evolution of the cosmic star formation density for different
simulations.  The solid line gives the star formation rate per
comoving volume for our fiducial $2\times 50^3$ simulation, with
parameters $t_0^\star=2.1\,{\rm Gyr}$, $\eta=2$, and $\chi=0.25$. The
dot-dashed line is for the same model, but at a higher resolution of
$2\times 100^3$.  The dashed line shows a model with a much more
energetic wind, specified by $\eta=0.5$ and $\chi=1$, while the dotted
line shows the result for a run without winds.
\label{figcosmicsfr} } \ec
\end{figure}

The discrepancy between the high efficiency of cooling and the low
abundance of stars suggests that any feedback model lacking a
mechanism for returning baryons from the condensed phase into the
extended halo or IGM will be unable to account for the
observations. For example, in the no-wind runs shown in
Fig.~\ref{figsfrtime}, more than 18.2\% of the baryons are locked up
in stars, which is already above current observational constraints,
even though this number represents an underestimate due to the limited
resolution of these simulations, and it does not yet include the
dense, neutral gas remaining in the simulation.

However, winds are able to significantly affect the cosmic star
formation history. In Figure~\ref{figcosmicsfr}, we show simulation
results for the comoving star formation density as a function of
redshift for different wind strengths. The dotted line shows a
$2\times 50^3$ run using our fiducial set of parameters and the
quiescent model of star formation without winds.  The solid line is
the same model, but this time including winds with a strength
specified by $\eta=2$ and $\chi=0.25$, i.e.~star-forming regions blow
a wind with speed $242\,{\rm km\,s}^{-1}$ at a mass-loss rate equal to
twice the star formation rate. This reduces the integrated star
formation rate such that 13.5\% of the baryons become bound in stars,
providing a better match to direct observational constraints of the
cosmic luminosity density, although perhaps still being uncomfortably
high.

Faster winds can escape from bigger haloes and provide larger heating
to the surrounding IGM. This can make them more effective in reducing
subsequent star formation, even if the mass-loss rate is smaller.  For
example, the dashed line in Fig.~\ref{figcosmicsfr} shows a rather
extreme wind model, with $\eta=0.5$ and $\chi=1$. The wind has thus an
initial speed of $968\,{\rm km\,s}^{-1}$, but a mass-loss rate four
times smaller than in our fiducial model. This energetic wind leads to
substantial heating of the IGM, and reduces the star formation rate
considerably, such that this model turns only 8\% of its baryons into
stars.

Finally, we also show the result for a $2 \times 100^3$ simulation of
our fiducial model. Here, the star formation rate is substantially
larger at high redshift. This is expected because the improved
resolution now allows much more of the star formation to be seen in
low-mass objects that begin to form abundantly at high redshift. It is
thus clear that much better resolution than was used in the $2\times
50^3$ runs is required to obtain converged results in the
high-redshift regime. However, this is not necessarily the case for
the integrated star formation rate. Since the elapsed time is rather
small at high redshift, the amount of stellar material formed then is
small compared to the total, despite the large star formation
rates. Therefore, the $2\times 100^3$ run produces only moderately
more stars by $z=0$; the fraction of baryons is 14.8\%, up from 13.5\%
for the corresponding $2\times 50^3$ run. Note that the slight upturn
seen in the $2\times 50^3$ and $2\times 100^3$ runs with slow winds at
redshifts below $\simeq 0.5$ is due to the small box size of these
simulations, which do not allow a proper sampling of the halo mass
function at low redshift, where in fact the fundamental modes of the
boxes already become non-linear. In particular, the star formation
rate begins to be dominated by the few most massive halos in the box,
and they are able to reaccrete at low redshift some of the gas ejected
by winds out of their progenitor halos. If larger cosmological volumes
are simulated, such biases in the estimated mean of the cosmic star
formation density can be avoided \citep{SprHerSFR}.

\begin{figure}
\bc
\resizebox{8cm}{!}{\includegraphics{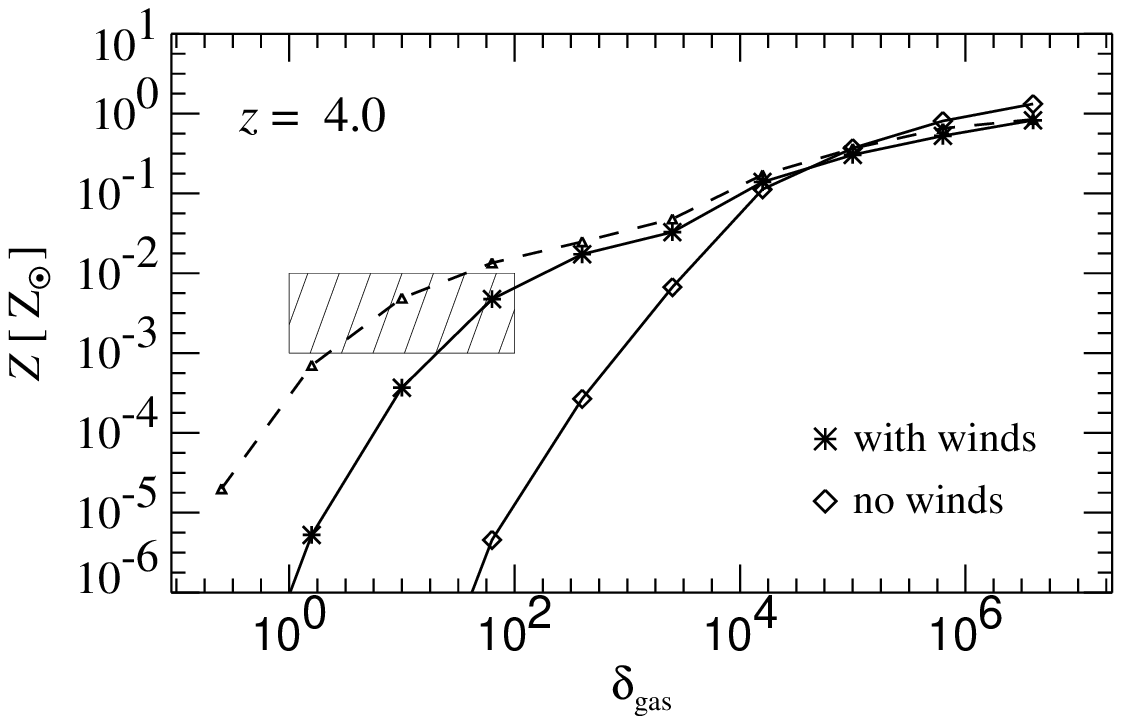}}\\%
\resizebox{8cm}{!}{\includegraphics{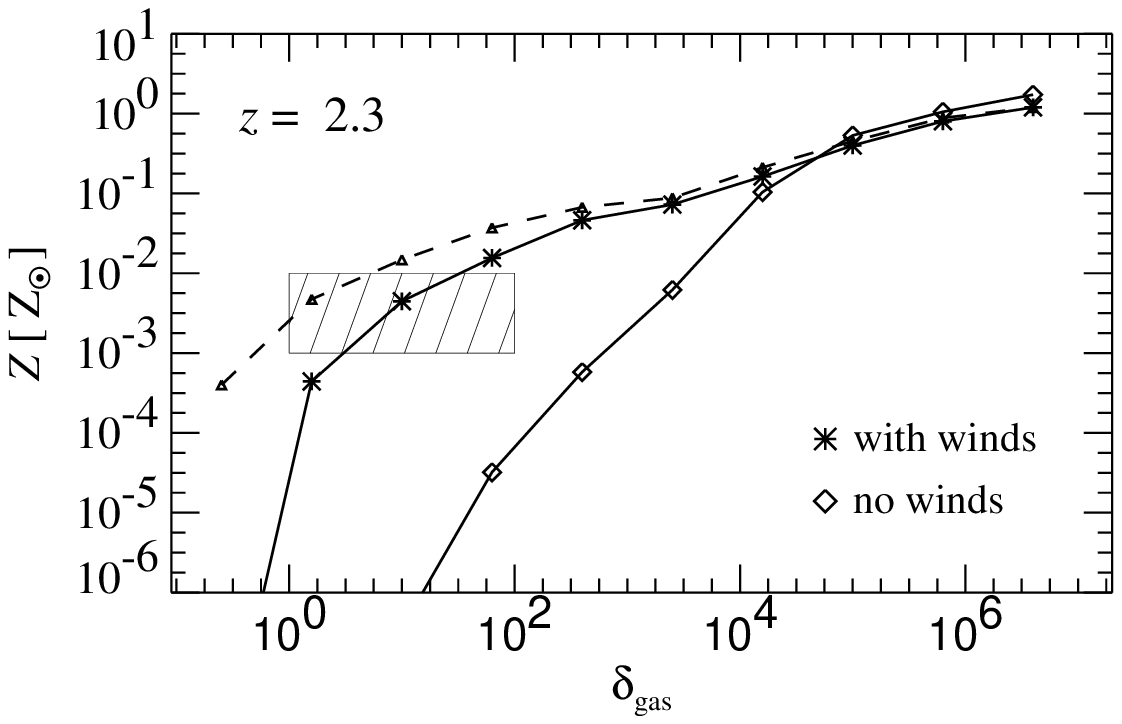}}\\%
\resizebox{8cm}{!}{\includegraphics{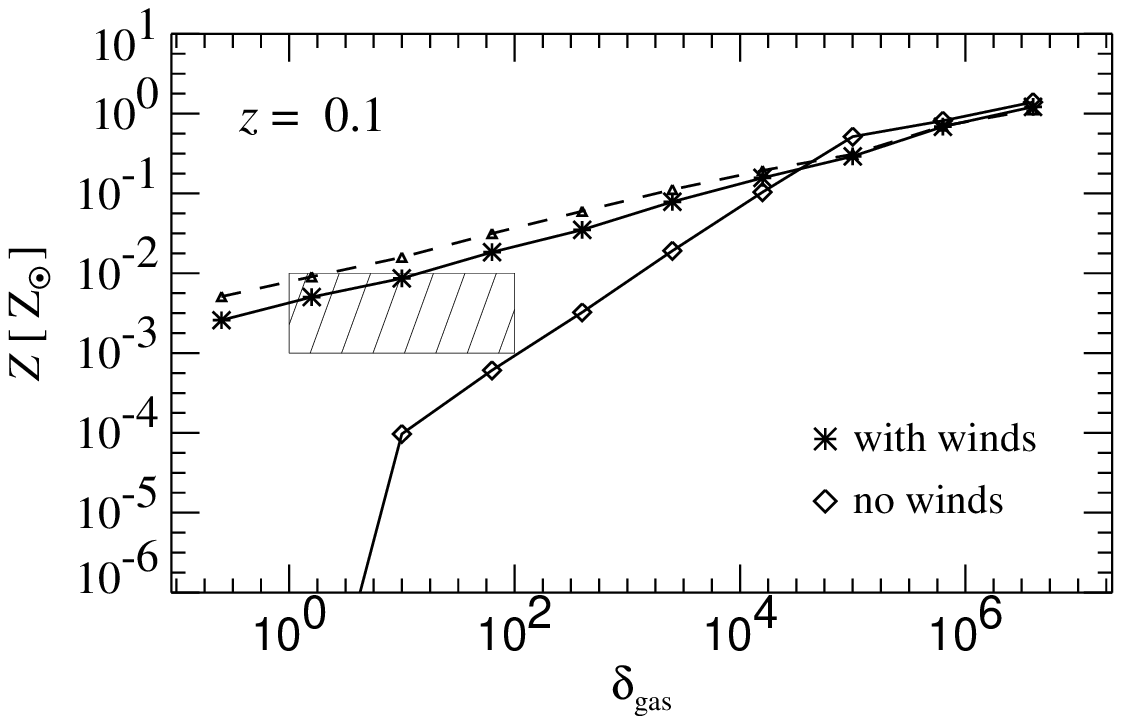}}%
\caption{Mean metallicity of the gas as a function of baryonic
overdensity. The three panels show results for redshifts $z=4$,
$z=2.3$, and $z=0$, both for simulations with and without winds. We
follow \citet{Agui01c} and use a shaded region to approximatively
indicate the range of metallicities observed in Ly-$\alpha$ absorption
line studies at $z\simeq 3$.
\label{figenrichment}}
\ec
\end{figure}

\subsection{Metal enrichment of the IGM}

As we discussed in Section~4, galactic winds may be of crucial
importance for the transport of metals into the IGM, where they have
been observed in absorption line systems down to very low column
densities.  In Figure~\ref{figenrichment}, we show the mean
metallicity of the gas as a function of baryonic overdensity in our
cosmological simulations. We give results at redshifts $z=4$,
$z=2.33$, and $z=0$, for runs carried out with our fiducial
parameters, both with and without winds.

\begin{figure*}
\bc
\resizebox{7.3cm}{!}{\includegraphics{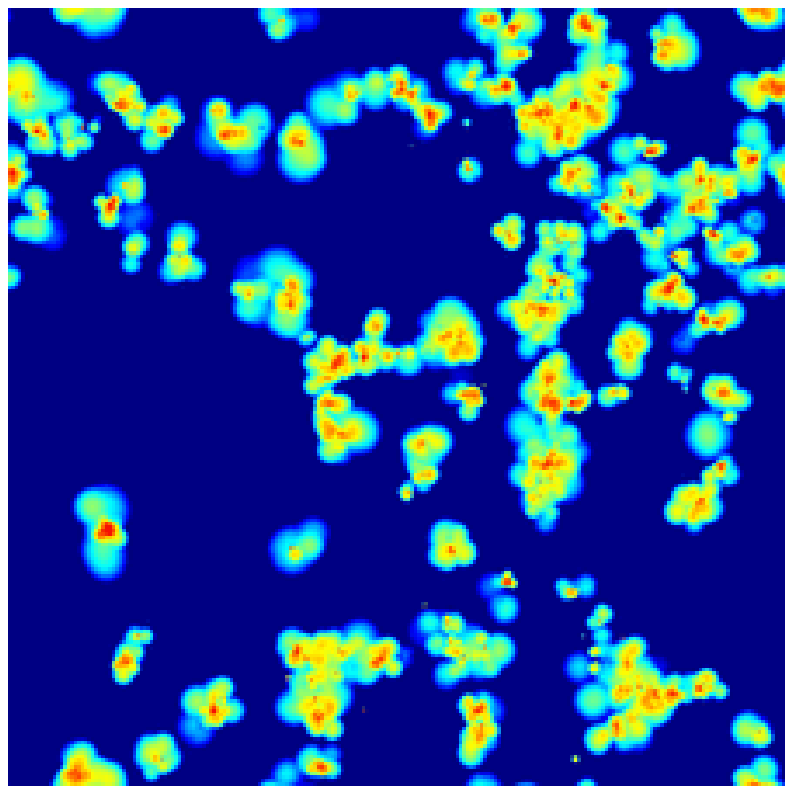}}\hspace{0.3cm}%
\resizebox{7.3cm}{!}{\includegraphics{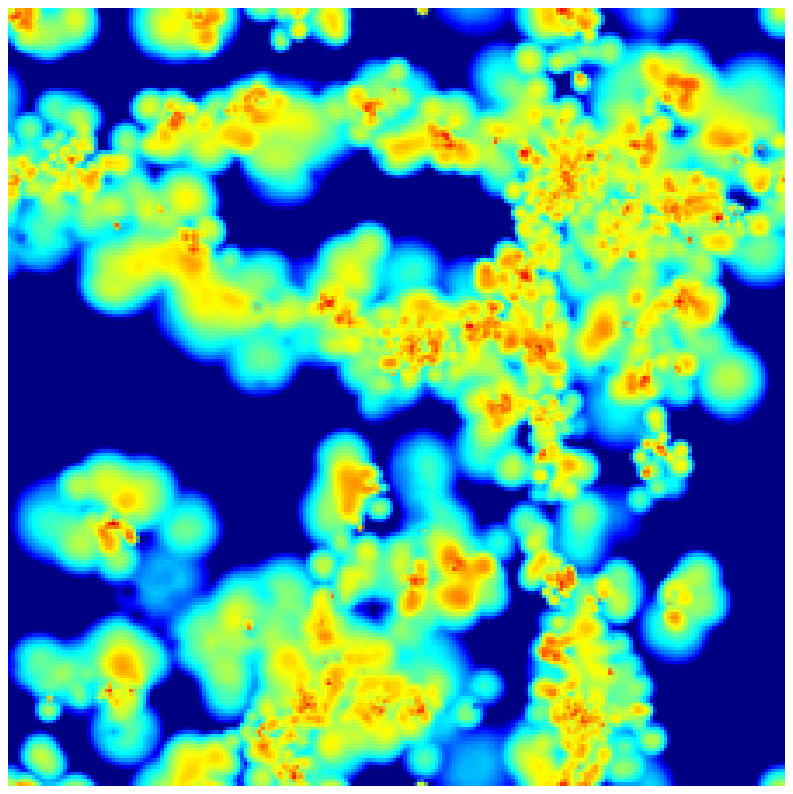}}\vspace{0.3cm}\\%
\resizebox{10.5cm}{!}{\includegraphics{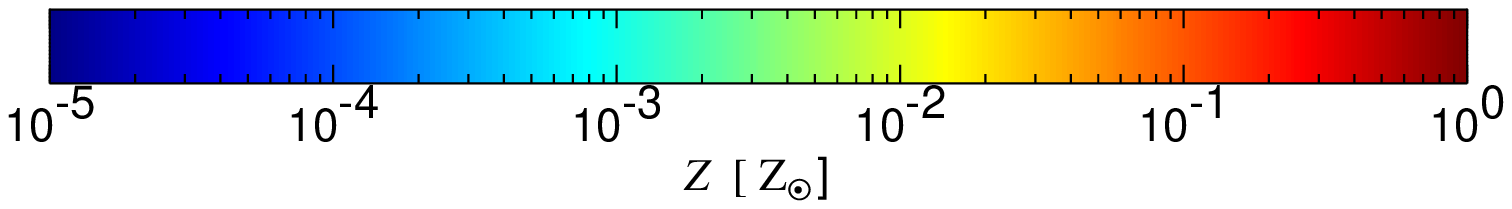}}\vspace*{-0.3cm}\\%
\caption{Projected mean metallicity of the gas in a $2\times 50^3$
simulation that includes galactic winds. The map is $11.3\,h^{-1}{\rm
Mpc}$ on a side and shows the full simulation box in projection at
redshift $z=2.3$ (left) and $z=0$ (right).  The mean metallicity of
the gas is indicated by a logarithmic colour-scale.
\label{figmetalmap}}
\ec
\end{figure*}

It is evident that in the simulation without winds, metals are nearly
confined to gas with overdensities $\delta>10$. At redshift $z=0$,
such gas reaches a metallicity of about $10^{-4} {\rm Z}_\odot$, where
${\rm Z}_\odot$ is the solar metallicity. However, at redshifts higher
than $z>2.3$ even gas at overdensities 100 is well below $10^{-4} {\rm
Z}_\odot$, and thus falls significantly short of the metallicities
that are observed in the Ly-$\alpha$ forest at these redshifts, as
indicated by the shaded region \citep[following][]{Agui01c}. Recall
that in our self-regulated model for star formation, stars begin to
form only at high physical densities, corresponding to baryonic
overdensities of about $\sim 10^6$ at $z=0$. If winds are not
included, only dynamical stripping may bring metal-enriched gas from
the ISM to the lower density environments. The results of
Fig.~\ref{figenrichment} suggest that these processes are not
efficient enough to explain the enrichment of the IGM.

However, including winds changes the enrichment pattern significantly,
as expected. Our fiducial model enriches the IGM to an interesting
metallicity of $Z\simeq 10^{-2.5}\,{\rm Z}_\odot$ for the relevant
densities. It is thus clear that a wind model like the one discussed
here can in principle account for the mean metallicity of the IGM.

Note that the distribution of metals in the gas is highly
inhomogeneous. This is seen in Figure~\ref{figmetalmap}, where we show
projected metallicity maps at redshifts $z=2.3$ and $z=0$. The
detailed metallicity distribution together with the inhomogeneous
heating pattern due to the winds potentially yield signatures in the
Ly-$\alpha$ forest that may be very constraining for the enrichment
model discussed here. We plan to address this question further in
future work. Note that at low velocities, winds will primarily be able
to escape from small haloes, pushing the epoch of enrichment of the
IGM to high redshift, when these haloes form abundantly, and when the
winds can propagate relatively far because of the small scale-size of
the universe at the epoch of ejection.

\section{Discussion}

We have presented a new model for the treatment of star formation and
feedback in cosmological simulations of galaxy formation.  Our approach
leads to the establishment of 
a tight self-regulation cycle for star formation which is
based on a rough, yet physically motivated model of the ISM. A crucial
ingredient of the model is the assumed multi-phase structure of the
ISM. At high density, much of the interstellar material becomes bound
in cold clouds, becoming sites for star formation. The density of the
ambient phase is lowered accordingly, which also reduces its radiative
losses and allows it to be heated by supernovae to a temperature high
enough to provide some pressure support for the ISM. The stabilising
effect of this pressure counteracts star formation and, together with
the cloud formation/evaporation cycle, leads to self-regulation.  Due
to the nature of the approximations made, it is clear that our model
is still phenomenological to a large degree.  In principle, however,
the physical approximations we used can be refined in the future,
thereby improving the faithfulness of the model.

The star formation timescale in the quiescent mode of star formation
can be directly determined from observations of local disk galaxies.
If this normalisation is adopted, we find that cosmological
simulations nevertheless lead to an overproduction of stars.  We have
invoked galactic winds as a heuristic extension of our model to remedy
this problem. While there is mounting observational evidence for the
existence of such winds, they are introduced in our model in a
phenomenological manner. However, winds clearly have a range of highly
interesting consequences. They reduce the global efficiency of star
formation, and this suppression is particularly strong in low-mass
haloes, thereby helping to explain the observed ``low'' values of the
luminosity density. Winds are also capable of accounting for the
enrichment of the low-density IGM with metals, and they may be crucial
to understanding the metal distribution in the intra-cluster gas of
clusters of galaxies.

Another important property of our technique is that it is numerically
well posed in the sense that the parameters of the model can be
determined directly based on physical arguments or observational
input, and need not to be changed when the mass resolution is
varied. The use of the sub-resolution technique introduces a
well-specified continuum limit in the hydrodynamic equations, despite
the inclusion of cooling.  Unlike in standard single-phase
simulations, an unphysical collapse of gas to a scale set by the
resolution limit of the simulation is thus prevented.  Combined with
improved formulations of SPH \citep{SprHe01} this makes it much easier
to reach numerical convergence even under conditions of only moderate
resolution.

It will thus be interesting to study the predictions of our model in
more detail.  As first steps, we have already carried out a detailed
analysis of the predictions of the model for the cosmic star formation
history in the $\Lambda$CDM cosmology \citep{SprHerSFR}, and we
analysed the impact of cooling, star formation, and winds on secondary
anisotropies of the cosmic microwave background \citep{WhiHerSpr02}.
In these studies, a new comprehensive set of simulations that included
strong galactic winds was used.  We plan to extend our analysis of
these simulations in future work.

\section*{Acknowledgements}

We are grateful to Simon White for highly useful comments on this
paper. This work was supported in part by NSF grants ACI 96-19019, AST
98-02568, AST 99-00877, and AST 00-71019.  The simulations were
performed at the Center for Parallel Astrophysical Computing at the
Harvard-Smithsonian Center for Astrophysics.

\bibliographystyle{mnras}
\bibliography{paper}

\end{document}